\documentclass[11pt,a4paper]{article}
\usepackage[utf8]{inputenc}\usepackage[T1]{fontenc}\usepackage{lmodern}\usepackage{microtype}
\usepackage[margin=2.5cm]{geometry}\usepackage{graphicx,longtable,booktabs,array,amsmath,amssymb,seqsplit,caption}
\usepackage[super,sort&compress,comma]{natbib}\usepackage[hidelinks]{hyperref}\usepackage{url}
\renewcommand{\figurename}{Fig.}
\title{Language-model agents map and edit the weights of climate emulators}
\author{Dmitrii Pantiukhin, Ivan Kuznetsov, Kacper Nowak, Boris Shapkin, Nikolay Koldunov\\[4pt]
{\small Alfred Wegener Institute, Helmholtz Centre for Polar and Marine Research, Bremerhaven, Germany}}
\date{}
\begin{document}
\maketitle
\begin{abstract}
Machine-learning weather and climate emulators run far faster than physics-based models, yet some long runs drift, develop biases or fail. Released weights are fixed, and retraining them requires computing resources. Here we map and edit model weights directly to reduce such errors. Under human supervision, language-model agents ran tests, recorded predictions beforehand and audited each other. Across 20 model components, scaling the original weights of a typical group changed the first forecast step more than a random change of the same size in 9, about as much in 6 and less in 5. Seven ACE-family atmosphere models shared one broad sensitivity ranking. Three targeted edits reduced instability or bias. Unedited 1.4$^\circ$ stochastic NeuralGCM failed after about four months in our initial runs; edited upper-level wind weights, together with the developers' pressure correction, prevented breakdown across 5--11-year runs. DLESyM v1.0 with its default atmosphere and ocean networks had an Arctic warm bias in our runs; edited ocean weights lowered its global near-surface temperature error by 35--38~\% across five multi-year runs within the training period. In Pangu-Weather, editing the output-layer weights that produce near-surface temperature reduced the mean cold bias after 60 days from $-$2.09 to $-$0.78 K across 24 training-period initial conditions. These edits are targeted fixes for specific errors, not general model improvements. Yet they show that emulator weights can be measured, tested and adjusted, a step towards explaining how these models work.
\end{abstract}

\section{Introduction}

Machine-learning emulators now simulate the atmosphere, ocean and coupled climate far faster than physics-based general circulation models. ACE2 reproduces atmospheric variability on timescales from days to decades\cite{r1}. The original DLESyM ran for 1,000 years with near-zero global drift\cite{r2}. NeuralGCM pairs fluid-flow equations (a dynamical core) with neural-network small-scale processes (learned physics)\cite{r3}. With sea-surface temperatures and sea ice prescribed, 22 of 37 simulations with the deterministic 2.8$^\circ$ NeuralGCM stayed stable for 40 years\cite{r3}. Their value for climate science therefore rests on how they behave over years and decades.

Some emulators become numerically unstable in long runs\cite{r3,r4,r5}. Even in 10-day forecasts, some emulators drift colder or warmer than their reference climate\cite{r6}. Many emulators are trained on reanalyses such as ERA5\cite{r7,r8}, which reconstruct past weather from observations and models. In a recent multi-model intercomparison, atmosphere emulators reproduced many historical climate features\cite{r8}. Some underestimated historical warming trends. When sea-surface temperatures were raised uniformly, the models gave different land-temperature responses, including cooling in some cases\cite{r8}. Which errors appear depends on the model version and setup.

In physics-based models, developers can relate errors to physical processes and tune uncertain parameters while checking the effect on the model as a whole\cite{r9}. An emulator instead stores its behaviour across 10$^{6}$ to 10$^{9}$ weights trained together, and the weights themselves carry no physical labels. Once trained, these saved weights and settings are released as a fixed model checkpoint. Retraining requires the training data, code, computing time and developer expertise. In recent years, researchers have edited the weights of trained language models directly\cite{r10,r11,r12}. For emulators, it is not yet known how changing a group of weights affects forecasts and climate, so it is unclear which weights to edit.

Earlier studies examined the signals passed between layers in GraphCast\cite{r13,r14,r15,r16} and Aurora\cite{r17,r18}. Two GraphCast studies also changed these signals during forecasts\cite{r14,r15}. One study trained an ACE2 model without a CO$_2$ input\cite{r1}. Another trained new models with different CO$_2$ training data and energy constraints\cite{r19}. These studies did not test edits to the weights of an already trained weather or climate emulator in multi-year runs. Researchers have edited weights to update stored facts in language models\cite{r10}, to combine tasks learned by copies of one pre-trained model\cite{r11}, and to reduce repetitive reasoning in one language model\cite{r12}. Because individual weights in an emulator carry no physical labels, we group them by network part, such as encoders, processing blocks or output layers. We therefore need to know which variables and regions respond when a group of weights is changed, and whether such changes destabilise long runs.

Compared with language models, weather and climate emulators add two further constraints. First, a model's climate emerges over years, so the change after one forecast step may not predict multi-year behaviour. Second, physical variables are coupled, so reducing an error in one variable or region can worsen another. Testing a weight edit therefore requires multi-year runs checked across variables and regions.

Here we ask whether the weights of existing weather and climate emulators can be mapped and edited directly to reduce instabilities, drifts and biases without retraining. Answering this question requires testing whether immediate effects on the first forecast step last over longer runs and whether edits harm other variables or regions. Human-directed language-model agents have been proposed for Earth-science data work\cite{r20}. We used them to probe weight groups, test edits and check results across millions to billions of weights.

\section{Results}

To see which network parts matter most, we grouped each model's weights by architectural component (weight groups) and built a weight atlas. The atlas shows how scaling each tested weight group changed the forecast and, where available, how a random change of the same size compared. In ACE2-ERA5, we mapped the temperature response to four weight groups and compared the response patterns of 28 groups across dates (Fig. 1). We surveyed 25 components across nine model families. The 26 atlas rows with complete counts total 5.38 billion weights (Supplementary Fig. 1; Supplementary Table 1; Methods, `Models and checkpoints'). Those families span ACE, SamudrACE, DLESyM, NeuralGCM, CAMulator\cite{r21}, GraphCast-small\cite{r13,r22}, Samudra2\cite{r23}, FloeNet-OM4\cite{r24} and Monthly diffusion (MD-1.5)\cite{r8,r25,r26}. In 20 of these components we compared scaling with random changes. Scaling a typical weight group by 0.9 changed the first forecast step more than a random change of the same size in 9 (all five DLESyM components, NeuralGCM 0.7$^\circ$ and 1.4$^\circ$, CAMulator and Samudra2; medians about 1.4 to 7 times as large), about as much in 6 and less in 5 (ACE and SamudrACE components, NeuralGCM 2.8$^\circ$ and GraphCast-small; Supplementary Figs. 1b and 2; Methods, `Weight scaling against random directions'). A random change of the same size is therefore a strict control in some model families and a weak one in others. Outside the atlas we tested edits in 24-hour Pangu-Weather\cite{r27}, FourCastNet v3\cite{r28} and DLWP\cite{r29}, and forecast skill in seven models including 6-hour Pangu-Weather.

\setcounter{figure}{0}\begin{figure}[tbp]\centering\includegraphics[width=\linewidth,height=0.72\textheight,keepaspectratio]{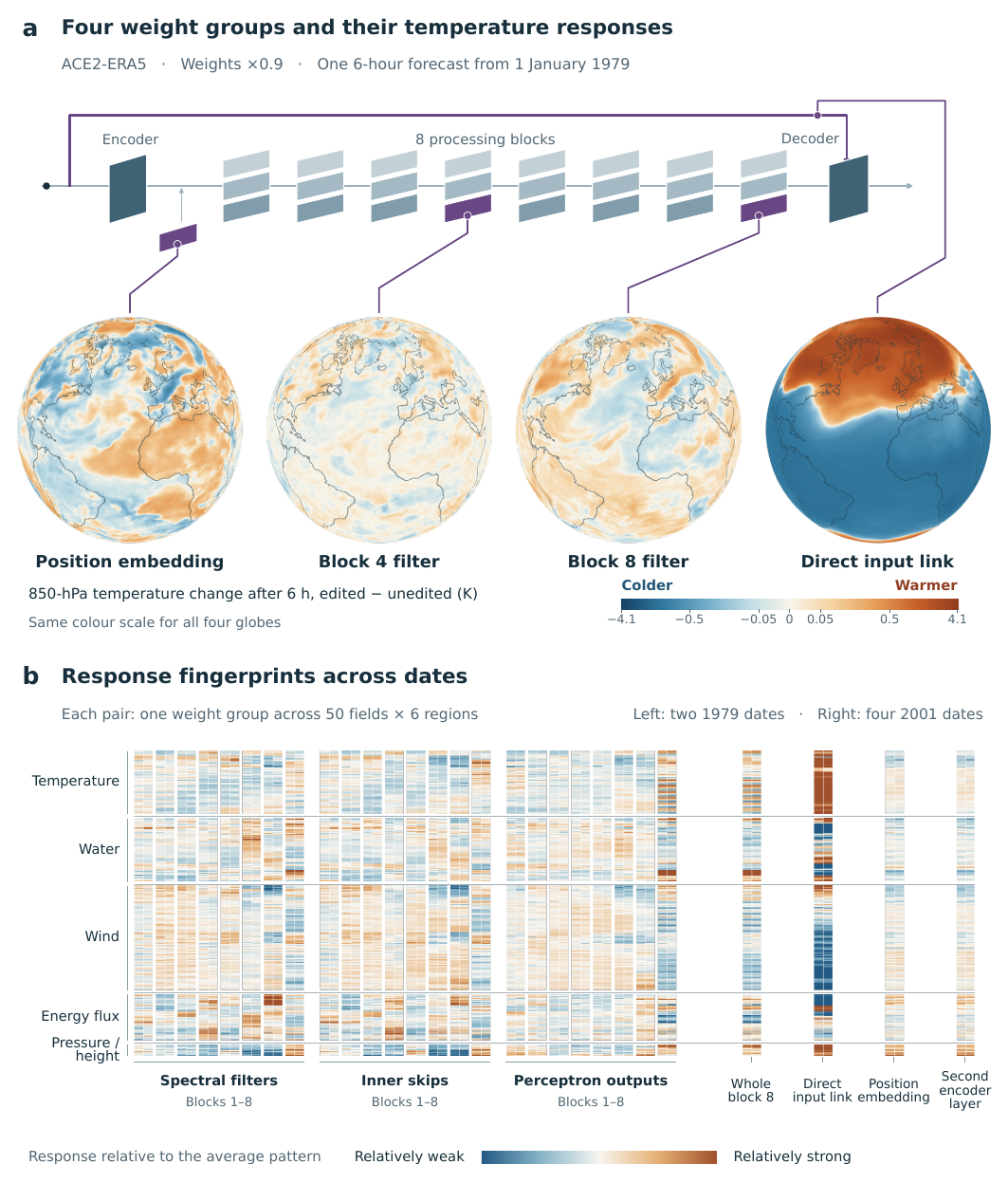}
\caption{\textbf{Forecast responses to scaling ACE2-ERA5 weight groups.} \textbf{a}, We scaled each group by 0.9 and mapped the change in 850-hPa temperature after one 6-hour step from 1 January 1979. The globes redraw the ACE2-ERA5 row of Fig. 3a on one nonlinear colour scale for all four globes ($\pm$4.1 K). Blue means the edited forecast is colder than the unedited forecast, and orange means it is warmer. \textbf{b}, Each pair of strands shows one group's response across 50 output fields and six regions at the same multiplier and lead time. We averaged normalised root-mean-square responses over two 1979 dates (left) or four 2001 evaluation dates (right), then centred their logarithms across groups and within each group. The horizontal lines separate the five classes of output field. After accounting for the pattern shared across groups, blue marks a weaker response than the group's average and orange a stronger one. All 28 groups matched their own fingerprint best across the two periods (median correlation 0.90). Fig. 2d and Supplementary Fig. 4 show the comparison at $\times$0.5. In year-long runs, four of six tested groups matched another group's fingerprint better than their own by month 1 (Supplementary Fig. 7b).}\label{fig1}\end{figure}

Because an immediate forecast change does not necessarily predict behaviour on longer timescales, we next tested whether these maps held across models and over time (Supplementary Fig. 1a). We compared forecast sensitivity across the 21 components we swept, mapped selected models in greater detail and ran year-long perturbation tests in three models. Where long runs drifted or failed, we tested targeted weight edits against random weight changes.

\subsection{The weight atlas shows which ACE2-ERA5 weights change the forecast most}

We mapped ACE2-ERA5's 6-hour forecast response on two 1979 initial conditions (1 January and 1 July). The model has an encoder, eight processing blocks and an output decoder (Fig. 2a). A shortcut, called the direct input link, passes the input fields straight to that decoder, bypassing the encoder and blocks. In each of the eight blocks, we tested its spectral filter, inner skip (a local shortcut), perceptron output (feature-mixing layer) and whole block, giving 32 weight groups. Adding the direct input link, position embedding and second encoder layer gave 35 weight groups in total. For each of the 50 normalised model output fields, we measured the area-weighted root-mean-square difference from the unedited forecast. We averaged those differences over the 50 output fields and both initial conditions (Methods, `Weight groups and multipliers'). Scaling the direct input link changed the 6-hour forecast more than scaling any other group. Along the main path, block 8, the final block before the decoder, had the largest effect. Blocks 1--7 had smaller effects that did not grow steadily from block 1 to block 7 (Fig. 2a, b). For the direct input link, forecast changes grew in proportion to the weight change (Fig. 2b; Supplementary Fig. 3c). To check the map on other dates, we used January, April, July and October 2001 initial conditions from the ACE2-ERA5 release\cite{r30}. These dates fall outside training and validation\cite{r1}. From those initial conditions, removing weights had 2.1--5.0 times the effect of doubling them in four encoder and final-block weight groups (Supplementary Fig. 3d).

\setcounter{figure}{1}\begin{figure}[tbp]\centering\includegraphics[width=\linewidth,height=0.72\textheight,keepaspectratio]{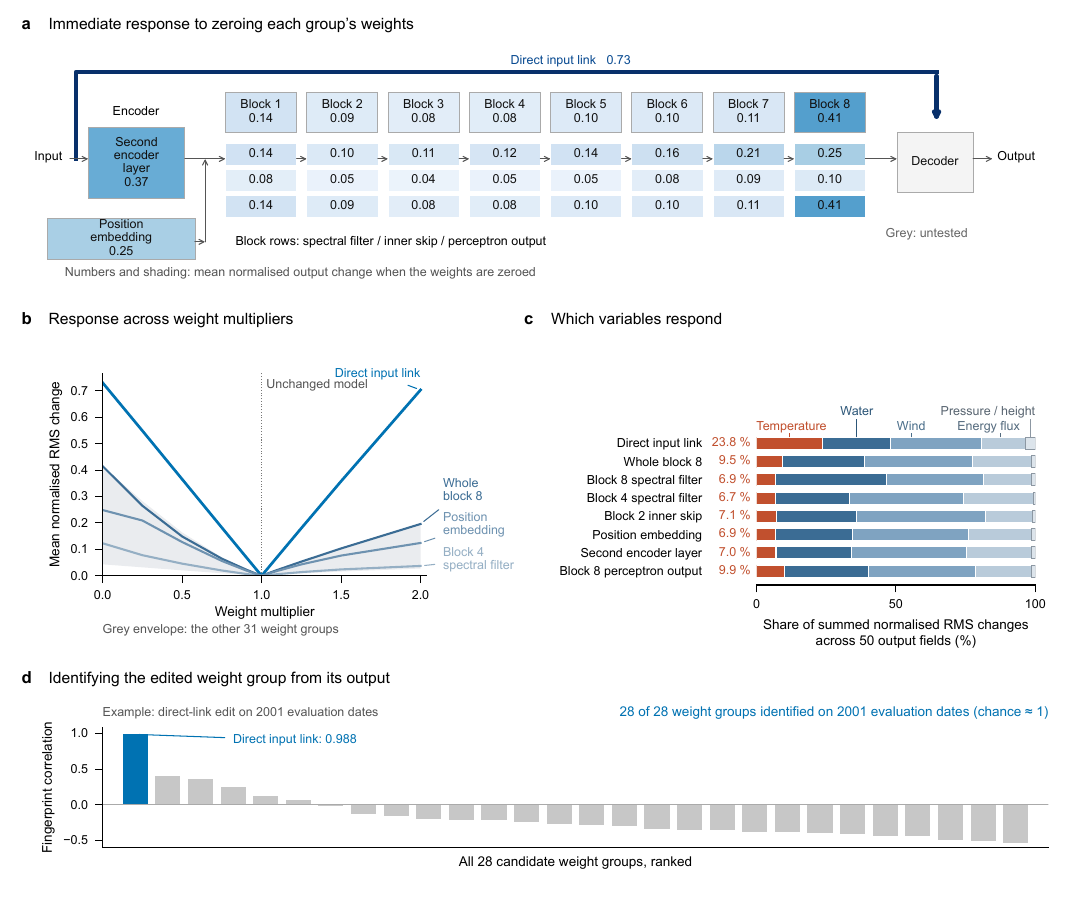}
\caption{\textbf{ACE2-ERA5's direct input link gives the largest immediate forecast change, and response fingerprints identify altered groups on 2001 evaluation dates.} \textbf{a}, Network diagram: mean normalised root-mean-square (RMS) change across 50 output fields after one 6-hour step when weight groups are zeroed (1 January and 1 July 1979; grey: untested). \textbf{b}, Mean normalised RMS change at each of 11 tested multipliers from 0 to 2. The reference at 1 is the unchanged model. Grey shading spans the other 31 groups. \textbf{c}, Share of summed normalised RMS changes across 50 fields in five classes at $\times$0.9 (temperature: 23.8~\% for direct input link versus 6.7--9.9~\% elsewhere). \textbf{d}, At $\times$0.5, fingerprints computed from responses averaged over four 2001 dates identified all 28 groups (one match expected by chance).}\label{fig2}\end{figure}

On those same four 2001 dates, all 35 weight groups kept almost the same 6-hour sensitivity order (rank correlation 0.998; Supplementary Fig. 3b). Scaling a weight group also changed the 50 output fields by different amounts across regions from the tropics to the poles after one 6-hour step (Fig. 2c). We call that field-by-region response pattern the group's fingerprint. In blocks 1--7, setting a whole block to zero gives the same response as zeroing its perceptron output, leaving 28 distinct weight groups for fingerprint matching. We compared the 6-hour response patterns from March, June, September and December 2001 against the 1979 reference fingerprints. At $\times$0.5, averaging the responses over those four dates matched all 28 altered weight groups to the right reference (Fig. 2d; Supplementary Fig. 4). At $\times$0.9 on each of those 2001 dates, the pattern still matched the right group for 26--28 of the 28 groups (Methods, `Fingerprint test on held-out states').

A repeatable 6-hour pattern still did not mean that a weight group represented a single physical process. When we applied a random change of the same size to a weight group, its 6-hour response pattern also repeated across those four 2001 dates. Yet that pattern differed from the pattern produced by scaling the original weights (Methods, `Fingerprint test on held-out states'). We could usually identify the group when the reference came from the same kind of weight change. The response pattern alone did not identify a physical process (Supplementary Table 2).

Longer forced runs showed a second limit of 6-hour maps. We ran 120-day ACE2-ERA5 simulations forced by warm El Ni\~no and cold La Ni\~na tropical Pacific sea-surface temperatures. In the La Ni\~na run, we replaced the output of one part of the network with the values that part produced in the El Ni\~no run (activation patching\cite{r31}). We measured how much of the El Ni\~no--La Ni\~na difference in tropical 850-hPa temperature (about 1.5 km altitude) each substitution reproduced. Replacing the output of the direct input link reproduced 55~\% of the difference, replacing the encoder output that feeds all eight processing blocks 38~\%, and replacing only the perceptron output of the last block 46~\%. Because the last block is one of the eight, these substitutions overlap and the percentages do not add up to a total (Methods, `Activation patching'). ACE2-ERA5 takes global-mean CO$_2$ as an input\cite{r1}. The AI Weather and Climate Model Intercomparison Project (AIMIP) Phase 1 protocol asks models to exclude CO$_2$ concentration as an input during training for its standard simulations\cite{r8}. Changing the CO$_2$ input barely affected one 6-hour ACE2-ERA5 forecast. In our ACE2-SHiELD run, holding CO$_2$ fixed reduced the 1995--2014 trend in 500-hPa geopotential by about 86~\% (Supplementary Methods, `Activation patching: details'). A 6-hour sensitivity map alone therefore could not reveal how forced responses build up over seasons and decades.

\subsection{Sensitivity rankings hold within the ACE family but differ in two ocean U-Nets}

To test whether this sensitivity pattern was peculiar to one checkpoint, we mapped six other ACE atmosphere models. For 10~\% and 50~\% weight reductions, the direct input link again ranked first despite having the fewest weights. Block 8 also had the largest effect of the eight processing blocks in all six models. Setting weights to zero broke this order in half of the models. Detailed fingerprints transferred poorly: on shared output fields, even the closest model pair matched at most 14 of 27 groups, about half (Methods, `Family transfer'). Related checkpoints of the same architecture thus share a broad sensitivity ranking, while detailed spatial patterns depend on the checkpoint.

We next compared atmosphere and ocean networks (Fig. 3). We repeated the weight scaling in the ocean component of SamudrACE-CM4\cite{r32} and in that of SamudrACE-E3SMv3\cite{r33}, each run without its atmosphere. Both are convolutional U-Nets that update each grid point from its neighbours and have no direct input link (Supplementary Fig. 5b). At $\times$0.9 (a 10~\% reduction), sensitivity was lowest in their middle stages, higher at the input and highest at the output (Methods, `Family transfer').

\setcounter{figure}{2}\begin{figure}[tbp]\centering\includegraphics[width=\linewidth,height=0.72\textheight,keepaspectratio]{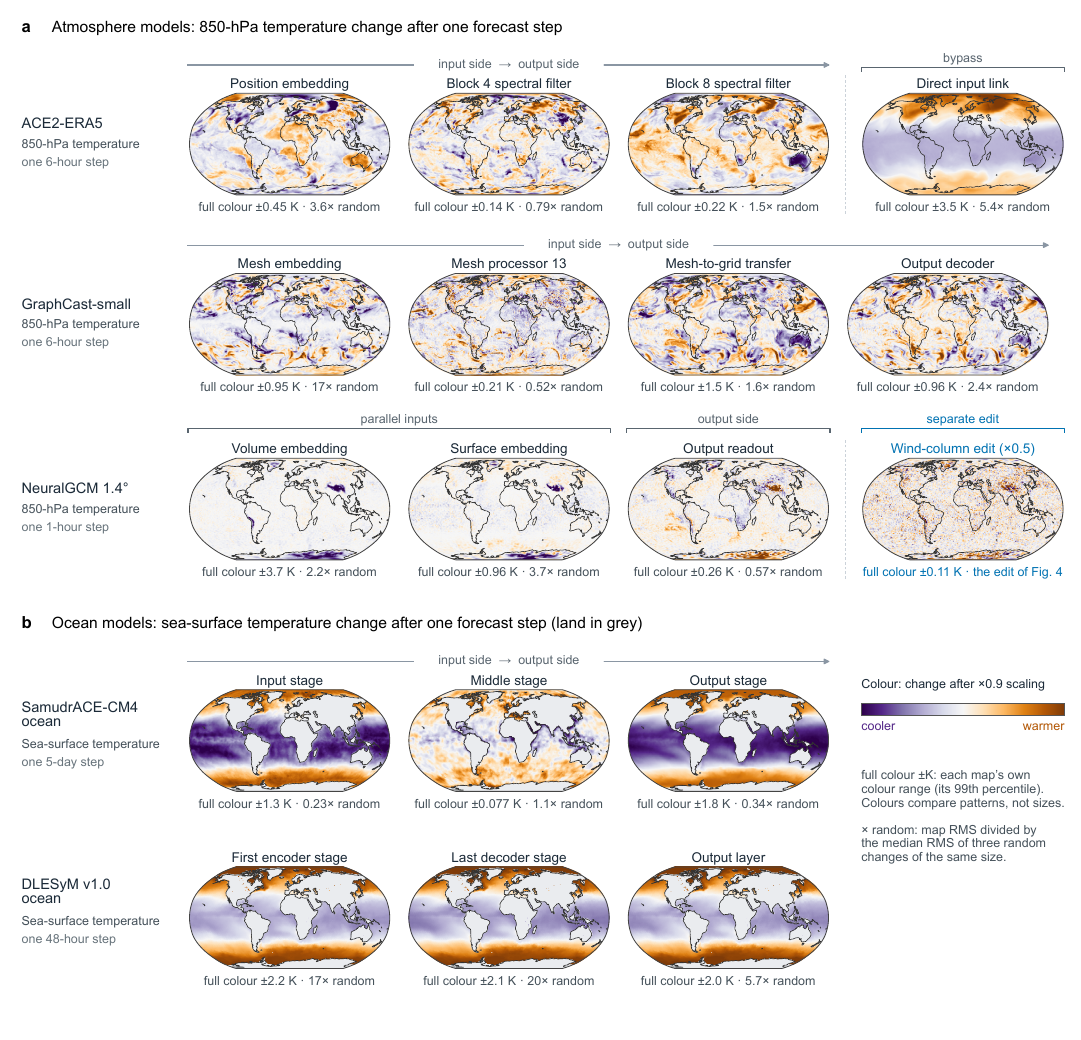}
\caption{\textbf{Scaling different network parts changes different regions of the forecast after one step across five models.} Maps show the change after one forecast step when a weight group is scaled by 0.9, from input side (left) to output side (right). \textbf{a}, 850-hPa temperature in three atmospheres. \textbf{b}, Sea-surface temperature in two oceans (grey land). Each map has its own colour scale, set by its 99th percentile and printed below in kelvin. Colours show the spatial pattern within each map. The second number is the root-mean-square forecast response to scaling divided by the median response to three random weight changes of the same size. NeuralGCM's fourth map shows the separate $\times$0.5 wind-column edit of Fig. 4 (no same-size immediate random control). Step lengths and initial conditions differ across models.}\label{fig3}\end{figure}

\subsection{Most ACE2-ERA5 response patterns change within a month}

Does the 6-hour atlas tell us what a weight edit does to the climate? To test this in ACE2-ERA5, we took six of the mapped weight groups, multiplied each by 0.9 or by 1.1, and ran the model for a year. Before the runs, we wrote down 41 predictions from the atlas. For example, we predicted that weakening the shortcut from the input to the decoder would cool the tropical lower atmosphere in the first month and that strengthening it would warm it. That prediction and 11 others came true. The atlas often predicted which fields would change most on the first day and in which direction, but it mostly failed to predict which regions and spatial scales would change most (Supplementary Figs. 6 and 7a, c; Methods, `Year-scale causal test').

The atlas also gives each edit a fingerprint: its pattern of changes across fields and regions after 6 hours. On the first day, each of the six $\times$0.9 edits (the direct input link, block 8 as a whole, the spectral filters in blocks 8 and 4, the shortcut inside block 2 and the position embedding) still matched its own fingerprint best. Averaged over the first month, four of them (block 8 as a whole, its spectral filter, the shortcut inside block 2 and the position embedding) matched another edit's fingerprint better than their own (Supplementary Fig. 7b). The change in climate itself did not fade. When we scaled each of the 35 mapped weight groups by 0.9, 33 (all except the edits to the feature-mixing output weights in blocks 4 and 5) shifted the annual mean by more than tiny perturbations of the unedited model (Supplementary Fig. 8a, b). For the four edits we followed through the year (the direct input link, block 8 as a whole, the spectral filter of block 8 and the position embedding), most regional changes seen in the first month also kept their direction and a similar size (Supplementary Fig. 7d).

Random weight changes of the same size typically shifted the annual climate about as much as scaling the original weights (Supplementary Fig. 8c). The atlas is therefore a guide to what an edit does in its first day, but a year later the climate mostly no longer shows which weights were changed.

\subsection{Weight edits reduce measured errors in three model checkpoints}

A map from one forecast step cannot tell us whether an edit will destabilise a model months later. In NeuralGCM, DLESyM and Pangu-Weather, we compared candidate edits with random weight changes of the same size or forecast effect and checked what happened to other variables and regions (Methods, `Edits and their evaluation'; Supplementary Table 3).

\subsubsection*{NeuralGCM}

The developers report that the unmodified 1.4$^\circ$ stochastic NeuralGCM is reliable for about six months\cite{r4}. Our initial runs from January 1995, with daily updates of sea-surface temperature and sea ice, broke down after about four months as winds in the upper atmosphere rapidly converged or spread apart. The network that represents small-scale physics amplified these wind disturbances. Six columns of weights in its final layer update east--west and north--south winds on the three highest model levels (Fig. 4c; Methods, `Localisation of the NeuralGCM instability').

\setcounter{figure}{3}\begin{figure}[tbp]\centering\includegraphics[width=\linewidth,height=0.72\textheight,keepaspectratio]{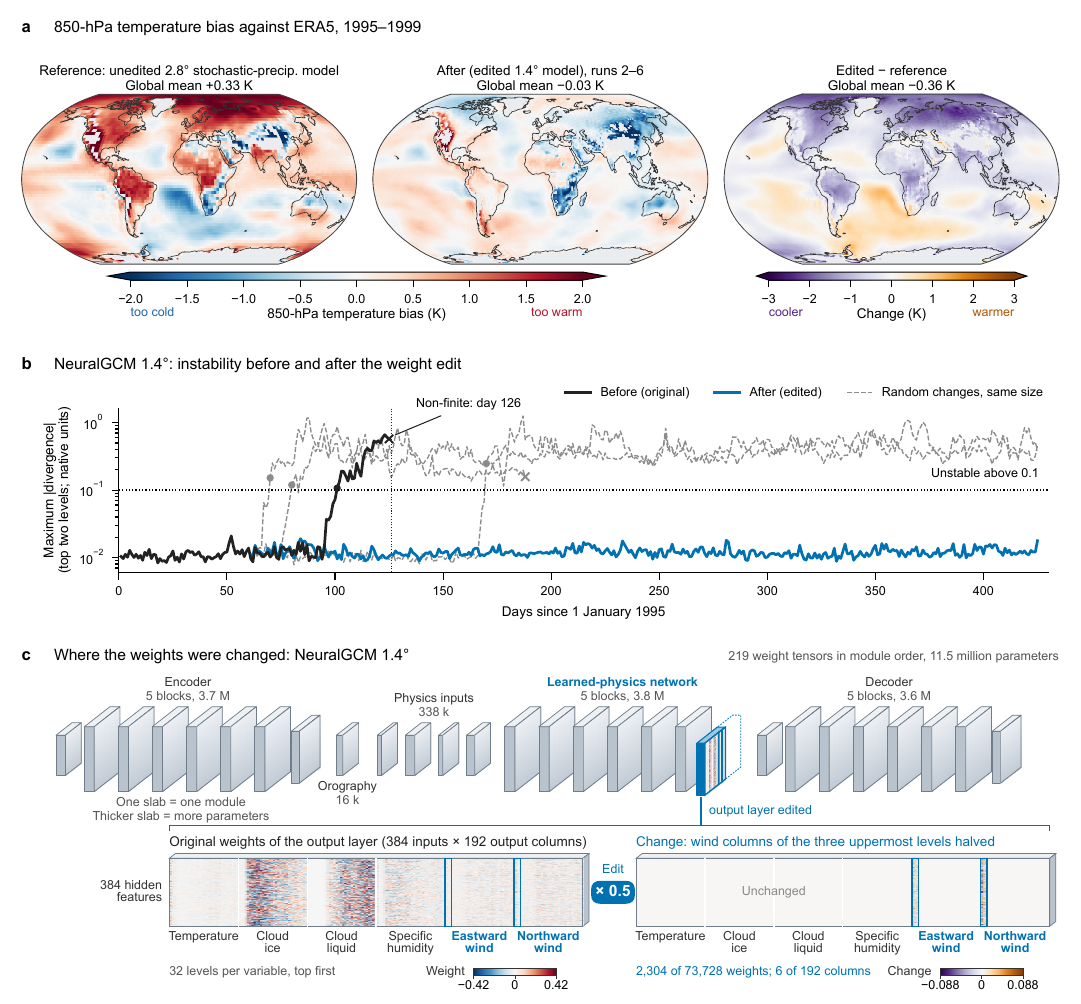}
\caption{\textbf{Halving six upper-level wind columns in 1.4$^\circ$ stochastic NeuralGCM suppresses breakdown, while regional biases remain.} \textbf{a}, Five-year (1995--1999) 850-hPa temperature bias against ERA5 ($\pm$2.0 K) with the pressure correction: 2.8$^\circ$ stochastic-precipitation reference (+0.33 K, left), edited 1.4$^\circ$ runs with seeds 2--6 ($-$0.03 K, middle; the cold seed-1 run is in Supplementary Fig. 9), and edited minus 2.8$^\circ$ ($-$0.36 K, $\pm$3 K, right; grey masks surface pressure < 850 hPa). \textbf{b}, Top-two-level daily maximum absolute divergence from 1 January 1995, with the edited run and three same-size random controls initialised from the state saved on 2 March (day 60). Dots mark the 0.1 threshold and $\times$ the last finite day. \textbf{c}, Edit site across 219 tensors (11.5 million weights): six halved columns in the 384 $\times$ 192 learned-physics output matrix.}\label{fig4}\end{figure}

Could weakening those wind updates prevent breakdown? Zeroing all 18 output columns for the three highest levels caused runaway temperatures. Zeroing just the six wind columns pushed wind speeds above our 150 m s$^{-1}$ limit. Halving those columns instead kept wind divergence below the instability threshold for a year, even without the developers' pressure correction. Three random changes of the same size to the same columns all crossed the instability threshold within 110 days (Fig. 4b; Supplementary Fig. 9a, b). Fig. 3a shows how the edit changed the first forecast of 850-hPa temperature.

The developers' pressure correction\cite{r4} holds global-mean log surface pressure fixed at each step. On its own it did not prevent breakdown in our initial runs. With the weight edit alone, pressure still drifted upward. We therefore combined the two in multi-year runs. Six runs from the same 1995 state with random seeds 1--6 for the stochastic physics, plus a separate 2015 run, stayed finite for 5--11 years (Supplementary Fig. 9c, d). A finite run can still drift into an unrealistic climate. In a separate 20-year test, we checked monthly global means against limits chosen after the runs, including 200--320 K for 850-hPa temperature. With the pressure correction, all six edited runs stayed within those limits for 20 years. None of the six unedited runs did so beyond four years. Four of six unedited runs without the correction stayed finite for two years (Supplementary Table 2; Methods, `Additional NeuralGCM evaluation').

How closely did the edited climate follow ERA5? Five-year 850-hPa temperature root-mean-square error (RMSE) was about 0.5 K for seeds 2--6 (Fig. 4a; Supplementary Methods, `Edits and their evaluation: details'). For seed 1 it reached 1.05 K as a persistently cold northern polar vortex developed. We ran seed 1 for ten years from both 1995 and 2015. The second run reached 2024, five years beyond the end of training for the stochastic model\cite{r3}. By year 10, global mean temperature in the lower stratosphere (50 hPa) had fallen to about 200 K in both runs, compared with about 211 K in ERA5 (Supplementary Fig. 9d). For seeds 2--6, temperatures stayed close to ERA5 over their available runs of 5--11 years. One edited run in the separate 20-year test also developed a cold northern polar cap (Supplementary Methods, `Additional NeuralGCM evaluation: details'). The risk of polar stratospheric cooling therefore depended on the noise seed.

\subsubsection*{DLESyM}

The original DLESyM ran for 1,000 years with almost no global drift\cite{r2}. The later DLESyM v1.0 release includes four atmosphere networks and four ocean networks\cite{r5,r34}. In its default configuration, atmosphere 0 coupled to ocean 0, Arctic near-surface air was 5.8--7.1 K warmer than ERA5\cite{r7} across five initial conditions. Replacing ocean 0 with ocean 1 removed most of the excess Arctic warming in a one-year test. Replacing the atmosphere alone had little effect. We therefore targeted the decoder of ocean 0, which produces the ocean forecast (Methods, `Perturbation map of DLESyM v1.0').

Using four saved time steps from one unedited 1990 run, we estimated new values for 12 weight tensors in the ocean decoder by least squares. We also subtracted 65~\% of the mean open-water sea-surface temperature (SST) tendency in that run from each of the two SST output biases (Fig. 5c; Supplementary Fig. 10b; Methods, `Offset-factor dose test (DLESyM)'). We kept the edited ocean with atmosphere 0 after seeing the results of a planned comparison with atmosphere 1 (Methods, `Edits and their evaluation').

\setcounter{figure}{4}\begin{figure}[tbp]\centering\includegraphics[width=\linewidth,height=0.72\textheight,keepaspectratio]{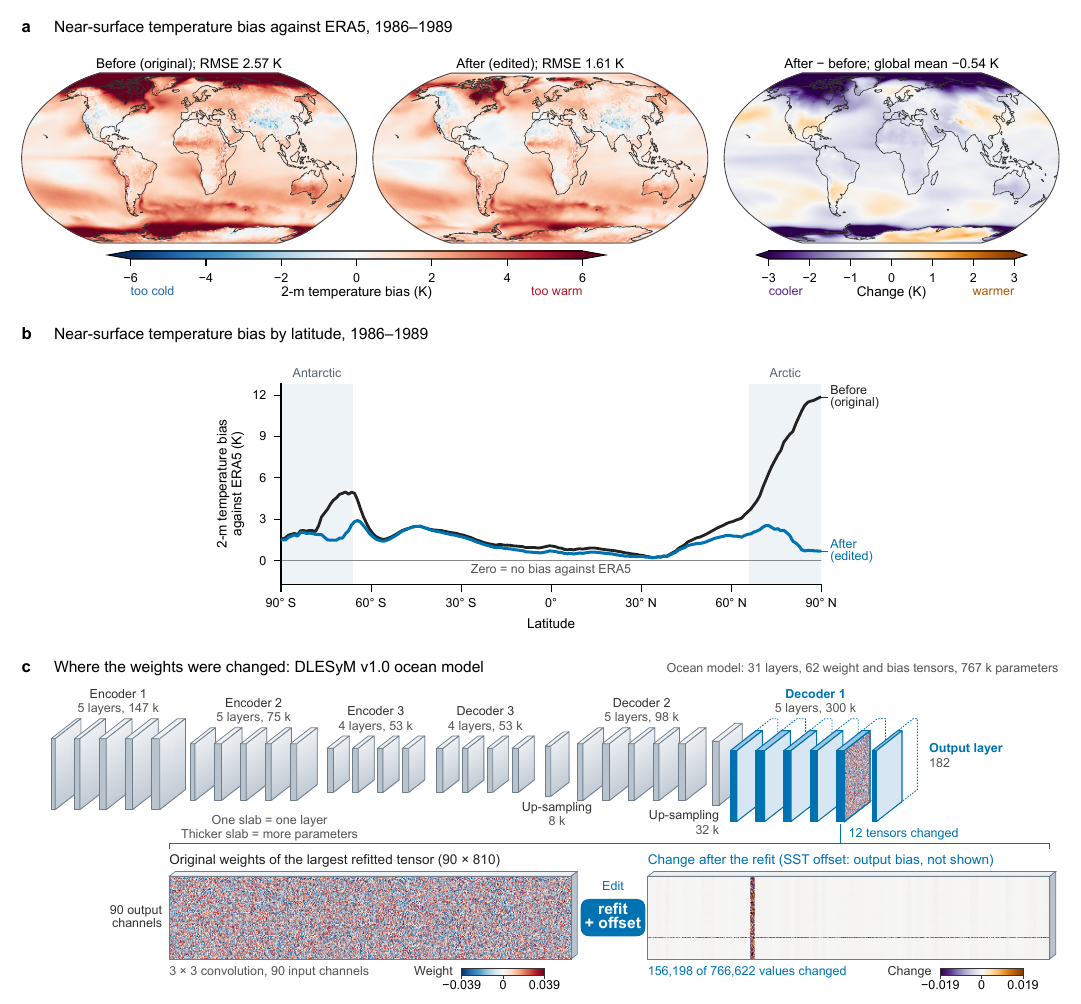}
\caption{\textbf{Recalibrating one DLESyM v1.0 ocean decoder reduces polar warm bias across five initial conditions, while regional seasonal errors remain.} \textbf{a}, Near-surface (2-m) temperature bias against ERA5 over 1986--1989 ($\pm$6 K) before (RMSE 2.57 K) and after (1.61 K) the edit, plus after minus before (mean $-$0.54 K, $\pm$3 K). \textbf{b}, Zonal-mean 2-m temperature bias against ERA5 (K) over 1986--1989 before and after the edit. The other four evaluation periods are in Supplementary Methods, `Edits and their evaluation: details'. \textbf{c}, Edit site across 31 ocean layers (62 tensors, 767 thousand weights; 12 edited tensors in blue, with 0.65 times the mean open-water SST tendency subtracted from the output biases).}\label{fig5}\end{figure}

Across five initial conditions, the Arctic warm bias fell to +0.49--2.03 K and global near-surface temperature root-mean-square error against ERA5 fell by 35--38~\% (Fig. 5a, b; Supplementary Methods, `Edits and their evaluation: details'). All five runs began in years other than 1990, but their evaluation periods were within the model's training years (Methods, `Edits and their evaluation').

Each edited run still failed 9--26 of 80 climate checks. Most checks allowed seasonal-cycle error to rise by at most 10~\% relative to the unedited run (Methods, `Edits and their evaluation'; Supplementary Methods, `Edits and their evaluation: details'). In four-year comparisons from eight initial conditions, near-surface air temperatures in selected polar regions fell by 5.6--8.1 K in winter and spring, closer to ERA5. European seas were 0.49 K cooler in summer than in the unedited runs, further from ERA5. The amplitude of the seasonal SST cycle decreased along the North American coast. Over the Sahel, June--August column water vapour fell by about 1.25 kg m$^{-2}$, further from ERA5 in all eight runs. Across the regional comparisons, 42~\% were inconclusive (Methods, `Regional effects of the DLESyM edit').

\subsubsection*{Pangu-Weather}

Rackow et al.\cite{r6} found that global-mean 2-m temperature in 10-day Pangu-Weather\cite{r27} forecasts for 1955 and 2023 drifted colder by about 0.03 K per day. Starting from ERA5 and using each 24-hour forecast as the next input, we reproduced the drift at $-$0.034 K per day. The model cooled more over the next day when it started from its own forecast than when it started from ERA5 for the same time. By day 10 its 2-m temperature forecasts had less fine-scale variability than ERA5. Smoothing the ERA5 inputs reproduced most of the extra cooling.

To reduce the extra cooling from smoother inputs, we tested edits in an internal layer and in the final layer that turns internal signals into surface forecasts. The internal edit had no measurable effect on drift. For the final layer, we calculated 384 weight adjustments from the difference between forecasts made with original and smoothed ERA5 inputs, without using observed forecast errors. We applied the same adjustments to each of the 16 rows that predict 2-m temperature, changing 6,144 weights (Fig. 6c). At day 60, absolute 2-m temperature bias fell on all 24 dates tested within the training years, and mean bias went from $-$2.09 to $-$0.78 K. The edit reduced bias more than three random edits matched for the size of the first-day temperature change. At 850 hPa, the median of the per-date reductions in absolute temperature bias was 47~\% (Fig. 6a, b; Supplementary Fig. 11a, b; Methods, `Pangu-Weather surface-head edit'). Changing the weights for surface temperature alone reduced bias both near the ground and higher in the atmosphere.

\setcounter{figure}{5}\begin{figure}[tbp]\centering\includegraphics[width=\linewidth,height=0.72\textheight,keepaspectratio]{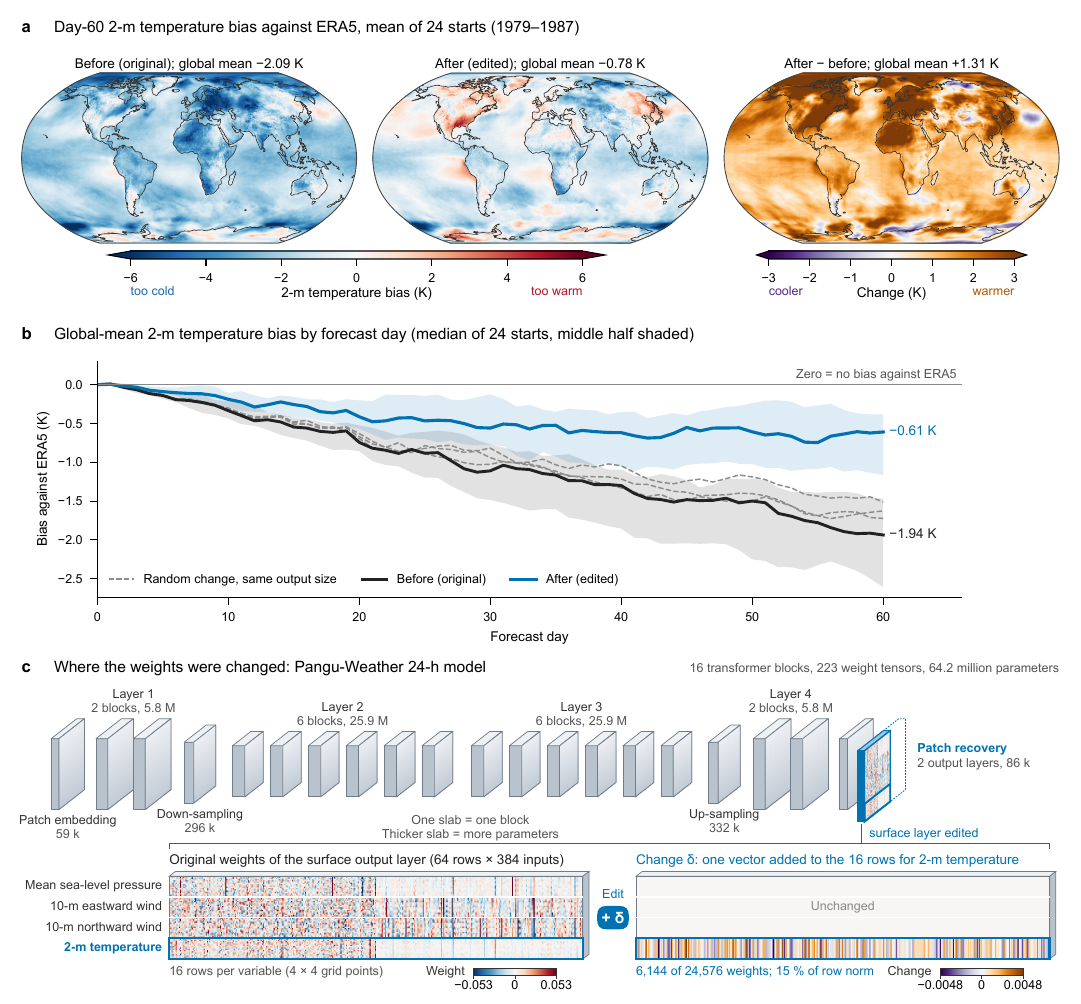}
\caption{\textbf{Adding one shared vector to Pangu-Weather's surface output layer reduces 60-day cold drift inside training years.} \textbf{a}, Mean day-60 2-m temperature bias against ERA5 across 24 initial conditions in 1979--1987 ($\pm$6 K) before ($-$2.09 K) and after ($-$0.78 K) the edit, plus after minus before (+1.31 K, $\pm$3 K). \textbf{b}, Daily global-mean 2-m temperature bias across 24 initial conditions (median, with the interquartile range shaded). By day 60 the unedited model reaches $-$1.94 K. The three random changes, matched to the edit for the size of the first-day temperature response, stay close to the unedited model. The edited model levels off near $-$0.61 K (out-of-sample tests in Supplementary Fig. 11). \textbf{c}, The edit site is shown among the 223 tensors of the 24-hour model (64.2 million weights). In the 64 $\times$ 384 surface output matrix, we added the same vector to each of the 16 rows for 2-m temperature (48 unchanged).}\label{fig6}\end{figure}

Could we reduce cold drift just as much outside the training years by correcting the forecast temperatures directly? We left the weights unchanged and added the same amount of warming to the 2-m temperature field after every forecast step. We chose the amount on development runs and tested both approaches in 10-day forecasts from 12 initial conditions per period. The weight edit reduced 2-m temperature drift by a median of 26.7~\% in 2020--2023 and 55.7~\% in 1957--1960. The constant correction reduced drift as much as or more than the weight edit at both 2 m and 850 hPa (Supplementary Fig. 11c; Supplementary Methods, `Pangu-Weather surface-head edit: details'). By editing the weights that produce surface temperature, we could also reduce cold drift within the model without retraining it.

\subsection{No weather-model edit met the 10-day forecast-skill rule}
We tested whether weight edits improve 10-day forecast skill in Aurora 1.5 and its ensemble\cite{r17,r35}, 6-hour Pangu-Weather\cite{r27}, AIFS Single 2.0\cite{r36,r37}, AIFS ENS v2\cite{r38,r39,r40}, WeatherNext 2\cite{r41} and operational GraphCast\cite{r13,r22}. No edit met the criteria set in advance for seven fields at days 3, 5, 7 and 10. Mean error had to fall by at least 1~\%, and error for any field at any lead time could rise by no more than 0.5~\%. The Aurora edit reduced mean error by 0.79~\% across 42 initial conditions. The GraphCast edit reduced it by 0.44~\% across 36. Relaxing unedited forecasts towards their initial states leaves the weights unchanged. On 12 held-out initial conditions, relaxation reduced error by 4.15~\% in Aurora and 2.70~\% in GraphCast (Methods, `Forecast edits in weather models').

\section{Discussion}

The weight atlas shows how changing a group of weights affects the forecast after one model step. It does not show the physical processes behind that change, or how the model's climate changes over years (Supplementary Figs. 1 and 5). In our year-long ACE2-ERA5 runs, most 6-hour fingerprints stopped identifying the altered weight group within a month, even though the climate stayed shifted. Across all 35 groups tested from one initial condition, weight scaling and sampled equal-size random changes also gave similar median year-mean response sizes (Supplementary Fig. 8c). Choosing candidate weight edits therefore requires multi-year runs and random weight controls, not projections from a single forecast step.

Three of the edits we tested reduced instability or temperature bias. The tests did not establish that any edit restored a missing physical process. Pangu-Weather shows this difference. Its weight edit reduced cold bias more than random controls, yet adding a constant correction to the forecast temperatures after every step did as well or better on 10-day tests. Both adjustments reduced drift, so lower bias alone does not show that the edit fixed the cause of cooling.

Four problems recurred when weight edits fell short. First, changing one pathway could miss a process spread across several pathways or absent from the model. Edits to individual network parts in ACE2-ERA5 increased climate error, and our calculation for DLESyM v0 showed that linear ocean edits sized to prolong Pacific heat anomalies would also shift mean sea-surface temperatures. Second, some errors arose outside the weights, in a dynamical core, a solar-time calculation or mismatched inputs. Third, an edit could improve one quantity while worsening others. FourCastNet v3\cite{r28} output edits delayed breakdown but worsened sea-level pressure, temperature drift or regional climate measures. Fourth, controls and natural variability limited what we could attribute to an edit. In DLWP\cite{r29}, one of two random edits removed almost as much second-year cold bias as the designed edit. In ACE2-ERA5, two of three random edits reduced precipitation error more than the designed edit, and the unedited Arctic summer bias lay within natural variability. We nudged 10-day weather forecasts back towards their starting state without changing the model weights. This correction reduced error several times as much as the weight edits alone in Aurora, GraphCast and AIFS Single. To show an advantage, a weight edit must outperform simple corrections and random controls while keeping errors in other fields within acceptable limits (Supplementary Tables 2--4; Methods, `Statistics').

In this study, we used language-model agents to map how changing the weights of trained weather and climate emulators altered their forecasts. For the atlas, we compared these responses with those from random weight changes of the same size. We then tested whether specific edits improved the models without retraining. In multi-year NeuralGCM runs, halving the weights in six upper-level wind columns together with the pressure correction suppressed breakdown. In DLESyM v1.0\cite{r34}, editing ocean network 0 reduced temperature bias when it ran with atmosphere 0, although some regional errors grew. In successive 24-hour Pangu-Weather forecasts, editing the surface output weights reduced the cold drift reported by Rackow et al.\cite{r6}. Other groups among the authors repeated the NeuralGCM and DLESyM edits. Under human direction, agents set up and ran experiments, tested predictions made in advance and checked one another's findings. Sharing these tasks among agents could speed up the testing of hypotheses about what emulator weights do.

The next step is to extend the atlas to more model families and develop ways to locate and edit weights beyond scaling whole groups. We could scale one channel or one principal direction in the weights at a time and measure how much of the single-step response falls in each class of output field, such as temperature, water, wind or pressure. If temperature accounted for 80--90~\% of the response, for example, we would flag that channel or direction as a candidate for a temperature edit. Controlled multi-year runs can then test which changes persist, where they alter other fields and how the affected weights shape the climate each emulator produces. For Pangu-Weather, a 60-day comparison could test whether smaller weight edits also reduce cooling and how they compare with the full edit and a constant correction to forecast temperatures. Adapting weight editing from language models to weather and climate emulators requires tracing changes through coupled physical fields, sometimes from ocean to atmosphere over years. As agents improve, modellers could guide and check their experiments to build complete weight atlases faster, then use those maps to improve forecast skill or reduce precipitation bias by editing weights without retraining.

\section{Methods}

The study has three parts, and the sections below follow that order. First, we built a single-step weight atlas by scaling groups of weights in 21 model components and measuring how their forecasts changed after one step. Second, we ran detailed tests of that atlas in ACE2-ERA5\cite{r1}, our reference model: on held-out dates, over one year, by activation patching\cite{r10,r31} and by spectral band. Third, we carried out longer runs and targeted weight edits in DLESyM v1.0\cite{r34,r42}, 1.4$^\circ$ stochastic NeuralGCM\cite{r3,r4} and 24-hour Pangu-Weather\cite{r27}, followed by 10-day forecast edits in seven weather models.

\subsection{Models and checkpoints}
We analysed published checkpoints of 20 model configurations across nine families without retraining them. Supplementary Table 5 lists each checkpoint with its settings and SHA-256 checksum. The models were:

\begin{itemize}\setlength\itemsep{1pt}
\item \textbf{ACE atmospheres} use spherical Fourier neural operators\cite{r43} in the fme PyTorch package\cite{r44}. We tested ACE2-ERA5\cite{r1}, ACE2-SHiELD\cite{r1}, ACE2-SOM\cite{r45}, ACE2-EAMv3\cite{r46}, ACE2S-SHiELD+\cite{r19}, HiRO-ACE v1\cite{r47}, and the atmospheres of SamudrACE-CM4\cite{r32} and SamudrACE-E3SMv3\cite{r33}.
\item \textbf{HiRO-ACE v2}\cite{r48}: the atlas catalogued its U-Net downscaler.
\item \textbf{Ocean components of SamudrACE}: we ran the convolutional U-Nets of SamudrACE-CM4\cite{r32} and SamudrACE-E3SMv3\cite{r33} independently of their atmospheres.
\item \textbf{NeuralGCM}\cite{r3,r4}: we mapped and edited the 1.4$^\circ$ stochastic model. The deterministic 0.7$^\circ$ and 2.8$^\circ$ models appear only in the single-step atlas. The stochastic 2.8$^\circ$ NeuralGCM precipitation model\cite{r49,r50} provided the climate reference for the edited 1.4$^\circ$ runs (Fig. 4a).
\item \textbf{DLESyM} atmospheres and oceans (the original model\cite{r2} released as v0\cite{r51}, and the later versions v1.0\cite{r5,r34,r42} and v1.1\cite{r42}), \textbf{CAMulator}\cite{r21}, \textbf{Samudra2} (a 1$^\circ$ ocean model)\cite{r23}, \textbf{GraphCast-small}\cite{r13,r22}, \textbf{FloeNet-OM4}\cite{r24} and \textbf{Monthly diffusion (MD-1.5 v0.9)}\cite{r8,r25,r26}.
\end{itemize}

Atmosphere-only models, GraphCast-small\cite{r13,r22} and two weather models outside the atlas (FourCastNet v3\cite{r28} and DLWP\cite{r29}; their edits are in Supplementary Table 3) received prescribed sea-surface temperature (SST) and sea ice where required. DLESyM\cite{r2,r42} ran as a freely coupled atmosphere--ocean model. Pangu-Weather\cite{r27} (24-hour version), also outside the atlas, fed its own 65 upper-air and four surface fields back in as the next input. Unless stated otherwise, ERA5\cite{r7} initial conditions and reference fields were regridded to each model's native grid.

Counting the two networks in each of five coupled models (the two SamudrACE models and the three DLESyM versions) separately gives 25 components. We swept 21 of them, in 497 weight groups in total (Supplementary Table 1). Four components contributed weight catalogues only (ACE2-SOM\cite{r45}, HiRO-ACE v2\cite{r48}, FloeNet-OM4\cite{r24} and MD-1.5\cite{r25,r26}). Further details are in Supplementary Methods (`Figure conventions' and `Models and checkpoints: details').

\subsection{Weight groups and multipliers}
In each component tested, we grouped weights by network part and measured how changing each group affected the forecast after one step (group counts in Supplementary Table 1).

\textbf{All swept components.} For each group, we multiplied its weights W by a factor $\alpha$ and kept everything else unchanged. The response is the change from the unedited forecast ($\alpha$ = 1) after one forecast step: the area-weighted root-mean-square (RMS) difference in each output field, combined over fields. The PyTorch models (ACE, SamudrACE, DLESyM, CAMulator and Samudra2) were scored on their native output fields. NeuralGCM\cite{r3} and GraphCast-small\cite{r13,r22} run in JAX. For these models, Supplementary Table 1 combines 65 shared pressure-level fields from January and July 1995. Supplementary Fig. 2 uses each model's native fields (`Weight scaling against random directions'). ACE2-ERA5, six other ACE atmospheres and the ocean components of the two SamudrACE models were swept over 11 multipliers (`Family transfer'). Comparisons across all swept components use $\alpha$ = 0.9.

\textbf{ACE2-ERA5, the reference model.} We mapped ACE2-ERA5\cite{r1,r30} in most detail. Each of its eight processing blocks\cite{r1} contributed four groups:

\begin{itemize}\setlength\itemsep{1pt}
\item its spectral filter, which mixes features in spherical-harmonic space\cite{r43,r44};
\item its inner skip, a local linear shortcut\cite{r44};
\item its perceptron output, which mixes features at each point;
\item the whole block.
\end{itemize}

These give 32 groups. Three more complete the set of 35 weight groups: the direct input link (a shortcut that passes the 44 input fields straight to the decoder\cite{r1,r44}), the position embedding\cite{r44} and the second encoder layer.

In ACE2-ERA5 we used $\alpha$ = 0, 0.25, 0.5, 0.75, 0.9, 0.95, 1.05, 1.1, 1.25, 1.5 and 2. Each test was one native 6-hour step\cite{r1} from the released ERA5\cite{r7} initial conditions\cite{r30} of 1 January and 1 July 1979.

For each of the 50 output fields, we calculated the cosine-latitude-weighted RMS difference between the edited and unedited forecasts and divided it by the model's normalisation standard deviation\cite{r1,r30} for that field. We did this in six regions: the globe, the tropics (20$^\circ$ S--20$^\circ$ N), the northern and southern mid-latitudes (20--66$^\circ$), the Arctic (north of 66$^\circ$ N) and the Antarctic (south of 66$^\circ$ S). To rank the groups, we averaged the global responses over the 50 output fields and both 1979 initial conditions (Fig. 2a, b; Supplementary Fig. 3). Sweep details are in Supplementary Methods (`Weight groups and multipliers: details').

\subsection{Family transfer}
We tested whether the sensitivity ranking in ACE2-ERA5 also appeared in related models. We applied the same 11 multipliers to six other ACE atmospheres (all those listed above except ACE2-SOM) and to the ocean components of both SamudrACE models\cite{r32,r33}. Each test covered one forecast step from one initial condition per model. The ocean components ran independently of their atmospheres.

At $\alpha$ = 0.9 and $\alpha$ = 0.5, the direct input link changed the forecast most in all six atmospheres, and block 8 changed it most among the eight blocks. At $\alpha$ = 0.9, block 8 responded 1.48--2.48 times as much as block 1. In the ocean components of the two SamudrACE models (10 stages each, 20 groups in all), the response at $\alpha$ = 0.9 was lowest in stages 3--6, higher at the input and highest at the output.

We also compared fingerprints between models; a fingerprint is a group's pattern of responses across output fields and regions (`Fingerprint test on held-out states'). Even the closest atmosphere pair (ACE2-SHiELD\cite{r1} and the SamudrACE-CM4 atmosphere\cite{r32}) matched only 14 of 27 shared groups on identically named output fields; we required at least 10 shared fields to compare a pair. ACE2S-SHiELD+\cite{r19} and HiRO-ACE v1\cite{r47} matched each other in 7--9 groups (Supplementary Methods, `Family transfer: details').

\subsection{Weight scaling against random directions}
In every swept component, we asked whether scaling the original weights changes the forecast more than a random change of the same size. For each group, we compared the response to $\times$0.9 (a change of size $\varepsilon$ = 0.1) with the responses to three random directions. Each random direction was drawn from a standard normal distribution with a fixed seed and used for every multiplier and initial condition.

For models scored on their native PyTorch output fields, the ratio is the $\times$0.9 response divided by the median of the three random responses.

NeuralGCM\cite{r3,r4} (1.4$^\circ$ stochastic, 0.7$^\circ$ and 2.8$^\circ$ deterministic, 1-hour step) and GraphCast-small\cite{r13,r22} (6-hour step) run in JAX. We tested them from ERA5\cite{r7} initial conditions in January and July 1995. Their responses were combined over 65 shared output fields on 13 pressure levels between 50 and 1000 hPa, each divided by its ERA5 climatological standard deviation. Here the ratio is the scaling response divided by the mean of the three random responses.

A ratio near 1 can occur even when both forecast responses are large. In 10$^{4}$--10$^{8}$ dimensions, a random direction is almost perpendicular to the weights\cite{r52}, and a ratio above 1 points neither to the worst possible direction nor to a physical process. Medians and stricter controls are in Supplementary Methods (`Weight scaling against random directions: details' and `Additional random controls, worst-case directions and local linearity').

\subsection{Fingerprint test on held-out states}
This section and the next three concern ACE2-ERA5\cite{r1} only. They test whether its single-step map holds on other dates and over a year, which pathways carry a forced response and how responses depend on spatial scale.

We tested whether the response after one forecast step can tell which weight group was changed, on dates not used to build the map. The map was built on two 1979 initial conditions. We used the released initial conditions for 1 March, 1 June, 1 September and 1 December 2001\cite{r30}. All four dates are in the model's evaluation period\cite{r1}.

At $\alpha$ = 0, zeroing all tested weights of each of blocks 1--7 gave the same response as zeroing its perceptron output. We therefore dropped those seven whole-block groups beforehand, which left 28 groups.

\textbf{Building a fingerprint.} For each group at $\alpha$ = 0.5, we took its regional RMS responses for 50 output fields in six regions (300 values). We averaged each value over initial conditions, added 10$^{-12}$ and took its logarithm. We then subtracted the mean across groups and the mean within the group (double centring). What remains is the group's fingerprint: its own pattern of which fields and regions respond most.

\textbf{Matching.} We correlated each 1979 fingerprint with each 2001 fingerprint, which gave a 28 $\times$ 28 matrix. A group counted as identified only if its own 2001 fingerprint had the single highest correlation; ties counted as failures. All 28 groups were identified. The median correlation was 0.908 for the same group and 0.000 between different groups, and none of 10,000 random relabellings reached the observed mean. At $\times$0.9, with responses again averaged over the four 2001 dates, all 28 groups matched their own 1979 fingerprint best (median correlation 0.898).

\textbf{Single dates and random directions.} A second test used $\times$0.9 and treated each 2001 date separately. Next to scaling, each group also received three random weight changes of the same size. For each type of weight change, we compared the response on each 2001 date with the mean response to that same change on the two 1979 dates. Scaling identified 26, 28, 28 and 28 of 28 groups on the evaluation dates, and each of the three random changes identified 26--28. Comparing scaling with random changes identified only 3--9 of 28 groups. Median correlations were 0.24--0.41 for scaling compared with random changes and 0.79--0.88 when the type of change was the same (Supplementary Methods, `Fingerprint test on held-out states: details').

\subsection{Year-scale causal test}
The fingerprint test covers one forecast step. We next tested whether the immediate responses lasted over a year of simulation. We changed six ACE2-ERA5 weight groups by $\times$0.9 and $\times$1.1, and compared each with two controls:

\begin{itemize}\setlength\itemsep{1pt}
\item a random change of the same size in the same tensor, W + 0.1$\|$W$\|$R/$\|$R$\|$, where R holds standard-normal values drawn with a fixed seed (groups with several tensors were matched tensor by tensor);
\item $\times$0.9 applied to a selected comparison group elsewhere in the network.
\end{itemize}

The groups and their controls are listed in Supplementary Methods (`Year-scale causal test: details'). The 25 configurations (the unedited model, 12 scaled edits and 12 controls) each ran for one year from 1 January and 1 July of both 1979 and 2001\cite{r30}. All 100 runs stayed finite.

\textbf{Predictions.} Before the runs, we wrote down 41 predictions (Supplementary Table 6) about the edits and five about the controls. A prediction passed only if every one of its checks fell inside the range set in advance. For the 41 predictions about the edits, 130 of 226 individual checks passed and 12 predictions passed every check.

\textbf{How responses were measured.} For each output field and region, we averaged the signed edited-minus-unedited difference over the four initial conditions first. We then took the spatial RMS and divided it by the model's normalisation standard deviation for that field. Year 1 is the plain average of the 12 monthly means. For checks that used a signal threshold, we kept field--region combinations with a mean at least two standard errors from zero: |mean| / (s.d./2) $\geq$ 2, using the sample standard deviation (n $-$ 1). This is a screen, not a corrected statistical test.

\textbf{Recognition over time.} At day 1, all six $\times$0.9 edits matched their own single-step fingerprint best, but only 4 of 12 day-1 checks with fixed thresholds passed. At month 1, a response had to meet the correlation threshold set in advance for its own group and match its own fingerprint more closely than any of the other candidates in the set of six. Even without the threshold, four groups no longer matched their own fingerprint most closely at month 1 (correlations $\leq$ 0.15).

\textbf{All 35 groups.} A second test, with 108 one-year runs from one 2001 initial condition, covered all 35 groups. Each group ran at $\times$0.9, at $\times$1.1 and with one random change of the same size (105 runs). Three further runs with tiny (10$^{-4}$) perturbations gave the response size expected without an edit (Supplementary Fig. 8b). Here each response was the global spatial RMS averaged over output fields, and we divided each scaling response by the response to its random change (Supplementary Methods, `Year-scale causal test: details').

\subsection{Activation patching}
The tests above change weights. To follow a forced signal through the network, we also wanted to know which internal pathways of ACE2-ERA5 carry its response to El Ni\~no. We ran the model for 120 days from four January 1979 initial conditions, once with an El Ni\~no SST pattern and once with a La Ni\~na pattern ($\pm$1 K in the Ni\~no-3.4 region).

For each pathway in turn, we replaced the cold run's intermediate layer outputs (activations) with those produced by the same pathway in the warm run (activation patching\cite{r10,r31}). We measured how much of the warm--cold difference in February--April tropical 850-hPa temperature this replacement reproduced. These fractions (mean $\pm$ standard error across four initial conditions) were 0.55 $\pm$ 0.12 for the direct input link, 0.38 $\pm$ 0.09 for the main trunk (blocks 1--8) and 0.46 $\pm$ 0.06 for block 8's perceptron output. The pathways overlap, so these fractions cannot be added (Supplementary Methods, `Activation patching: details').

\subsection{Scale sensitivity of the ACE2-ERA5 spectral filters}
The groups above treat each ACE2-ERA5\cite{r1,r43,r44} spectral filter as a whole. We scaled the weights in one spherical-harmonic degree band of one filter at a time by 0.9. For each band, we measured the change after one forecast step from the initial conditions of 1 January and 1 July 1979. The three bands were large scales (degrees 0--10, 6~\% of the filter's parameters), middle scales (11--40, 17~\%) and small scales (41--179, 77~\%). Each band edit was compared with three isotropic random changes of the same size (Frobenius norm) within that band.

This test measures how sensitive the forecast is after one step. It says nothing about forecast skill, about what information the filter stores, or about a repair (Supplementary Methods, `Scale sensitivity of the ACE2-ERA5 spectral filters: details').

\subsection{Perturbation map of DLESyM v1.0}
The remaining sections concern longer runs and edits. Where DLESyM v1.0, 1.4$^\circ$ stochastic NeuralGCM and Pangu-Weather drifted or failed, we looked for the source and then tested targeted weight edits.

We looked for fragile weights in the coupled DLESyM v1.0\cite{r34,r42}. The release has four atmosphere and four ocean networks, numbered 0--3\cite{r42}. In the notation a/o, a denotes the atmosphere network and o the ocean network. The inference package\cite{r53} defaults to 0/0, the first atmosphere network coupled to the first ocean network. In the default configuration, we scaled each of its 63 weight tensors by 0.9 and by 1.1. From 1 January 1995 we ran 126 scaled cases (two per tensor), five unedited baseline runs and eight runs in four pairs with tiny opposite perturbations ($\pm$10$^{-5}$), each for up to 365 days (139 coupled runs). We recorded a failure on the first day any scored field crossed its physical limit. Increasing a tensor by 10~\% took at least one field outside its physical limits for 23 of the 63 tensors. Decreasing a tensor by 10~\% did so for two. For one tensor, both changes caused a field to cross its limit, giving 24 distinct tensors in total (Supplementary Fig. 10a). For the other 39, both runs stayed within the limits for 365 days. In the 100 year-long ACE2-ERA5 runs, the stability check asked only whether the outputs stayed finite (`Year-scale causal test'). A finite value can exceed a physical limit, so these tests cannot rank the models by stability.

In the five multi-year evaluation runs described below (`Edits and their evaluation'), the unedited default model was 5.8--7.1 K too warm in the Arctic; extending the run from 2005 from ten to twenty years reduced its mean bias from 5.79 to 5.70 K (Supplementary Methods, `Edits and their evaluation: details'). Before the sweep, we exchanged either the atmosphere or the ocean between configurations 0/0 and 1/1. Each coupled model ran for one year from 1 January 1990. Replacing ocean 0 removed most of the extra Arctic warming. Replacing atmosphere 0 had little effect on that warming. In the default configuration, the main source of the drift was therefore ocean 0 (drift table in Supplementary Methods, `Perturbation map of DLESyM v1.0: details').

\subsection{Localisation of the NeuralGCM instability}
In our initial unedited runs of the 1.4$^\circ$ stochastic NeuralGCM\cite{r3,r4} from 1 January 1995 (Supplementary Table 5), the last finite output occurred on days 113--129. In these runs and the tests from saved states below, we updated SST and sea ice daily from ERA5. To find where the breakdown starts, we used 40-day runs from the state saved on day 90 and 24-hour power iterations of the Jacobian, which estimate how fast small perturbations grow. The learned-physics network, which represents small-scale processes alongside the dynamical core\cite{r3}, amplified the upper-level perturbations.

Zeroing subsets of its outputs identified six columns to test in its bias-free final layer. We number levels downwards from the top of the model, and these six columns update both wind components on levels 1--3.

We measured instability as M, the daily maximum absolute wind divergence on the top two levels. Onset was the first day with M > 0.1; non-finite output was recorded separately. All one-year tests began from the state saved on 2 March 1995 (day 60), and we followed each run for up to 365 days (Supplementary Methods, `Localisation of the NeuralGCM instability: details').

\subsection{Edits and their evaluation}
We treated every candidate weight edit as a climate experiment with four rules:

\begin{enumerate}\setlength\itemsep{1pt}
\item For confirmatory tests we set pass criteria before the runs. We identify exploratory choices and criteria set after the runs where we report them.
\item Random changes of the same size served as controls wherever possible.
\item Multi-year runs were scored against ERA5\cite{r7}.
\item Supplementary Table 3 gives the criteria and results for the tested edits.
\end{enumerate}

A change that shifted the output without altering how the forecast evolved counted as post-processing, not as a weight edit.

The edits below are in NeuralGCM's learned-physics output layer and in ocean 0 of DLESyM v1.0, the parts identified in the two preceding sections. The Pangu-Weather and 10-day forecast edits have their own sections.

\textbf{NeuralGCM 1.4$^\circ$.} We specified two sets of edits in advance and tested both from the state saved on day 60 (2 March 1995), with the developers' documented pressure correction\cite{r4} active. This correction holds global-mean log surface pressure fixed at every step.

\begin{itemize}\setlength\itemsep{1pt}
\item Zeroing all 18 output columns of the top three levels (six variables per level) made the model run away.
\item Zeroing only the six wind columns pushed stratospheric wind above our limit (153 against 150 m s$^{-1}$ at 20 hPa).
\item Halving the six wind columns kept divergence and stratospheric wind within their one-year limits, both with and without the pressure correction.
\end{itemize}

Without the correction, however, global-mean log surface pressure still drifted upward (+2.4 $\times$ 10$^{-4}$ d$^{-1}$, against +1.9 $\times$ 10$^{-4}$ d$^{-1}$ unedited). Multi-year runs therefore combined the $\times$0.5 edit with this pressure correction. In these runs, we updated SST and sea ice daily from ERA5.

We made six such runs from the same 1995 initial condition, with noise seeds 1--6. All six stayed finite for five years. One of the six (seed 1) settled into a cold polar-vortex state; for one case in six, the exact binomial 95~\% interval\cite{r54} for that fraction is 0.004--0.64 (Supplementary Methods, `Edits and their evaluation: details').

\textbf{Additional NeuralGCM evaluation.} A 20-year ensemble from January 1995 compared six edited and six unedited runs, all with the pressure correction, plus three random controls. Monthly plausibility checks, set after the runs, required each monthly global mean to stay inside a fixed range (for example 200--320 K for 850-hPa temperature). All six edited runs met those limits for 20 years. None of the six unedited runs still met them beyond four years (one-sided Fisher p = 1/924). In a separate ensemble with 12-hourly updates of SST and sea ice and no pressure correction, four of six unedited runs passed the monthly checks through 730 days, so the unedited lifetime depends on the setup and the seed (Supplementary Methods, `Additional NeuralGCM evaluation: details').

\textbf{DLESyM v1.0.} We used the released v1.0 ERA5 version\cite{r5,r34,r42} and ran four coupled configurations individually (0/0, 1/1, 2/2, 3/3). These runs used a different setup from the published 1,000-year v0\cite{r2} and 90-year\cite{r5} experiments. For comparison, our 20-year (240-month) run of the newer DLESyM v1.1 release\cite{r42} stayed finite, with a seasonal 2-m root-mean-square error (RMSE) of 1.73 K and an interannual standard deviation of 2-m temperature 3.48 times that of ERA5\cite{r7}.

We specified two sets of candidate edits for ocean 0 in advance. Three rank-one edits of the output layer broke in August--September 1990. Three edits that built a lower bound on SST near freezing into the ocean network failed 27, 13 and 22 of 80 climate checks in three-year runs. These checks compare an edited run with the unedited run: the seasonal-cycle RMSE of four fields (2-m and 850-hPa temperature, column water vapour and 500-hPa geopotential) over the globe and 18 regions may rise by at most 10~\% (76 checks), and four more cover variability and the 2-m climatology.

The edit we kept had two parts (156,198 elements in total). First, we estimated new values for 12 tensors in the final decoder, skip and output layers, including 544 biases. We used least squares with four saved time steps (steps 16, 31, 61 and 76) from one unedited 1990 run. We reserved winter and summer time steps for validation. Second, we subtracted 0.65 times the 1990 mean open-water SST tendency from each of the two SST output biases. We chose the factor from development runs.

We chose to couple the edited ocean to atmosphere 0 after seeing the results of a preregistered comparison with atmosphere 1 (runs from 1986 and 1993). We used the same edited ocean weights with both atmospheres. The edited and unedited 0/0 configurations then ran from 1 January of 1986, 1993, 1995, 2003 and 2005, all inside the training period\cite{r42}. The first table of Supplementary Methods (`Edits and their evaluation: details') gives the period evaluated for each of these five runs.

\textbf{DLESyM evaluation details.} In a separate 48-month test from a 1998 initial condition, three random changes, matched in size tensor by tensor, altered all 300,602 weights of the 12 recalculated tensors (the edit itself changed 156,198). The three control runs exceeded physical limits in months 1, 2 and 6, respectively. Only runs within those physical limits qualified for climate scoring.

Separately, the unedited coupled configurations 0/0, 1/1, 2/2 and 3/3 from 1 January 1995 all ran away after 4.6--6.8 years. From the same date, the edited 0/0 configuration ran for 20 finite years, while from nine other initial conditions the unedited 0/0 configuration stayed finite for the available runs: seven for eight years, one for 12 years and one for 20 years. The edited and unedited runs from 1995 lacked matched run-time settings, so this comparison cannot establish a general stability gain (Supplementary Methods, `DLESyM evaluation: control and runaway details').

\subsection{Regional effects of the DLESyM edit}
To see where the DLESyM edit helps and where it hurts, a second group among the authors compared 48-month edited and unedited 0/0 runs from eight initial conditions in the training period\cite{r42} (1986, 1990, 1993, 1995, 2000, 2003, 2005 and 2010). The regions overlap, and eight initial conditions are too few to resolve year-to-year variability or the amplitude of the El Ni\~no--Southern Oscillation (ENSO).

We evaluated 2,751 combinations of variable, region and climate statistic across five variables and 56 regions. Each combination is one cell in the results below. Paired t-tests with Benjamini--Hochberg correction\cite{r55} (false-discovery rate 0.10) found 1,023 cells that differed: 744 better, 49 worse and 230 mixed. Another 572 cells met the fixed equivalence margins in two one-sided tests\cite{r56}, and 1,156 (42~\%) were inconclusive. Near-surface temperature met the regional improvement rule: improved area exceeded twice the worsened area, and no flagged cell worsened by more than twice its margin. The other four variables had mixed outcomes (Supplementary Methods, `Regional effects of the DLESyM edit: details').

\subsection{Offset-factor dose test (DLESyM)}
We tested how the DLESyM v1.0 edit depended on the SST offset by keeping the recalculated decoder weights fixed and changing only the offset factor. Four-year coupled runs from 1 January in 1998 and 2008 tested eight factors (0, 0.5, 0.6, 0.65, 0.7, 0.8, 0.9 and 1). The response curve did not meet the conditions set in advance, so the test did not identify an optimal factor.

For DLESyM v1.1\cite{r42}, we tested the same adjustment to SST. We estimated the two offsets again for this release and kept the factor at 0.65. From January 1998, 2003, 2008 and 2013, the unedited model, the edit and three same-size random weight changes each used the same three seeds for noise in the model's conditional normalisation layers\cite{r42,r53}. The edit increased the absolute annual SST bias of the Southern Ocean by 1.18, 1.10, 0.57 and 1.43 K (Supplementary Methods, `Offset-factor dose test (DLESyM): details').

\subsection{Pangu-Weather surface-head edit}
Pangu-Weather lies outside the atlas. Our chained 24-hour forecasts drift cold, as reported for Pangu-Weather's 10-day forecasts\cite{r6}, and we tested whether a calibration of its output layer, fitted without using forecast errors, can reduce that drift. We ran the 24-hour checkpoint\cite{r27} (Supplementary Table 5) through the inference package\cite{r53} as chained daily 0.25$^\circ$ forecasts from ERA5\cite{r7}, covering 69 fields.

\textbf{Effect of smoothing the inputs.} After one forecast day, Pangu-Weather cooled more over the next 24 hours when started from its own forecast than from ERA5 at the same valid time. We measured the change in global-mean 2-m temperature over that step from each input. By day 10, the model's forecasts are smoother than ERA5: 2-m temperature variance within 4 $\times$ 4 patches was 0.73 of ERA5 across 12 initial conditions. We therefore tested whether smoothing alone reproduces the extra cooling. The test used 12 development and 12 evaluation dates; each set was split into two folds of six by taking alternate dates. Smoothing ERA5 alone, on its original grid with a 0.5$^\circ$ Gaussian width, reproduced 87--89~\% of the extra cooling in the two evaluation folds and 76--89~\% in the two development folds. As a control, we shifted the forecast's deviation from ERA5 by 90$^\circ$ in longitude. The magnitude of the temperature response exceeded the limit set in advance, so the test did not identify a local cause of the cooling.

\textbf{The edit.} We estimated the weight change by ridge regression\cite{r57}, comparing forecasts from original ERA5 inputs with forecasts from the same inputs smoothed at widths of 0.5$^\circ$ or 1.5$^\circ$. The result is one shared vector of 384 elements (norm 0.036), added to the 16 rows of the output layer that produce 2-m temperature. The edit changed 6,144 weights. The RMS change in the first forecast was 0.096 K.

\textbf{Inside the training period\cite{r27}.} Across 24 initial conditions from 1979--1987, the mean 2-m bias at day 60 fell from $-$2.09 to $-$0.78 K. The median fraction of absolute bias removed was 0.59, positive in all 24 cases. Three random vectors matched for the size of the first-step temperature response gave 0.09--0.19, and the negated edit $-$0.98 (Supplementary Fig. 11a).

\textbf{Outside the training period\cite{r27}.} A 216-run test used 12 initial conditions in 2020--2023 and 12 in 1957--1960, with a 10-day rule of five criteria set in advance. We measured the slope of the global-mean temperature error over the 10 days. For comparison, we added a constant warming after every forecast step while leaving the weights unchanged (fixed warming). We tuned this constant on 24 development runs to match the edit's median fractional reduction in 10-day near-surface temperature drift (Supplementary Methods, `Pangu-Weather surface-head edit: details'). The edit had to:

\begin{enumerate}\setlength\itemsep{1pt}
\item cut the magnitude of the 10-day near-surface temperature error slope by a median of at least 30~\%, with improvement on at least 8 of 12 dates;
\item reduce the day-10 latitude-band bias pattern more than this fixed warming, after removing the global mean and allowing for uncertainty across dates;
\item reduce the 850-hPa temperature error slope more than the same fixed warming;
\item beat each random edit by at least 0.15 in fractional near-surface drift reduction, with the negated edit worsening the drift;
\item keep the increase in day-1 and day-10 errors within their limits.
\end{enumerate}

Neither period passed. In 2020--2023, four of the five criteria failed: the near-surface reduction missed 30~\%, and the latitude-band test, the 850-hPa comparison and the margin over random edits all fell short. In 1957--1960, only the 850-hPa comparison failed. Fixed warming also matched or exceeded the near-surface reduction in both periods (Supplementary Fig. 11c; Supplementary Methods, `Pangu-Weather surface-head edit: details').

\subsection{Forecast edits in weather models}
The edits above target longer runs that drift or fail. We also tested whether weight edits lower 10-day forecast error in seven released weather models outside the atlas: Aurora 1.5\cite{r17,r35} (0.25$^\circ$, 1.26 billion parameters), the 6-hour Pangu-Weather\cite{r27} model run alone for 40 steps, AIFS Single 2.0\cite{r36,r37}, operational GraphCast\cite{r13,r22} (0.25$^\circ$, 13 levels, 36.0 million parameters) and three ensemble models, Aurora 1.5 Ensemble\cite{r17,r35} (four members), AIFS ENS v2\cite{r38,r39,r40} and WeatherNext 2\cite{r41} (eight members each). Supplementary Table 5 identifies all seven checkpoints.

\textbf{Forecasts and score.} For the primary benchmark we used ERA5\cite{r7} initial states from 2025. The standard development and held-out sets comprised 12 dates each at 00 UTC (the first and the 15th of each month); further dates served for confirmation. We scored 500-hPa geopotential, 850-hPa temperature, 2-m temperature, both 10-m wind components, mean sea-level pressure and 700-hPa specific humidity at days 3, 5, 7 and 10 against ERA5, which gives 28 numbers. Deterministic models were scored by cosine-latitude-weighted RMSE. Ensembles were scored by the fair continuous ranked probability score\cite{r58} (the form corrected for ensemble size), with the same noise seeds in the edited and unedited runs. An edit's score is the mean relative change of the 28 numbers from the unedited model on the same initial conditions. Intervals are 90~\% paired bootstrap intervals\cite{r59} over initial conditions.

\textbf{Rule set in advance.} An edit passed only if its score was $-$1~\% or lower on both development and held-out initial conditions, with an interval below zero and no single number more than 0.5~\% worse. For ensembles, the ensemble-mean RMSE also had to stay within +0.2~\%.

\textbf{Edits.} All edits were written in closed form, without gradients or an optimiser. They scaled attention logits, normalisation gains, branch outputs, input-time weights, output weights or noise paths, or shifted output biases by values computed from mean development errors. We set the primary skill rule in advance. In the investigation of the AIFS Single 2.0\cite{r37} inputs absent from our ERA5\cite{r7} fields (Supplementary Methods, `Forecast edits in weather models: details'), the smoothing prediction was recorded after its run and the input-selection rule was revised after two results were known.

\textbf{Post-processing comparators.} We compared three post-processing approaches separately, each applied to the unedited forecasts with the weights unchanged. The first added a constant per field and lead time. The second smoothed each field with a Gaussian whose width was chosen per field and lead time on the development dates. The third relaxed each forecast towards its initial state, y = x + $\lambda$(x$_0$ $-$ x), with one $\lambda$ per field and lead time. For Aurora's relaxation-only score, we estimated the coefficients on development dates. To measure the additional effect of editing the weights, we estimated coefficients separately for edited and unedited forecasts, leaving out the date being evaluated each time. An edit counted as more than post-processing only if it lowered the score further when edited and unedited forecasts received the same step.

\textbf{Outcome.} No edit passed the rule. Two edits also reduced error after forecast relaxation. Before relaxation, their scores were $-$0.79~\% in Aurora 1.5 (across 42 initial conditions not used to select it) and $-$0.44~\% in operational GraphCast (across 36 initial conditions). On 12 held-out initial conditions, relaxation alone gave $-$4.15~\% and $-$2.70~\%, and the edits added $-$0.73~\% and $-$0.24~\% on top. Results for each model are in Supplementary Methods (`Forecast edits in weather models: details').

\subsection{Workflow, preregistration and verification}
\textbf{Roles of researchers and agents.} We set the research questions, proposed most of the hypotheses and candidate weight edits and specified which layers or weight groups to examine. We used Claude Opus 5.5 (Anthropic) in Claude Code to supervise the other agents, run simulations and analyse the results. CLI Agent Orchestrator\cite{r60} connected the agents' terminals.

Within those directions, the agents chose which weights to test and examined how changing them affected the forecasts. GPT-6.2 Astra (OpenAI) in Codex and Gemini 4 Neon (Google) in Agy searched the literature and project records, evaluated hypotheses, reviewed results, and drafted and checked the manuscript. Claude Fable 5.2 in Claude Code made independent predictions and reviewed the tests and results. For each confirmatory test, different agents fixed the predictions and scored the results. A third reviewed the predictions and scoring procedure before launch. Other groups among the co-authors repeated calculations from stored outputs and reran selected simulations on separate computers. We checked the protocol, code, numbers and figures and decided what to report (Supplementary Table 4 and Supplementary Methods, `Other weight edits' and `Workflow, preregistration and verification: details').

\subsection{Statistics}
The signal-to-noise screen for year-scale predictions is defined in `Year-scale causal test'.

Two negative tests in ACE2-ERA5\cite{r1} are cited in the Discussion (Supplementary Table 3). In the first, a $\times$0.75 edit of the first block cut precipitation error by 0.094 mm d$^{-1}$, but two random controls cut it more, by 0.177 and 0.172 mm d$^{-1}$, and the edit added 0.95 K to the Antarctic winter error. In the second, the model's unedited Arctic summer 850-hPa temperature bias was 0.299 K, below the 0.451 K needed to detect an improvement, so the measured bias was too small to resolve an improvement (Supplementary Methods, `Statistics: details').

\begin{itemize}\setlength\itemsep{1pt}
\item \textbf{Correlations:} Pearson.
\item \textbf{Area means:} cosine-latitude weighted on the native grid.
\item \textbf{Standard deviations across initial conditions:} sample standard deviation, denominator n $-$ 1.
\item \textbf{Permutation p-values:} (k + 1)/(n + 1)\cite{r61}, where n is the number of permutations and k counts those with a result at least as extreme as the observed result.
\item \textbf{Kinds of statement}, kept apart throughout:
\item confirmatory tests set in advance: the held-out fingerprint criteria and the year-scale predictions;
\item descriptive counts: the 12 of 41 predictions, survival counts and stability counts;
\item exploratory checks added during review: random subsets of output fields and weight groups, the wrong-group benchmark (prediction sets reassigned across weight groups; Supplementary Methods, `Statistics: details') and month-to-year identification.
\end{itemize}

\subsection{Data availability}
Multiplier-response tables, fingerprints, year-scale scores and maps are archived in the repository below. ERA5\cite{r7} is available from the Copernicus Climate Data Store. Model checkpoints are available from their developers and are not redistributed here.

\subsection{Code availability}
Analysis code, fixed prediction files, scoring scripts and results for the tested edits are at github.com/CliDyn/emulator-weight-atlas.

\subsection{Acknowledgements}
This work was supported by the CLAIMA project (grant agreement No 101290971), funded by the European Union. Views and opinions expressed are however those of the author(s) only and do not necessarily reflect those of the European Union or the European Research Executive Agency (REA). Neither the European Union nor the granting authority can be held responsible for them. This work used resources of the Deutsches Klimarechenzentrum (DKRZ). Google provided access to Gemini 4 models in Google Antigravity at no cost through its programme for scientists.

\renewcommand{\refname}{References}

\clearpage
\appendix
\renewcommand{\thesection}{S\arabic{section}}
\begin{center}{\Large\bfseries Supplementary Information}\end{center}
\section*{Supplementary Methods}

The scripts and Source Data in the repository give the full experimental setup and values (Methods, `Code availability').

\subsection*{Figure conventions}

Supplementary Fig. 1 has 29 rows for 25 components and four additional DLESyM v1.0 networks\cite{r42}. Excluding buffers, static fields and three rows with partial counts gives 5,380,076,258 parameters across 26 rows. Scaling covered 89.5~\% of that total. The counts in Supplementary Table 1 sum to 4.83 billion and cover the 21 swept components, including three partial counts.

For Fig. 3 we chose groups by their earlier responses, network position and edit location before viewing the maps. Each map covers one native model step.

\subsection*{Models and checkpoints: details}

Catalogues range from 766,622 weights in DLESyM v1.0 ocean members 1 and 3\cite{r42} to 874,382,736 in CAMulator\cite{r21}, a latitude--longitude CrossFormer\cite{r62}. Supplementary Table 5 identifies our releases. Configurations in the AI Weather and Climate Model Intercomparison Project (AIMIP)\cite{r8} can differ.

Two baseline results bear on the El Ni\~no tests. The Ni\~no-3.4 standard deviation in DLESyM v0\cite{r2,r51} was 0.20 K, compared with 0.90 K in ERA5\cite{r7}. El Ni\~no warming of tropical 850-hPa air was about 31~\% weaker in ACE2-ERA5\cite{r1} than in ACE2S-SHiELD+\cite{r19}.

\subsection*{Tensor catalogue}

Corresponding layers in related models had different weights. In DLESyM v1.0\cite{r42}, matching hidden channels gave cosines of 0.03--0.08 within each family of networks. Across three ACE atmospheres\cite{r1,r46}, median cosines were 0.006--0.028, against 0.004--0.015 for randomly permuted channels.

\subsection*{Weight groups and multipliers: details}

In ACE2-ERA5\cite{r1}, zeroing the direct input link gave a normalised RMS response of 0.73. Zeroing block 8 gave 0.41, against 0.08--0.14 for blocks 1--7 and 0.10--0.21 for filters 1--7. Bypassing block 8 gave 0.51 because it removed the bias and input normalisation as well (Supplementary Fig. 3).

\subsection*{Family transfer: details}

We compared 27 shared groups across related atmospheres: three paths in each of eight blocks, the second encoder layer, the position embedding and the direct input link. Seven pairs shared at least ten output fields. ACE2-SHiELD\cite{r1} and the SamudrACE-CM4 atmosphere\cite{r32} identified 14 groups in one direction and 13 in the other. ACE2S-SHiELD+\cite{r19} and HiRO-ACE v1\cite{r47} identified 7--9. The two SamudrACE oceans\cite{r32,r33} shared only two output-field names, too few to compare fingerprints. Within each ocean, the output stage responded 1.26--1.28 times as much as the input stage at $\times$0.5.

In each of the six additional atmospheres, the direct input link ranked first at $\times$0.9 and $\times$0.5 despite having the fewest parameters. At $\times$0.9 the median inner-skip to filter response ratios were 0.60--0.65 in the three atmospheres without detected conditional normalisation and 0.90--0.96 in the three with it. These ratios describe an association. The comparison does not isolate the effect of normalisation.

\subsection*{Weight scaling against random directions: details}

Supplementary Fig. 2 uses native fields and the mean of three random responses as its denominator. Supplementary Table 1 uses their median for the native-field PyTorch models and their mean on 65 shared pressure-level fields for NeuralGCM\cite{r3,r4} and GraphCast-small\cite{r13,r22}. The figure markers are medians across groups. Spectral normalisation\cite{r63}, found in eight of nine CAMulator\cite{r21} groups, partly cancels scaling. The table below gives median scaling-to-random ratios for the native-field PyTorch models, including five subsequent measurements.

{\footnotesize\setlength\tabcolsep{3pt}\begin{longtable}{>{\raggedright\arraybackslash\hspace{0pt}}p{0.386\linewidth}>{\raggedright\arraybackslash\hspace{0pt}}p{0.160\linewidth}>{\raggedright\arraybackslash\hspace{0pt}}p{0.386\linewidth}}\toprule
\textbf{Component} & \textbf{Groups with a finite ratio} & \textbf{Median ratio of scaling to random at $\times$0.9 [interquartile range]} \\ \midrule\endhead
ACE2-ERA5\cite{r1} & 35 & 1.00 \\
ACE2-SHiELD\cite{r1} & 35 & 1.01 \\
ACE2-EAMv3\cite{r46} & 35 & 1.08 \\
ACE2S-SHiELD+\cite{r19} & 43 & 0.77 \\
HiRO-ACE v1\cite{r47} & 43 & 0.49 \\
SamudrACE-CM4\cite{r32} atmosphere & 35 & 1.08 \\
SamudrACE-E3SMv3\cite{r33} atmosphere & 43 & 0.69 \\
Ocean component of SamudrACE-CM4\cite{r32} & 10 & 0.65 \\
Ocean component of SamudrACE-E3SMv3\cite{r33} & 10 & 0.92 \\
DLESyM v0\cite{r2,r51} ocean & 9 & 2.37 \\
DLESyM v1.0\cite{r5,r34,r42} atmosphere (network 0, default in inference package\cite{r53}) & 12 & 2.58 \\
DLESyM v1.0 ocean (default) & 9 & 4.54 \\
DLESyM v1.1\cite{r42,r53} atmosphere & 12 & 1.40 \\
DLESyM v1.1 ocean & 9 & 6.71 \\
Samudra2\cite{r23} & 10 & 2.38 \\
CAMulator\cite{r21} & 7 of 9 & 2.61 \\
DLESyM v0 atmosphere (later run) & 12 & 2.341 [1.190, 3.099] \\
DLESyM v1.0 atmosphere, member 2 (later run) & 12 & 1.835 [1.206, 2.568] \\
DLESyM v1.0 atmosphere, member 3 (later run) & 12 & 2.899 [1.653, 3.920] \\
DLESyM v1.0 ocean, member 1 (later run) & 9 & 5.932 [3.220, 8.551] \\
DLESyM v1.0 ocean, member 3 (later run) & 9 & 5.948 [3.508, 8.050] \\
\bottomrule\end{longtable}}

Preserving singular values in the random controls gave median scaling-to-control ratios of 0.92 for ACE2-ERA5\cite{r1} and 2.60 for the DLESyM v1.0 atmosphere\cite{r5,r34,r42}, compared with 1.00 and 2.49 with isotropic controls. In multi-matrix groups, the median energy-preserving to isotropic response ratios were 0.77 and 1.00, respectively.

\subsection*{Additional random controls, worst-case directions and local linearity}

We estimated which weight changes of a given norm would alter the forecast most. The search converged in 50 of 70 group--initial-condition pairs in ACE2-ERA5\cite{r1} and 8 of 24 in the DLESyM v1.0 atmosphere\cite{r5,r34,r42}. The largest responses found are therefore lower bounds on the maxima. The median scaling response divided by the largest found was 0.019 and 0.0027, respectively.

We measured the error of a linear approximation as the norm of its difference from the forecast response divided by the norm of that response. At $\times$0.9 the median relative error was 0.08 in ACE2-ERA5 and 0.87 in DLESyM. At $\pm$0.1~\% it was below 0.01 in all 47 groups. The linear approximation still gave similar rankings to scaling (rank correlations 0.999 and 1.0, respectively). Four of the 13 scored predictions for these random-control, worst-case and linearity tests passed.

\subsection*{Fingerprint test on held-out states: details}

An earlier test on four other 2001 dates could not distinguish the groups. Predicting the same fields and regions for every group passed its rule, so we use those dates only for rankings (Supplementary Fig. 3b). On the separate evaluation dates at $\times$0.5, the mean same-group correlation was 0.90. Half-samples of fields and regions identified a median of 28 groups (minimum 26).

At $\times$0.9, correlations were 0.79--0.88 for matched interventions and 0.24--0.41 for scaling versus random changes. All 16 matched comparisons had permutation p = 0.0001 (Holm-adjusted\cite{r64} p = 0.0016). Eight of nine predictions for this per-date test passed.

\subsection*{Year-scale causal test: details}

Each of the six groups had one random weight change shared across the four initial conditions. Day 1 averages the first four forecast steps. We centred the regional fingerprints across the six $\times$0.9 edits and their six comparison-group controls. The table below gives each comparison group, the two day-1 correlations for the same-group random change, and the minimum correlations set for the scaled group.

{\footnotesize\setlength\tabcolsep{3pt}\begin{longtable}{>{\raggedright\arraybackslash\hspace{0pt}}p{0.108\linewidth}>{\raggedright\arraybackslash\hspace{0pt}}p{0.245\linewidth}>{\raggedright\arraybackslash\hspace{0pt}}p{0.182\linewidth}>{\raggedright\arraybackslash\hspace{0pt}}p{0.145\linewidth}>{\raggedright\arraybackslash\hspace{0pt}}p{0.224\linewidth}}\toprule
\textbf{Weight group} & \textbf{Comparable $\times$0.9 control} & \textbf{Random r with single-step fingerprint} & \textbf{Random r with day-1 scaled run} & \textbf{Pre-set minimum correlation (day 1 / month 1)} \\ \midrule\endhead
Direct input link & Decoder weights on the first 44 internal channels & 0.42 & 0.38 & 0.8 / 0.5 \\
Whole block 8 & Whole block 3 & $-$0.02 & 0.02 & 0.8 / 0.5 \\
Block-8 spectral filter & Block-3 spectral filter & 0.07 & 0.17 & 0.8 / 0.5 \\
Block-4 spectral filter & Block-7 spectral filter & 0.33 & 0.53 & 0.7 / 0.3 \\
Block-2 inner skip & Block-6 inner skip & 0.64 & 0.63 & 0.7 / 0.3 \\
Position embedding & Block-5 perceptron output & 0.06 & 0.08 & 0.8 / 0.5 \\
\bottomrule\end{longtable}}

Among detectable month-1 changes whose signs we had predicted, the year mean kept the same sign and 0.5--3 times the magnitude in 11/11 cases for the direct input link, 14/15 for block 8, 8/10 for its filter and 17/19 for the position embedding. At year 1, three groups matched their own single-step fingerprint most closely, below the four predicted. Month-1 fingerprints still identified all six edits and all six random controls at year 1. Supplementary Table 6 lists all 41 predictions. In the separate 35-group test from one initial condition, median year-mean scaling-to-random ratios were 0.988 at $\times$0.9 and 0.994 at $\times$1.1 (0.994 pooled). These ratios describe the sample and do not establish equivalence.

\subsection*{Activation patching: details}

The RMS difference between duplicate ACE2-ERA5\cite{r1} runs was 0.16 times the tropical 850-hPa temperature contrast between warm and cold forcing. The response to CO$_2$ was small over one ACE2-ERA5 forecast step. In ACE2-SHiELD\cite{r1}, holding CO$_2$ fixed from 1979 to 2014 reduced the 1995--2014 trend in 500-hPa geopotential from 32.51 to 4.48 m$^{2}$ s$^{-2}$ per decade, a reduction of about 86~\% (Supplementary Table 3). Holding CO$_2$ fixed also suppressed near-surface warming in earlier work\cite{r1,r19}. The models and durations differ, so this comparison cannot isolate a timescale effect in ACE2-ERA5.

\subsection*{Scale sensitivity of the ACE2-ERA5 spectral filters: details}

Across the eight ACE2-ERA5\cite{r1} filters\cite{r43,r44}, editing weights at the same spatial scale as the output band gave the largest share of response in that band in all 24 comparisons. Conversely, middle- and small-scale edits produced their largest share at the edited scale in every block, but large-scale edits did so in only two. Per unit weight change, the large-scale band produced 4--16 times the response of the small-scale band. Small-scale random changes concentrated 0.83--0.97 of their response in that band, compared with 0.44--0.66 for scaling.

\subsection*{Perturbation map of DLESyM v1.0: details}

In the default DLESyM v1.0\cite{r42} configuration (atmosphere 0 with ocean 0), scaling each of 63 tensors separately by $\times$1.1 drove fields outside physical limits for 11 atmosphere and 12 ocean tensors. At $\times$0.9 this happened for two tensors. A field could leave its physical range while remaining finite. Scaling the first ocean encoder by $\times$0.9 raised Arctic sea-surface temperature by 2.2--2.6 K from one initial condition.

The table below compares the Arctic drift in 1990 with the three-year Arctic mean bias north of 66.5$^\circ$ N. In 1990 the largest change on swapping a partner came from replacing ocean 0. Configurations 2/2 and 3/3 also drifted warm.

{\footnotesize\setlength\tabcolsep{3pt}\begin{longtable}{>{\raggedright\arraybackslash\hspace{0pt}}p{0.273\linewidth}>{\raggedright\arraybackslash\hspace{0pt}}p{0.385\linewidth}>{\raggedright\arraybackslash\hspace{0pt}}p{0.273\linewidth}}\toprule
\textbf{Atmosphere / ocean network numbers\cite{r5,r34,r42}} & \textbf{1990 Arctic Dec--Jan drift minus ERA5\cite{r7} seasonal change (K)} & \textbf{1990--1992 Arctic mean bias (K) and status} \\ \midrule\endhead
0 / 0 (default) & +13.65 & +6.41 (finite) \\
1 / 1 & +0.18 & $-$0.77 (finite) \\
2 / 2 & +8.80 & failed on 31 July 1992 \\
3 / 3 & +7.40 & +4.08 (finite) \\
0 / 1 (crossed) & +1.75 & +1.17 (finite) \\
1 / 0 (crossed) & +13.69 & +5.24 (finite) \\
\bottomrule\end{longtable}}

\subsection*{Localisation of the NeuralGCM instability: details}

In the 1.4$^\circ$ stochastic NeuralGCM\cite{r3,r4}, over 24 hours, perturbations of the top three levels grew by a factor of 2.29 with learned physics and 1.27 in the dynamical core alone. From 1 January 1995 the two unedited runs with the pressure correction crossed the divergence threshold on days 94 and 101 and stayed finite through days 113 and 125. Without the correction the corresponding days were 100 and 129.

From the state saved on 2 March 1995, three size-matched random edits crossed the threshold after 10, 20 and 110 days. Halving the top six wind columns kept maximum divergence at 0.022 without the pressure correction and 0.019 with it. Halving the corresponding columns on levels 4--6 also kept divergence below 0.1 for 365 days. Halving those on levels 11--13 gave onset on day 39, against day 41 unedited (Supplementary Fig. 9a, b). A weaker multiplier of $\times$0.75 on levels 1--3 failed in month 20 in a run initialised on 1 January 1995.

\subsection*{Edits and their evaluation: details}

\textbf{NeuralGCM.} Zeroing all 18 output columns of the top three levels caused runaway on day 53. With the six wind columns halved, five-year 850-hPa temperature RMSE was 0.48--0.53 K for seeds 2--6, compared with 0.84 K for the 2.8$^\circ$ stochastic precipitation model\cite{r49,r50}. Seed 1 reached 1.05 K. Its ten-year runs from 1995 and 2015 both cooled to about 200 K at 50 hPa, against a 1995--2014 ERA5\cite{r7} mean of 211.0 K. At the extrapolated 10-hPa level the unedited model's day-1 global mean was about 11 K below the January 1995 ERA5 mean.

\textbf{DLESyM.} The decoder calculation used a regularisation penalty in place of the term that fits the intercept to the data. The timing of the fitting choices is uncertain, so we treat them as exploratory. We evaluated the weights as calculated. Ten preregistered runs from 1986 and 1993 compared two atmosphere networks with the edited ocean. Atmosphere 1 missed the criteria. We selected atmosphere 0 after that comparison. All five climatological RMSEs fell with this pairing in both evaluation periods, but 16 and 23 of 80 climate checks failed.

The first table below gives global 2-m temperature RMSE and the number of failed climate checks for the five evaluation periods. Arctic bias fell from +7.14 to +2.03 K in 1986--1989. For the 2005 initial condition, extending the evaluation from ten to twenty years changed the unedited mean Arctic bias from +5.79 to +5.70 K. With the edit the twenty-year bias was +0.39 K.

{\footnotesize\setlength\tabcolsep{3pt}\begin{longtable}{>{\raggedright\arraybackslash\hspace{0pt}}p{0.179\linewidth}>{\raggedright\arraybackslash\hspace{0pt}}p{0.224\linewidth}>{\raggedright\arraybackslash\hspace{0pt}}p{0.201\linewidth}>{\raggedright\arraybackslash\hspace{0pt}}p{0.314\linewidth}}\toprule
\textbf{Evaluation period} & \textbf{Unedited 2-m RMSE (K)} & \textbf{Edited 2-m RMSE (K)} & \textbf{Failed climate checks (of 80)} \\ \midrule\endhead
1986--1989 & 2.57 & 1.61 & 16 \\
1993--1996 & 2.47 & 1.57 & 23 \\
1995--2001 & 2.53 & 1.59 & 26 \\
2003--2014 & 2.34 & 1.52 & 9 \\
2005--2014 & 2.34 & 1.48 & 12 \\
\bottomrule\end{longtable}}

The edited and unedited runs from 1995 lacked matched run-time settings (Methods, `Edits and their evaluation').

The table below shows the Arctic warm bias in 48-month unedited 0/0 runs with two inference-package versions\cite{r53}, each using its own initial conditions procedure.

{\footnotesize\setlength\tabcolsep{3pt}\begin{longtable}{>{\raggedright\arraybackslash\hspace{0pt}}p{0.314\linewidth}>{\raggedright\arraybackslash\hspace{0pt}}p{0.333\linewidth}>{\raggedright\arraybackslash\hspace{0pt}}p{0.285\linewidth}}\toprule
\textbf{Initial condition (0/0, 48 months)} & \textbf{Arctic bias, 0.19.0a0 vs 0.8.0\cite{r53} (K)} & \textbf{2-m RMSE, 0.19.0a0 vs 0.8.0 (K)} \\ \midrule\endhead
January 1986 & +7.14 vs +7.15 & 2.571 vs 2.572 \\
July 1989 & +7.45 vs +7.28 & 2.623 vs 2.565 \\
July 2019 & +5.54 vs +5.52 & 2.127 vs 2.159 \\
\bottomrule\end{longtable}}

\subsection*{Additional NeuralGCM evaluation: details}

From January 1995, all six edited runs of the 1.4$^\circ$ stochastic NeuralGCM\cite{r3,r4} and one of six unedited runs stayed finite for 20 years (one-sided Fisher p = 7/924). On the monthly physical checks, all six edited runs passed throughout and no unedited run passed beyond four years (p = 1/924). Both comparisons used the pressure correction. The year-1 predictions of non-finite output failed. One of six unedited runs and two of three random controls became non-finite that year. None of the controls passed the monthly checks after July 1995. With SST and sea ice updated every 12 hours and the developers' full checkpoint modification, which includes the pressure correction, all six unedited runs failed the monthly checks between May 1995 and October 1998. In a separate 150-day comparison with the pressure correction alone and the same 12-hour update interval, both the unedited and edited runs from January 1995 with seed 1 stayed finite. Peak absolute divergence on the top two levels was 0.049 unedited and 0.017 edited, below the 0.1 threshold in both runs. In a second pair with the same initial state, seed and settings, the unedited run crossed the divergence threshold on day 96, first produced non-finite output on day 118 and failed the April 1995 monthly check. The edited run stayed finite for 730 days and passed the monthly checks in all 24 months, with peak divergence of 0.021.

A second group among the authors recalculated the regional checks from the same months and regridded fields. At 850 hPa north of 60$^\circ$ N, seed 1 averaged 4.70 K colder than seeds 2--6 in 2010--2014. Its regional cold error against ERA5\cite{r7} exceeded 3 K and was hidden by the six-seed median. For the deterministic 1.4$^\circ$ model, all 37 initial conditions met the 850-hPa temperature error limits against the published archive\cite{r65} at initialisation and 48 h. Kochkov et al.\cite{r3} reported instability on day 389 from 18 September 2019. Our runs from that date and the other archive failure date\cite{r65}, 20 June 2019, stayed finite for two years.

\subsection*{DLESyM evaluation: control and runaway details}

From the 1998 initial condition, 48-month 2-m temperature RMSE was 2.41 K unedited, 1.40 K with the full SST offset and 1.48 K with the 0.65 offset. Three size-matched random controls produced all 48 months of output but exceeded physical limits in months 1, 2 and 6. From this initial condition the full-offset edit stayed within physical limits longer than all three controls. Six conditional predictions were not scored because they required at least two controls to stay within those limits.

From 1 January 1995, unedited DLESyM v1.0\cite{r5,r34,r42} configuration 0/0 ran away in the Southern Ocean on 14 July 2001. Including 1/1, 2/2 and 3/3, the four runaways occurred after 4.6--6.8 years. The 20-year edited run used different run-time settings. Other initial conditions gave finite unedited runs.

\subsection*{Regional effects of the DLESyM edit: details}

Each of eight initial conditions supplied one paired sample in this deterministic model. A difference from the unedited model counted as better or worse only if the paired test flagged it and at least seven initial conditions agreed on the change in error against ERA5\cite{r7}. For unflagged differences, equivalence margins were 0.3 K for air temperature, 0.5 kg m$^{-2}$ for column water vapour, 10 m for 500-hPa height, 0.2 K for SST and a factor of 1.25 for variability.

Only 2-m temperature met the area rule across the 2,751 comparisons (Methods, `Regional effects of the DLESyM edit'). Its improved area exceeded twice the worsened area, with no flagged cell worsening by more than twice its margin. In selected polar winter and spring regions the edit lowered 2-m air temperature by 5.6--8.1 K, closer to ERA5 on all eight initial conditions.

Five measures moved further from ERA5 on seven or eight initial conditions. Relative to the unedited runs, European summer SST fell 0.49 K and the North American coastal SST seasonal-cycle amplitude fell 0.78 K. Summer column water vapour in the Sahel fell 1.25 kg m$^{-2}$ and autumn column water vapour in the eastern tropical Pacific fell 1.19 kg m$^{-2}$. Southern Ocean autumn SST interannual standard deviation rose to 1.8 times its unedited value. Arctic 500-hPa geopotential bias over 2005--2024 shifted from +128 to $-$175 m$^{2}$ s$^{-2}$.

\subsection*{Offset-factor dose test (DLESyM): details}

We compared the DLESyM v1.0\cite{r5,r34,r42} SST offset with rotations of the same offset and a size-matched change in another tensor. In 1998, raising the factor from 0.65 to 0.70 raised error by 0.08 K, exceeding the allowed 1.5~\% of unedited error. In 2008, unedited RMSE was 2.18 K, below the 2.2 K minimum for scoring. Recalculating the decoder weights alone supplied about two thirds of the reduction in global 2-m RMSE. In v1.1\cite{r42,r53} the 0.65 factor increased absolute annual Southern Ocean SST bias by 1.18, 1.10, 0.57 and 1.43 K from the four initial conditions. The random controls increased bias by 0.74--2.01 K.

\subsection*{Pangu-Weather surface-head edit: details}

Across four ERA5\cite{r7} initial conditions in 1955 and 2023, the 24-hour Pangu-Weather model\cite{r27} cooled at $-$0.034 K d$^{-1}$ relative to ERA5 over ten days, close to the published $-$0.03 K d$^{-1}$ for both years\cite{r6}. Shifting the forecast deviation from ERA5 by 90$^\circ$ in longitude reversed the extra cooling. The test could therefore not identify a local cause.

We fitted the output edit by ridge regression\cite{r57} on 12 dates and screened it on 12 others. We used 24 development runs to choose two constant corrections to the forecast temperatures. Adding 0.001242 normalised units per step matched the edit's fractional reduction of 0.431 in the 10-day error slope. Adding 0.002884 brought the slope to +0.0002 K d$^{-1}$. Across 24 evaluation dates in 1979--1987, 17 met the minimum absolute smoothing response of 0.005 K. Across all 24 dates, the fractional reduction in median extra cooling was 1.33 for the edit and $-$0.01 to +0.10 for random edits (Supplementary Fig. 11b, right).

The table below gives median fractional reductions in 10-day drift outside the training years. Eligible dates were 12 and 11 at 2 m, and 11 and nine at 850 hPa, respectively. In 2020--2023 four criteria failed, including the random-control margin (0.147 against 0.15) and latitude-band check (upper bound +0.154). In the earlier period only the 850-hPa comparison missed. Neither period established an advantage over a constant correction.

{\footnotesize\setlength\tabcolsep{3pt}\begin{longtable}{>{\raggedright\arraybackslash\hspace{0pt}}p{0.192\linewidth}>{\raggedright\arraybackslash\hspace{0pt}}p{0.165\linewidth}>{\raggedright\arraybackslash\hspace{0pt}}p{0.165\linewidth}>{\raggedright\arraybackslash\hspace{0pt}}p{0.192\linewidth}>{\raggedright\arraybackslash\hspace{0pt}}p{0.192\linewidth}}\toprule
\textbf{Intervention} & \textbf{2-m temperature, 2020--2023} & \textbf{2-m temperature, 1957--1960} & \textbf{850-hPa temperature, 2020--2023} & \textbf{850-hPa temperature, 1957--1960} \\ \midrule\endhead
Weight edit & 0.267 & 0.557 & 0.024 & 0.072 \\
Response-matched fixed warming & 0.422 & 0.587 & 0.077 & 0.133 \\
Zero-drift fixed warming & 0.741 & 0.591 & 0.206 & 0.306 \\
Three random weight changes & 0.002--0.120 & 0.010--0.147 & $-$0.004 to 0.016 & 0.001--0.024 \\
Negated edit & $-$0.346 & $-$0.668 & 0.015 & $-$0.055 \\
\bottomrule\end{longtable}}

\subsection*{Forecast edits in weather models: details}

\textbf{Aurora 1.5\cite{r17,r35}.} Scaling attention in the finest-resolution encoder and decoder and offsetting the resulting pressure drift reduced mean error by 0.79~\% across 42 dates. On development dates mean error fell by 0.85~\%, short of the 1~\% required, and two of the 28 numbers worsened by more than the 0.5~\% limit (0.74~\% and 0.61~\%). On 12 held-out dates, relaxing the unedited forecast towards its initial state reduced error by 4.15~\%. The weight edit contributed a further 0.73~\% after relaxation. Day-10 tropical 250-hPa height bias rose from +43 to +72 m.

\textbf{Operational GraphCast\cite{r13,r22}.} Scaling the 2-m temperature output weights by 0.97 reduced error by 0.44~\% on 36 dates against ERA5\cite{r7} and 0.49~\% against operational ECMWF analyses\cite{r66}. Those dates had already tested a three-output version of the edit. On 12 held-out dates, relaxation alone reduced error by 2.70~\% and the edit added 0.24~\%.

\textbf{AIFS Single 2.0\cite{r36,r37}.} Scaling the output increment by 0.993 reduced error by 0.38~\% on development dates and 0.66~\% on held-out dates. Relaxation alone reduced development error by 2.39~\%, and the edit then changed it by $-$0.27~\% (90~\% interval $-$0.57 to +0.02~\%). For six wave-height fields absent from our ERA5 inputs, the first substitute inflated unedited error by 19~\%. Two apparent improvements disappeared with consistent inputs and worsened forecasts from operational analyses\cite{r66} by 1.5--3.6~\%.

\textbf{Pangu-Weather, 6-hour model\cite{r27}.} Moving the output biases towards the development correction reduced mean error by 1.31~\% on 12 held-out dates, but day-10 meridional wind error rose by 1.04~\%, above the 0.5~\% limit. A constant per field and lead time, estimated on development dates and added to the unedited forecasts, reduced it by 0.98~\%.

\textbf{Ensembles.} Both tested noise scalings increased error in Aurora 1.5 Ensemble\cite{r17,r35}. In WeatherNext 2\cite{r41}, scaling the noise-encoder weights by 1.4 reduced the score by 2.86~\% on development dates and 0.26~\% on held-out dates (90~\% interval $-$1.19 to +0.64~\%). A size-matched random change to the same weights reduced the development score by 1.7~\%. In AIFS ENS v2\cite{r38,r39,r40}, an edit that reduced error on forecasts initialised from ERA5 increased it by 1.17~\% on forecasts from six operational ensemble initial conditions evaluated at 24, 48 and 72 h against operational analyses\cite{r66}.

\subsection*{Other weight edits}

Apart from the NeuralGCM, DLESyM and Pangu-Weather cases described in the main text, every other tested edit fell short of its pre-specified criteria. Supplementary Table 3 gives the comparisons with controls and the diagnostic tests for errors outside the weights.

\subsection*{Workflow, preregistration and verification: details}

Under our direction, language models from three developers produced the measurements, analyses and most of the code. Recalculations used stored outputs, while independent simulations produced new forecasts. The findings apply to the workflow as a whole.

\subsection*{Statistics: details}

We report the ACE2-ERA5\cite{r1} year-scale prediction tally without adjustment for multiple comparisons (Supplementary Fig. 6). When we reassigned prediction sets across weight groups, 2 of 720 permutations reached 12 passes (mean 4.7). The groups had unequal numbers of predictions, so this comparison was exploratory.

\clearpage
\setcounter{figure}{0}\renewcommand{\figurename}{Supplementary Fig.}
\section*{Supplementary Figures}
\begin{figure}[!ht]\centering\includegraphics[width=\linewidth,height=0.62\textheight,keepaspectratio]{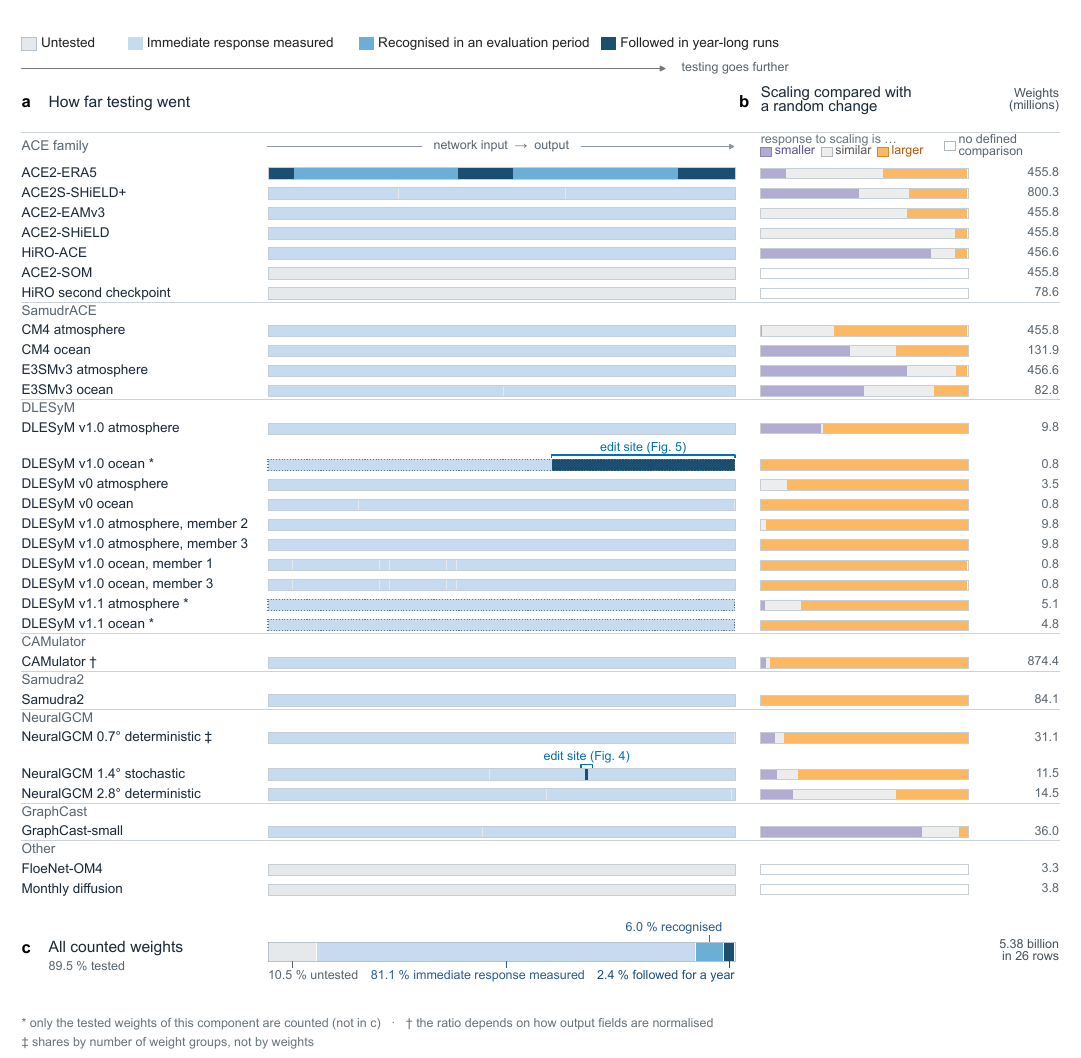}
\caption{\textbf{About nine tenths of counted weights have had their immediate response measured, while few have been followed in long runs or edited.} \textbf{a}, One strip per model component, from input (left) to output (right). Width shows a weight group's share of the component's weights (smallest direct-link strips widened for visibility), and colour shows how far its testing went. Brackets mark the two edit sites in the atlas, in NeuralGCM and DLESyM; Pangu-Weather was evaluated separately. \textbf{b}, Share of weights whose scaling response was smaller than, similar to or larger than a same-size random change (ratio below 0.8, 0.8--1.25, above 1.25; white: no defined comparison). \textbf{c}, Test levels from \textbf{a} across all 5.38 billion weights in the 26 rows with complete weight counts.}\label{sfig1}\end{figure}
\begin{figure}[tbp]\centering\includegraphics[width=\linewidth,height=0.72\textheight,keepaspectratio]{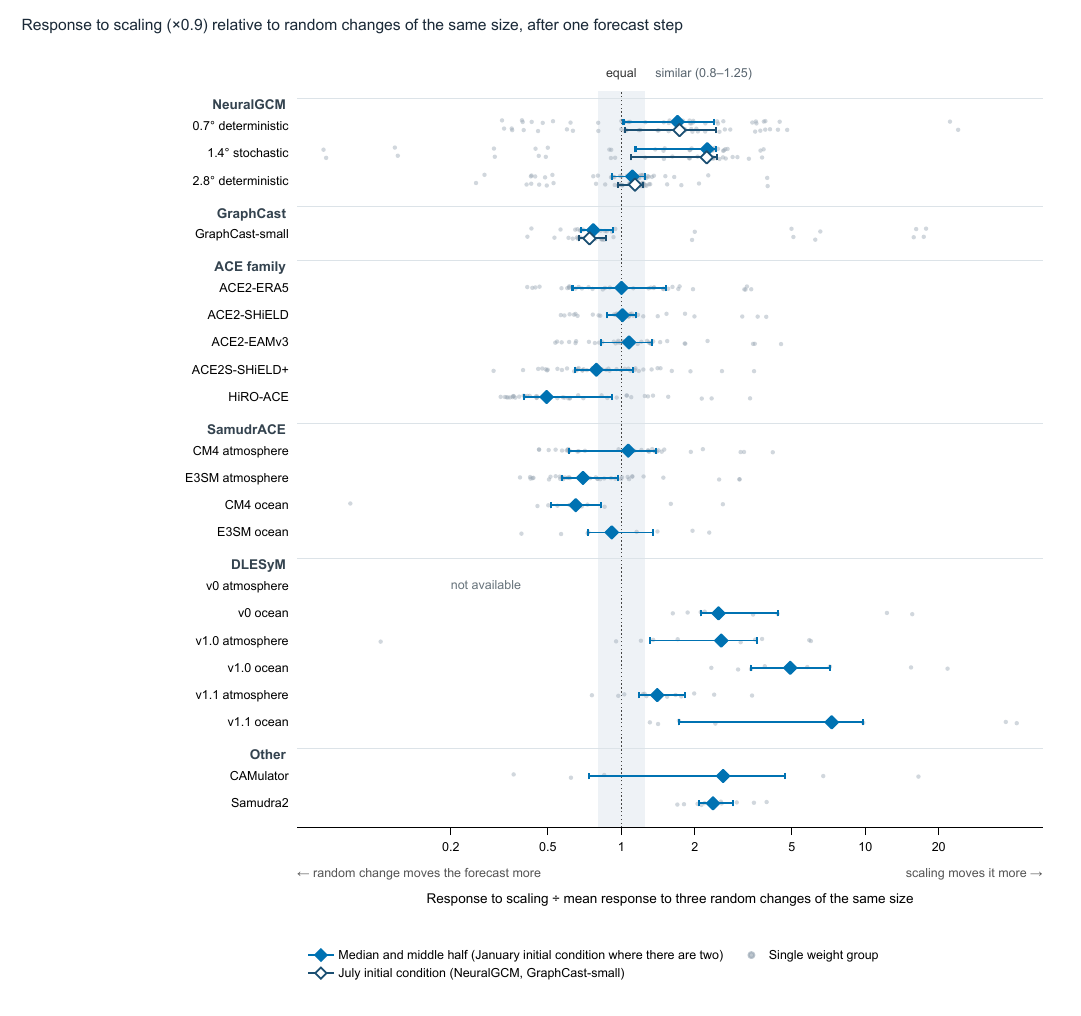}
\caption{\textbf{Scaling model weights changes a forecast about as much as a random change in some model families, and several times more in others.} Twenty model components have an estimate (DLESyM v0\cite{r2,r51} atmosphere: not available). For each group, its single-step $\times$0.9 response is divided by the mean response to three same-size random changes. Markers show the median over weight groups, bars span the interquartile range and dots mark single groups. NeuralGCM\cite{r3,r4} and GraphCast-small\cite{r13,r22} show separate markers for January and July 1995 ERA5\cite{r7} initial conditions on native output fields; other rows use one marker. ACE\cite{r1,r19,r46,r47}, SamudrACE\cite{r32,r33} and GraphCast-small sit near or below 1, whereas NeuralGCM, DLESyM\cite{r2,r5,r34,r42,r51}, CAMulator\cite{r21} and Samudra2\cite{r23} sit above 1. With the bands of Supplementary Fig. 1b, the median ratio is above 1.25 in 9 components, between 0.8 and 1.25 in 6 and below 0.8 in 5.}\label{sfig2}\end{figure}
\begin{figure}[tbp]\centering\includegraphics[width=\linewidth,height=0.72\textheight,keepaspectratio]{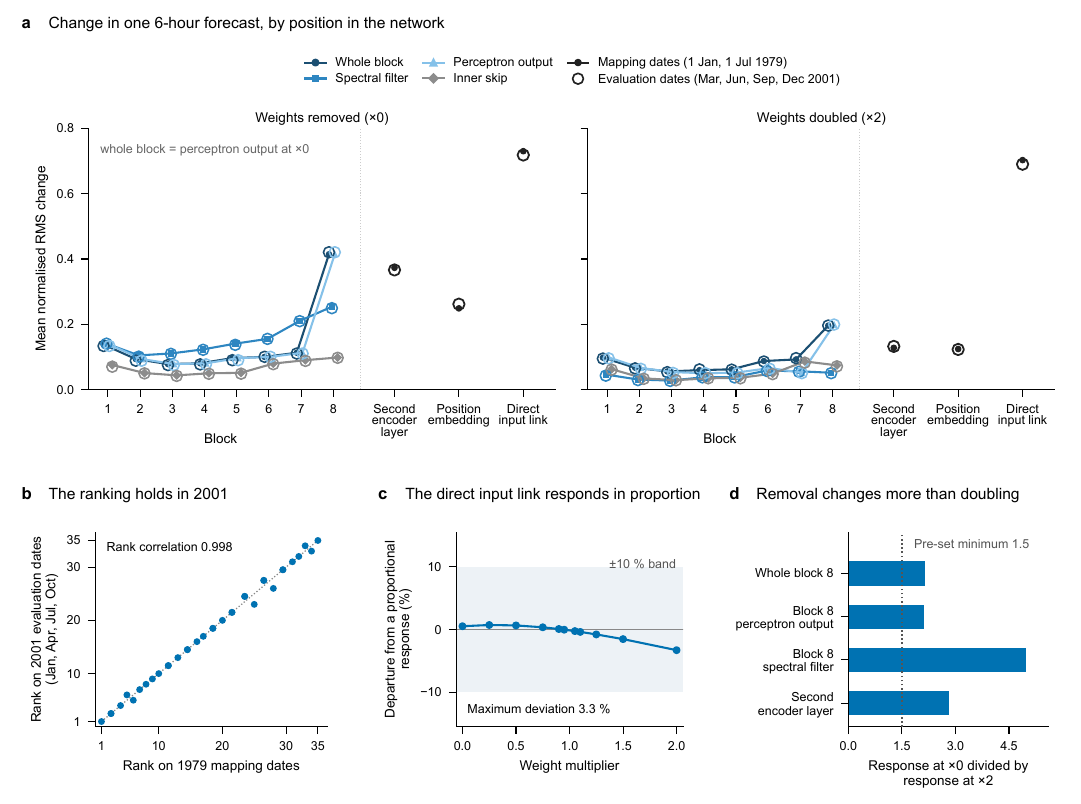}
\caption{\textbf{ACE2-ERA5's immediate sensitivity map holds on 2001 evaluation initial conditions not used to build it.} \textbf{a}, Forecast change in ACE2-ERA5\cite{r1} after one 6-hour step when each of 35 weight groups is removed ($\times$0, left) or doubled ($\times$2, right). Source Data lists 385 paired values. Filled symbols show 1 January and 1 July 1979\cite{r7,r30}; open circles show 1 March, June, September and December 2001\cite{r1,r30}. Effects did not rise steadily from block 1 to block 7. \textbf{b}, Rank on 1 January, April, July and October 2001\cite{r1,r30} against 1979 rank (correlation 0.998). \textbf{c}, Direct-input-link response stays proportional to weight change within 3.3~\%. \textbf{d}, Across four preselected groups, removing weights changes forecasts 2.1--5.0 times as much as doubling them (minimum threshold of 1.5 set in advance).}\label{sfig3}\end{figure}
\begin{figure}[tbp]\centering\includegraphics[width=\linewidth,height=0.72\textheight,keepaspectratio]{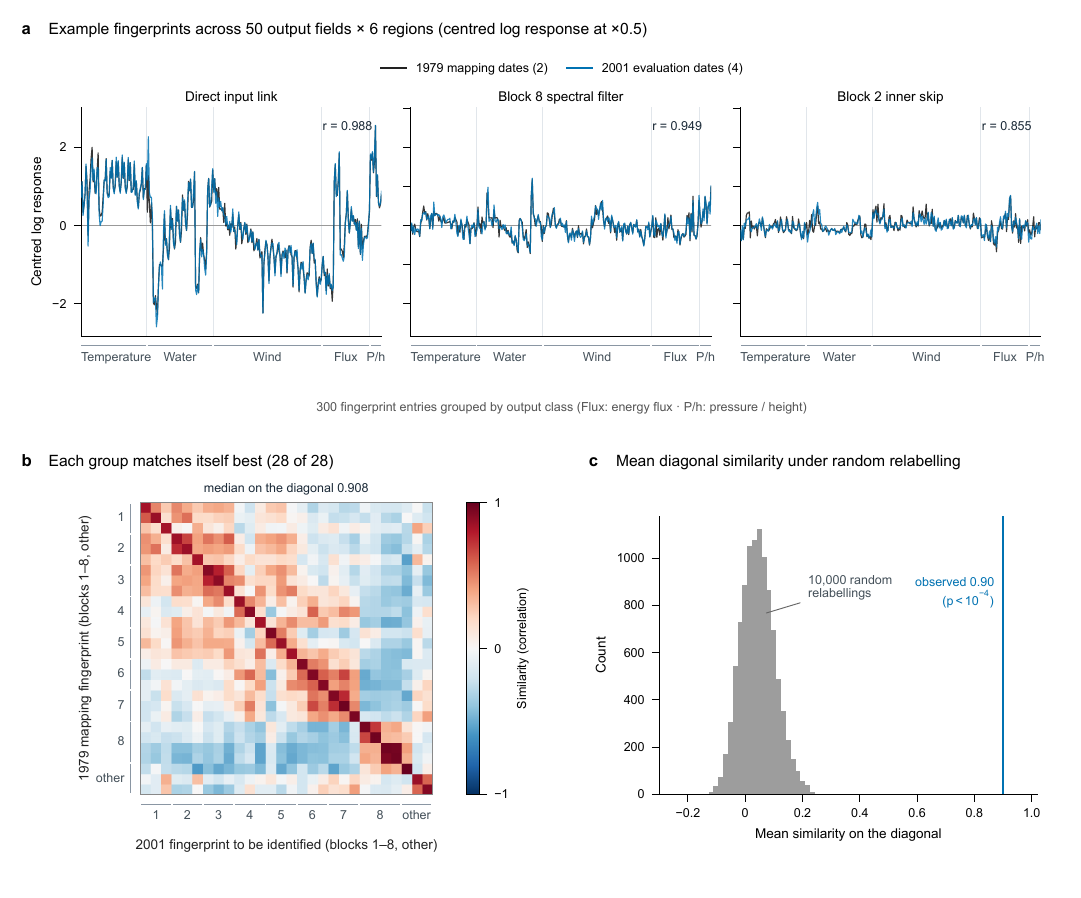}
\caption{\textbf{A fingerprint is a group's response pattern across output fields and regions, and it repeats on 2001 evaluation dates.} \textbf{a}, Fingerprints of three weight groups at $\times$0.5 after one forecast step across 50 output fields in six regions, centred across all 28 groups first. For each period, we averaged ACE2-ERA5\cite{r1} responses over dates before computing the fingerprints (two 1979 dates\cite{r7,r30} in black, four 2001 dates\cite{r1,r30} in blue). \textbf{b}, Correlation matrix between 1979 mapping fingerprints (rows) and 2001 evaluation fingerprints (columns) across 28 groups. Every group matches itself best on the diagonal (median 0.908). \textbf{c}, Mean diagonal correlation across 10,000 random relabellings of group names (grey) versus the observed mean of 0.90 (blue, p < 10$^{-4}$).}\label{sfig4}\end{figure}
\begin{figure}[tbp]\centering\includegraphics[width=\linewidth,height=0.72\textheight,keepaspectratio]{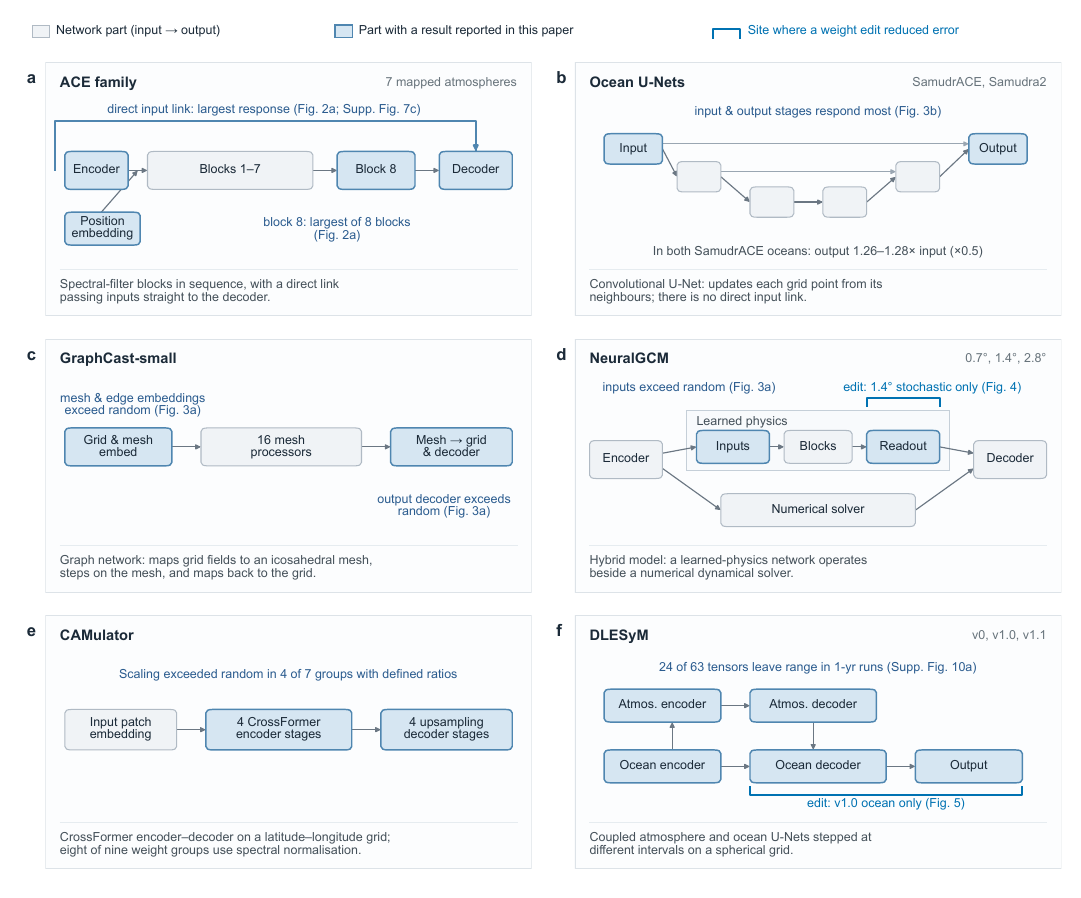}
\caption{\textbf{Guide to the six network designs studied and where reported results sit.} Tiles show model families from input (left) to output (right). Pale blue marks parts with reported results, and brackets mark the two atlas edit sites. \textbf{a}, ACE family\cite{r1,r43,r44}, with a direct input link to the decoder. \textbf{b}, The ocean components of SamudrACE-CM4\cite{r32} and SamudrACE-E3SMv3\cite{r33} and the Samudra2\cite{r23} ocean model are U-Nets without a direct link. In the two SamudrACE components, the $\times$0.5 output-to-input ratio is 1.26--1.28. \textbf{c}, GraphCast-small\cite{r13,r22}. \textbf{d}, NeuralGCM\cite{r3,r4}, pairing learned physics with a dynamical solver (edit in 1.4$^\circ$ stochastic only). \textbf{e}, CAMulator\cite{r21}, a CrossFormer\cite{r62} encoder--decoder on a latitude--longitude grid, where scaling exceeded random in 4 of 7 groups. \textbf{f}, Coupled DLESyM\cite{r2} (edit in v1.0\cite{r42} ocean only). Evidence grades appear in Supplementary Table 2.}\label{sfig5}\end{figure}
\begin{figure}[tbp]\centering\includegraphics[width=\linewidth,height=0.72\textheight,keepaspectratio]{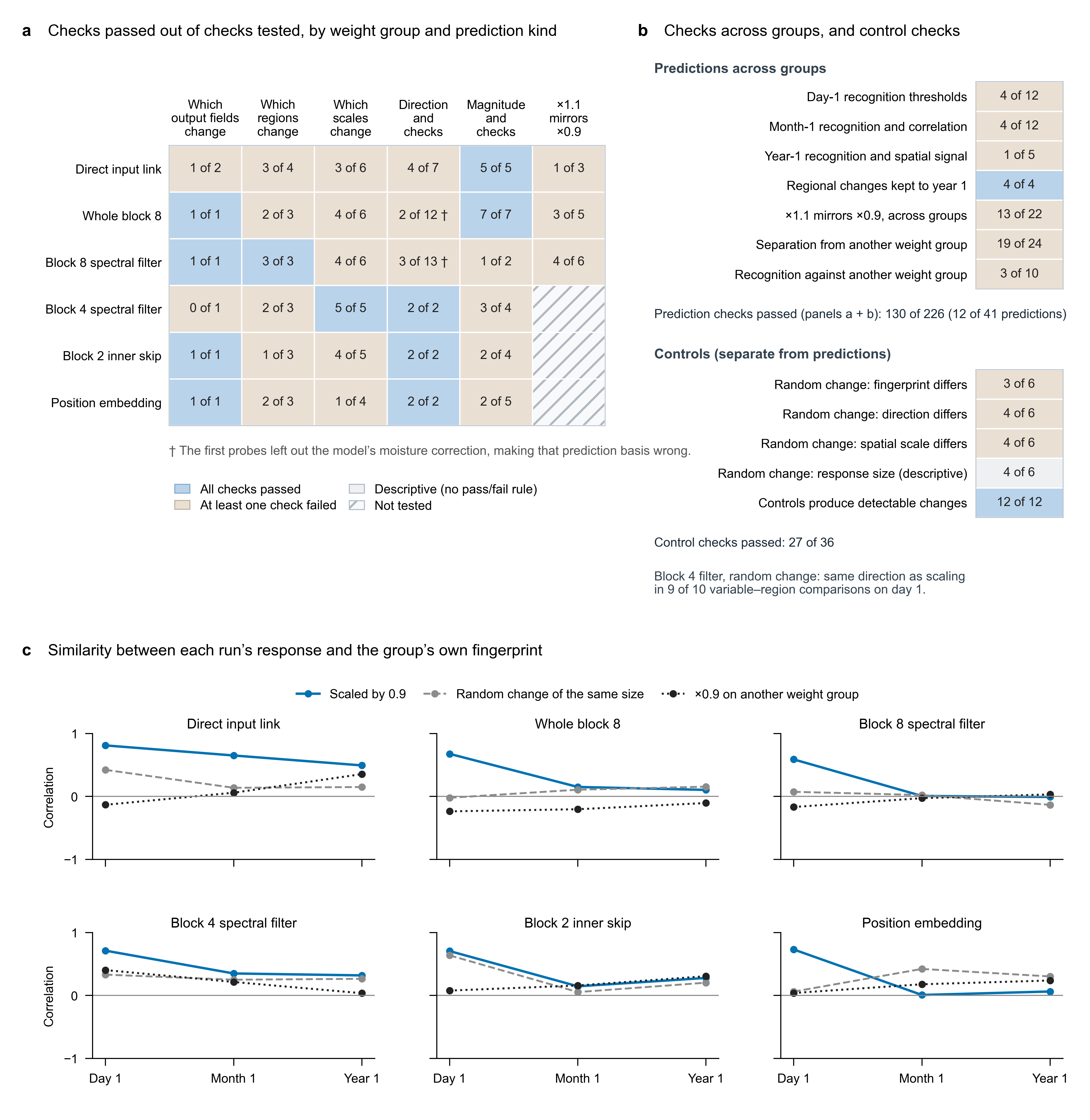}
\caption{\textbf{Most predictions in the year-long ACE2-ERA5 test passed some checks without passing all.} \textbf{a}, For each of six weight groups and six kinds of prediction, the tiles show how many checks passed out of the number tested. Blue indicates that every check for a prediction passed. A dagger (\textsuperscript{\dag}) marks ACE2-ERA5\cite{r1} predictions based on preliminary runs without the model's moisture-budget correction\cite{r1,r44}. The direct input link's `Direction and checks' tile combines two predictions, one about temperature, humidity, height and heat flux and one about wind and precipitation. \textbf{b}, Seven cross-group predictions (4 of 12 day-1 fingerprint checks with fixed correlation thresholds passed) and five control rows. For the block-4 filter, a random change matched scaling's day-1 direction in 9 of 10 variable--region comparisons. \textbf{c}, Correlation with each group's own fingerprint for $\times$0.9 and both controls. Overall, 130 of 226 prediction checks (12 of 41 predictions, Supplementary Table 6) and 27 of 36 control checks passed.}\label{sfig6}\end{figure}
\begin{figure}[tbp]\centering\includegraphics[width=\linewidth,height=0.72\textheight,keepaspectratio]{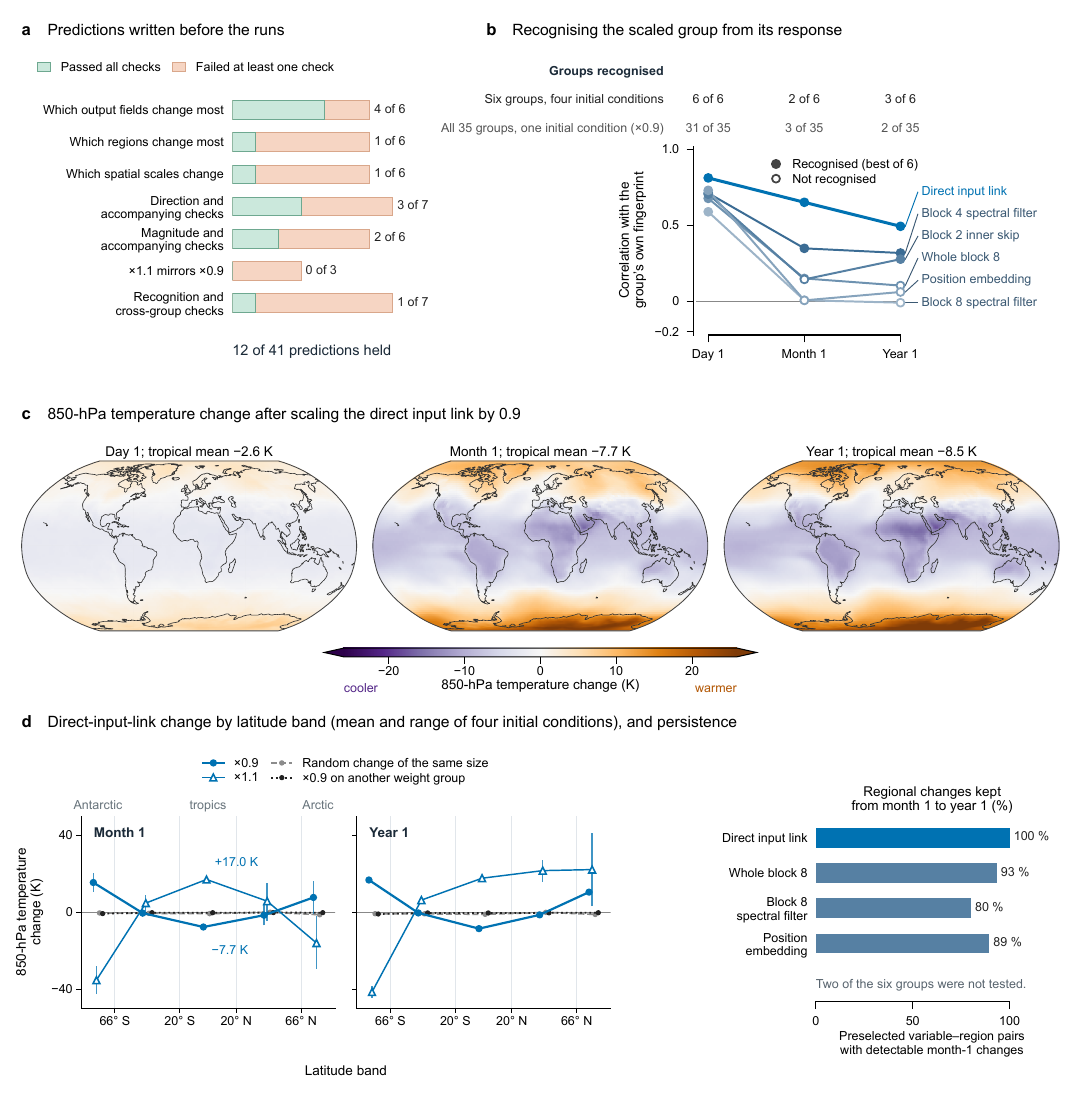}
\caption{\textbf{Immediate sensitivity maps of ACE2-ERA5 fail most year-long climate predictions, while first-month shifts persist.} Six weight groups were scaled by 0.9 or 1.1 from four initial conditions (100 one-year runs with controls and baselines). \textbf{a}, Predictions that passed all their checks, by kind of prediction (12 of 41). Five of six groups met the direction thresholds alone. \textbf{b}, Correlation between each group's response and its own fingerprint. Filled symbols mark the groups still recognisable. The counts above the plot cover six groups tested from four initial conditions and all 35 groups tested from one initial condition. \textbf{c}, 850-hPa temperature change for direct input link $\times$0.9 (tropical means printed). \textbf{d}, Same change by latitude band, with controls (mean and range of four initial conditions), and share of preselected regional changes kept to year 1.}\label{sfig7}\end{figure}
\begin{figure}[tbp]\centering\includegraphics[width=\linewidth,height=0.72\textheight,keepaspectratio]{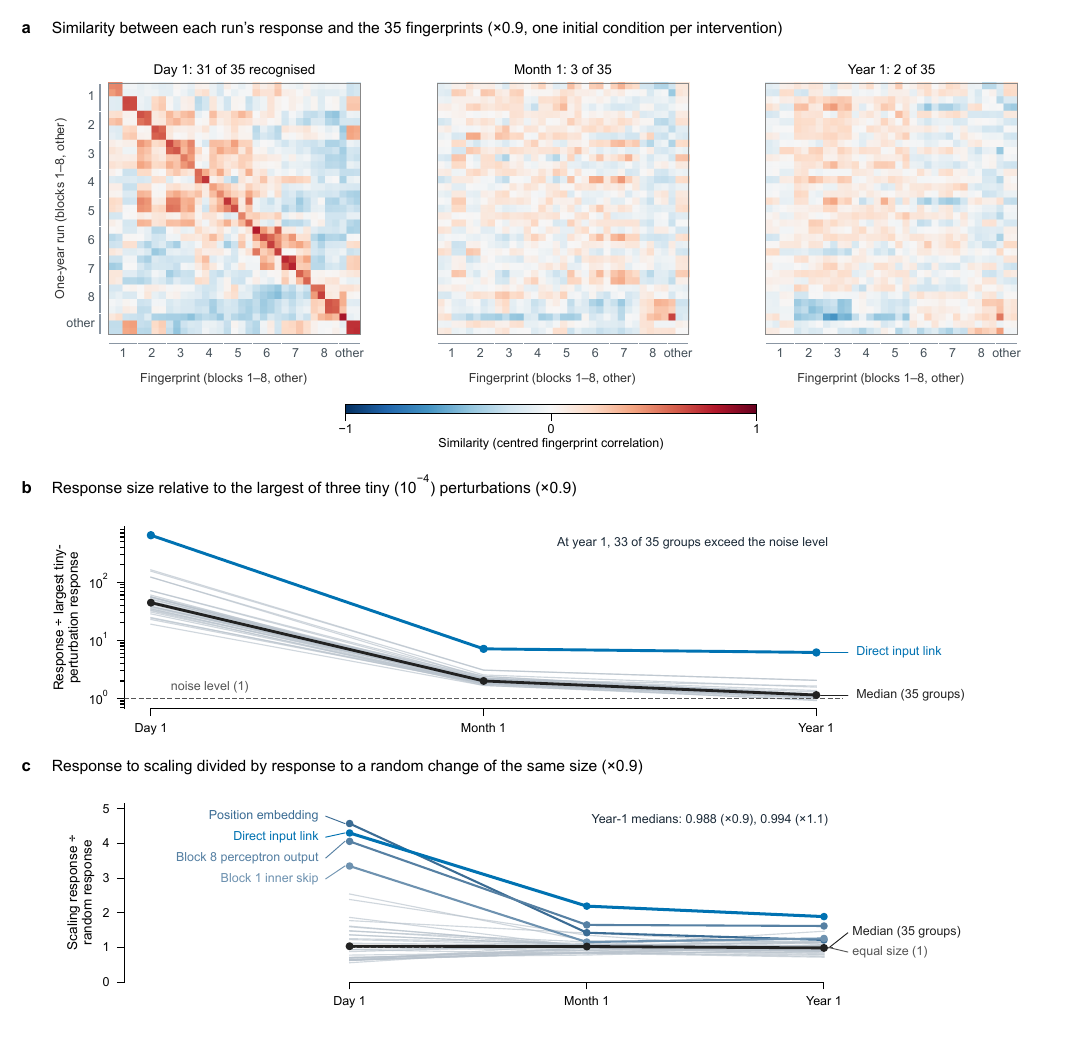}
\caption{\textbf{Most fingerprints were lost by month 1, although most responses remained detectable at year 1.} One year-long ACE2-ERA5\cite{r1} run per intervention across all 35 weight groups, from one 2001 initial condition\cite{r1,r30}. \textbf{a}, Similarity between each run's response pattern (rows) and the 35 single-step (6-hour) fingerprints (columns). At $\times$0.9, fingerprints identified 31 of 35 groups on day 1, three at month 1 and two at year 1. \textbf{b}, Response size relative to the largest response among three tiny (10$^{-4}$) perturbations; at year 1, 33 of 35 groups exceeded it. \textbf{c}, Forecast response to scaling divided by the forecast response to a random weight change of the same size. The year-1 medians are 0.988 ($\times$0.9) and 0.994 ($\times$1.1). Thin lines show single groups.}\label{sfig8}\end{figure}
\begin{figure}[tbp]\centering\includegraphics[width=\linewidth,height=0.72\textheight,keepaspectratio]{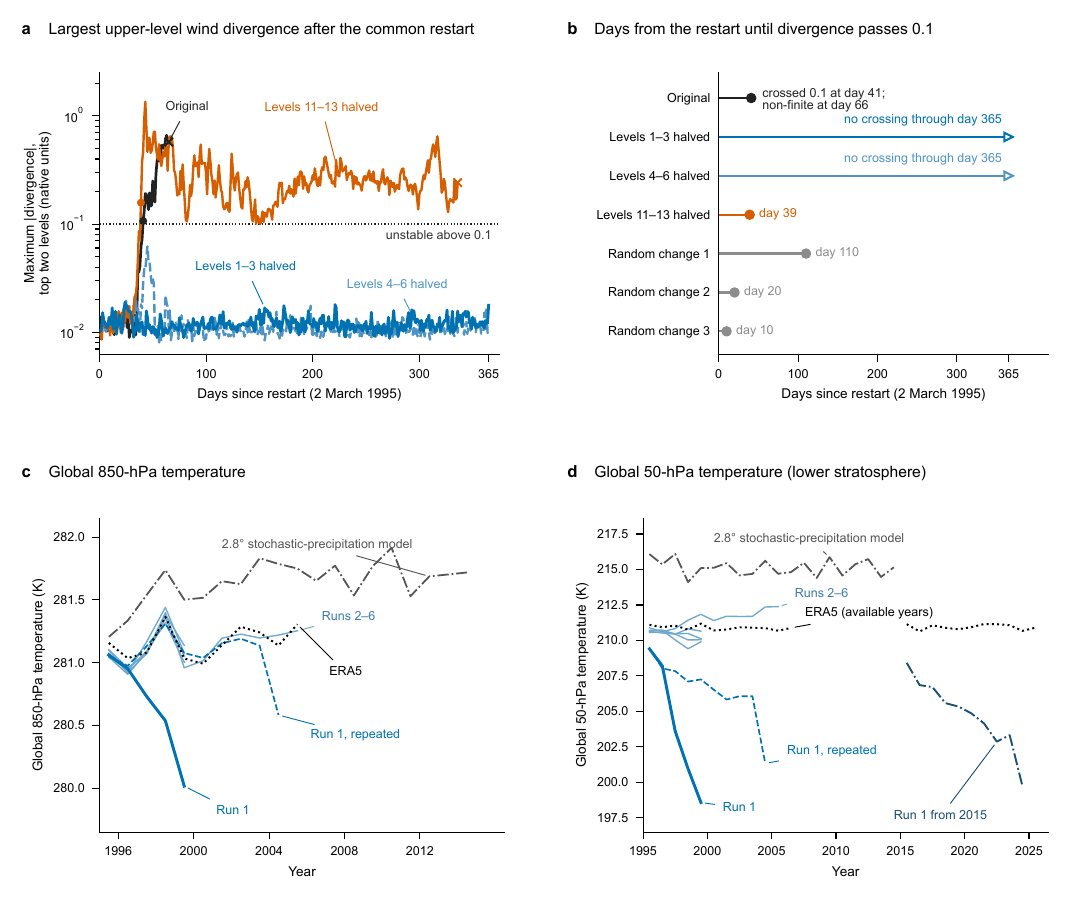}
\caption{\textbf{Halving NeuralGCM's upper-level wind columns suppresses breakdown in runs from the same saved state, while the edited run with seed 1 cools in the stratosphere.} \textbf{a}, Top-two-level daily maximum wind divergence in 1.4$^\circ$ stochastic NeuralGCM\cite{r3,r4} runs from the state saved on 2 March 1995 for unedited weights and halved wind columns on levels 1--3, 4--6 or 11--13 (0.1 dotted). \textbf{b}, Days from 2 March until divergence crosses 0.1, alongside three same-size random changes. The unedited run crosses 0.1 at day 41 and turns non-finite at day 66; levels 1--3 and 4--6 stay below 0.1 through day 365. \textbf{c}, Annual global 850-hPa temperature with the pressure correction\cite{r4} against ERA5\cite{r7} and 2.8$^\circ$ stochastic-precipitation NeuralGCM\cite{r49,r50}. \textbf{d}, Corresponding 50-hPa temperature against ERA5 (available years). `Original' marks the unedited model. `Run 1' is the cold ensemble member (noise seed 1). `Run 1, repeated' uses the same seed and initial condition in a separate job. `Run 1 from 2015' uses the same seed with the 2015 initial condition.}\label{sfig9}\end{figure}
\begin{figure}[tbp]\centering\includegraphics[width=\linewidth,height=0.72\textheight,keepaspectratio]{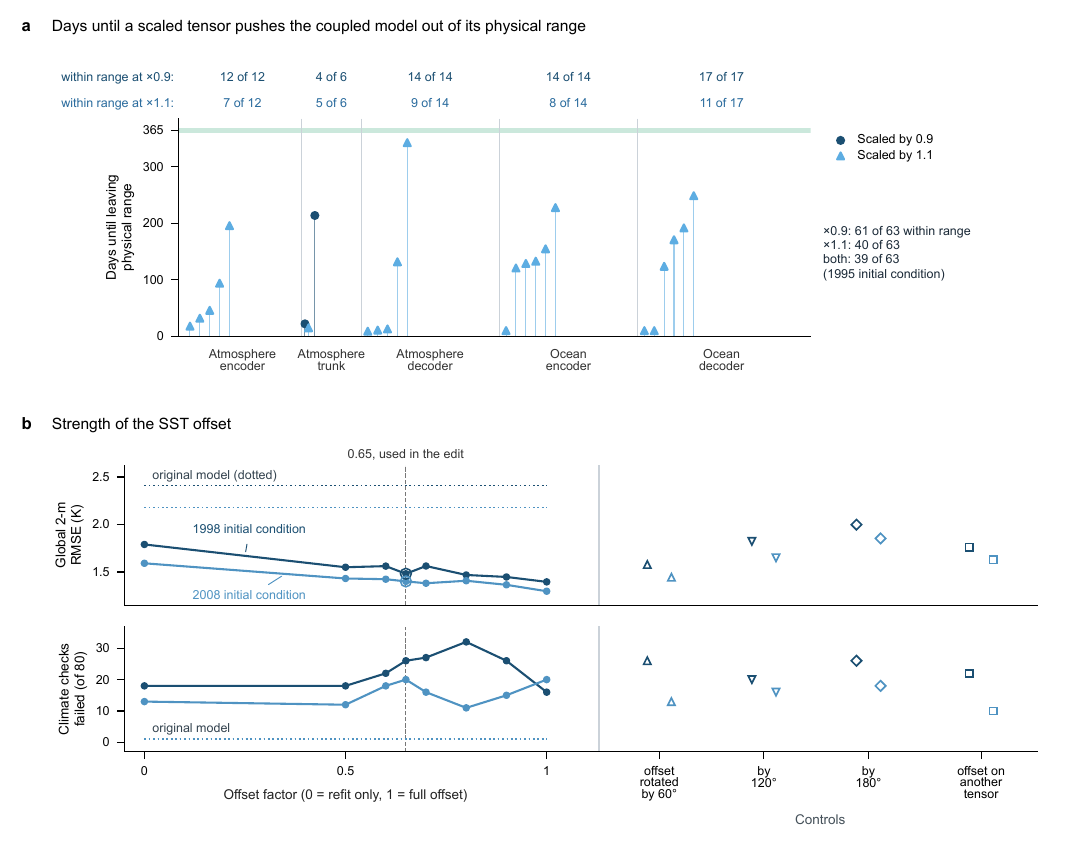}
\caption{\textbf{Increasing individual DLESyM v1.0 weight tensors by 10~\% produces values outside physical limits in many year-long runs. The tested SST offsets identify no optimal intermediate value.} \textbf{a}, Days to the first value outside physical limits after scaling each of 63 tensors in the default pairing\cite{r53} for DLESyM v1.0\cite{r42} (atmosphere network 0 with ocean network 0) in runs from the 1995 ERA5\cite{r7} initial condition. At $\times$0.9 (circles), 61 of 63 stayed within limits. At $\times$1.1 (triangles), 40 stayed within limits (39 at both multipliers). \textbf{b}, Four-year global 2-m temperature RMSE (top) and failed climate checks out of 80 (bottom; Methods, `Edits and their evaluation') across SST offset factors from 0 to 1. Curves show 1998 and 2008 initial conditions (0.65 used in the edit), three rotated-offset controls and a same-size change on another tensor (dotted: unedited model).}\label{sfig10}\end{figure}
\begin{figure}[tbp]\centering\includegraphics[width=\linewidth,height=0.72\textheight,keepaspectratio]{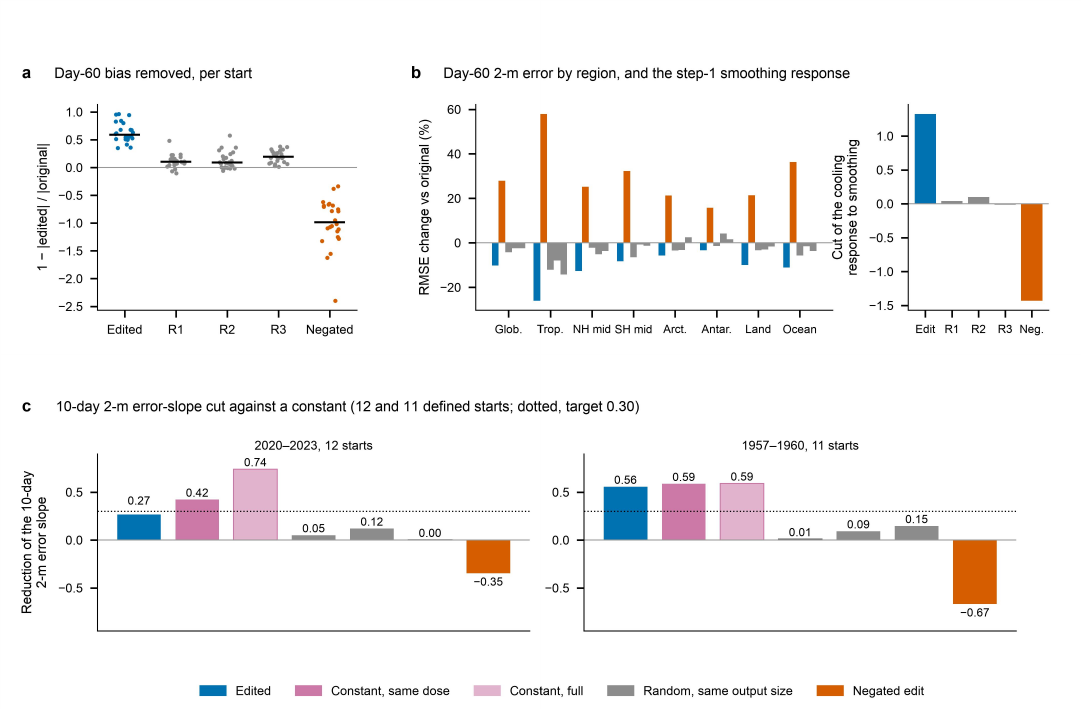}
\caption{\textbf{Inside training years the Pangu-Weather edit removes more cold bias than random changes, while outside them fixed warming matches or exceeds it.} \textbf{a}, Share of absolute day-60 2-m temperature bias removed in 24-hour Pangu-Weather\cite{r27} across 24 ERA5\cite{r7} initial conditions in 1979--1987 (medians: 0.59 edited, 0.09--0.19 for three random changes, $-$0.98 negated edit). \textbf{b}, Left, change in day-60 2-m RMSE by region (\%). Right, fractional reduction in the median extra cooling on day 1 when the model is initialised with smoothed fields. \textbf{c}, Median reduction in 10-day 2-m error slope across 12 initial conditions per out-of-sample period, where fixed warming (a constant added after each step, weights unchanged) matches or exceeds the edit. The unedited error slope was large enough to score the fractional reduction for 12 of 12 dates in 2020--2023 and 11 of 12 in 1957--1960 (dotted target 0.30).}\label{sfig11}\end{figure}

\clearpage
\section*{Supplementary Tables}
\subsection*{Supplementary Table 1. Coverage by model component}

This table lists the 21 model components with immediate multiplier sweeps across 497 weight groups (Methods, `Models and checkpoints'). For the 16 PyTorch components with a ratio, we used each model's native output fields. For each weight group, we averaged the response to $\times$0.9 scaling across the initial conditions used for mapping and divided by the median response to three random changes of the same size. For NeuralGCM\cite{r3,r4} and GraphCast-small\cite{r13,r22} (four JAX components), we used 65 shared pressure-level fields and the mean response to three random changes. Their percentages count pairs of weight group and initial condition across January and July 1995 (ERA5\cite{r7}), such as 45/54 = 83~\% and 11/48 = 23~\%. Separate counts for each initial condition appear in parentheses where they differ. A group is below the numerical floor when its scaling or random response gave no measurable change, so no ratio is defined. In `Weight-count scope', the three `Tested weights only' rows are the partial counts of Supplementary Methods (`Figure conventions'). Individual group responses, ratios and flags appear in Source Data.

{\tiny\setlength\tabcolsep{2pt}\begin{longtable}{>{\raggedright\arraybackslash\hspace{0pt}}p{0.122\linewidth}>{\raggedright\arraybackslash\hspace{0pt}}p{0.092\linewidth}>{\raggedright\arraybackslash\hspace{0pt}}p{0.072\linewidth}>{\raggedright\arraybackslash\hspace{0pt}}p{0.062\linewidth}>{\raggedright\arraybackslash\hspace{0pt}}p{0.059\linewidth}>{\raggedright\arraybackslash\hspace{0pt}}p{0.082\linewidth}>{\raggedright\arraybackslash\hspace{0pt}}p{0.072\linewidth}>{\raggedright\arraybackslash\hspace{0pt}}p{0.152\linewidth}>{\raggedright\arraybackslash\hspace{0pt}}p{0.072\linewidth}>{\raggedright\arraybackslash\hspace{0pt}}p{0.055\linewidth}}\toprule
\textbf{Component} & \textbf{Counted weights} & \textbf{Weight-count scope} & \textbf{Full tensor catalogue} & \textbf{Weight groups tested} & \textbf{Groups with finite scaling-to-random ratio} & \textbf{Groups below numerical floor} & \textbf{Share of finite-ratio groups > 1 (pairs of weight group and initial condition for JAX)} & \textbf{Groups in year-long tests} & \textbf{Edit site} \\ \midrule\endhead
ACE2-ERA5\cite{r1} & 455,831,040 & Full & Yes & 35 & 35 & 0 & 49~\% (17/35) & 35 & No \\
ACE2-SHiELD\cite{r1} & 455,831,040 & Full & Yes & 35 & 35 & 0 & 51~\% (18/35) & 0 & No \\
ACE2-EAMv3\cite{r46} & 455,824,896 & Full & Yes & 35 & 35 & 0 & 57~\% (20/35) & 0 & No \\
ACE2S-SHiELD+\cite{r19} & 800,320,000 & Full & Yes & 43 & 43 & 0 & 35~\% (15/43) & 0 & No \\
HiRO-ACE v1\cite{r47} & 456,605,184 & Full & Yes & 43 & 43 & 0 & 23~\% (10/43) & 0 & No \\
SamudrACE-CM4\cite{r32} atmosphere & 455,831,808 & Full & Yes & 35 & 35 & 0 & 60~\% (21/35) & 0 & No \\
Ocean component of SamudrACE-CM4\cite{r32} & 131,865,600 & Full & Yes & 10 & 10 & 0 & 20~\% (2/10) & 0 & No \\
SamudrACE-E3SMv3\cite{r33} atmosphere & 456,617,088 & Full & Yes & 43 & 43 & 0 & 23~\% (10/43) & 0 & No \\
Ocean component of SamudrACE-E3SMv3\cite{r33} & 82,832,864 & Full & Yes & 10 & 10 & 0 & 40~\% (4/10) & 0 & No \\
DLESyM v0\cite{r2,r51} atmosphere & 3,510,738 & Full & Yes & 12 & 0 (not available*) & 0 & not available* & 0 & No \\
DLESyM v0 ocean & 768,422 & Full & Yes & 9 & 9 & 0 & 100~\% (9/9) & 0 & No \\
DLESyM v1.0\cite{r5,r34,r42} atmosphere & 9,838,188 & Full & Yes & 12 & 12 & 0 & 83~\% (10/12) & 12 & No \\
DLESyM v1.0 ocean & 763,605 & Tested weights only & No & 9 & 9 & 0 & 100~\% (9/9) & 9 & Yes \\
DLESyM v1.1\cite{r42,r53} atmosphere & 5,093,016 & Tested weights only & No & 12 & 12 & 0 & 83~\% (10/12) & 0 & No \\
DLESyM v1.1 ocean & 4,821,560 & Tested weights only & No & 9 & 9 & 0 & 100~\% (9/9) & 0 & No \\
CAMulator\cite{r21} & 874,382,736 & Full & Yes & 9 & 7 & 2 & 57~\% (4/7) & 0 & No \\
Samudra2\cite{r23} & 84,066,474 & Full & Yes & 10 & 10 & 0 & 100~\% (10/10) & 0 & No \\
NeuralGCM 0.7$^\circ$ deterministic\cite{r3,r4} & 31,123,612 & Checkpoint metadata total & No & 38 & 38 & 0 & 79~\% (30/38) & 0 & No \\
NeuralGCM 1.4$^\circ$ stochastic\cite{r3,r4} & 11,519,783 & Full & Yes & 27 & 27 & 0 & 83~\% (22/27 Jan, 23/27 Jul) & 1 & Yes \\
NeuralGCM 2.8$^\circ$ deterministic\cite{r3,r4} & 14,518,180 & Full & Yes & 37 & 37 & 0 & 73~\% (27/37) & 0 & No \\
GraphCast-small\cite{r13,r22} & 35,979,347 & Full & Yes & 24 & 24 & 0 & 23~\% (6/24 Jan, 5/24 Jul) & 0 & No \\
\textbf{Total (21 swept components)} & \textbf{---} & \textbf{17 full, 1 metadata total, 3 tested-only} & \textbf{17 of 21 yes} & \textbf{497} & \textbf{20 components (483 groups)} & \textbf{2} & \textbf{---} & \textbf{4 components (57 groups)} & \textbf{2 edit sites} \\
\bottomrule\end{longtable}}

*The initial DLESyM v0\cite{r2,r51} atmosphere random sweep did not complete and is excluded from the 20-component summary. A later run completed all 12 groups with median ratio 2.341 (Supplementary Methods, `Weight scaling against random directions: details'). Four further surveyed components have full tensor catalogues without multiplier sweeps. The 24-hour Pangu-Weather\cite{r27} surface output-layer calibration was evaluated outside this 21-component atlas.

\subsection*{Supplementary Table 2. What each mapped layer supports: evidence, out-of-sample status, grade and use}

High: the result repeated on independent evaluation initial conditions or separated from controls. Medium: the immediate response pattern observed across several dates recurred on evaluation dates, or repeated runs reproduced the effect of an edit in one model setup. Low: the test located an effect descriptively or did not separate it from controls. Weight scaling multiplies original weights by one scalar, whereas random controls apply a change of the same size. In the grade column, the word before each grade names what is graded: `Signature' is identification by fingerprint, `Role' a physical role of the weights and `Ratio' the scaling-to-random ratio. `Own r' is the correlation with that group's single-step (6-hour) fingerprint at day 1, month 1 and year 1. These grades rate measurement support, not whether an edited model is safe for general use.

{\footnotesize\setlength\tabcolsep{3pt}\begin{longtable}{>{\raggedright\arraybackslash\hspace{0pt}}p{0.177\linewidth}>{\raggedright\arraybackslash\hspace{0pt}}p{0.142\linewidth}>{\raggedright\arraybackslash\hspace{0pt}}p{0.122\linewidth}>{\raggedright\arraybackslash\hspace{0pt}}p{0.332\linewidth}>{\raggedright\arraybackslash\hspace{0pt}}p{0.132\linewidth}}\toprule
\textbf{Weight group or edit} & \textbf{Kind of evidence} & \textbf{Out-of-sample status} & \textbf{Grade and numbers} & \textbf{Supported use} \\ \midrule\endhead
ACE2-ERA5\cite{r1}, 29 weight groups outside the year test from four initial conditions (blocks 1--8 and second encoder layer) & Immediate transfer to 2001 dates; year test of all 35 groups; three types of random control and a worst-case direction. & Four 2001 dates\cite{r1,r30}, averaged ($\times$0.5) and separate ($\times$0.9, three seeds); one 2001 date for 1 year; 1 January and 1 July 1979\cite{r7,r30}. & Signature: Medium. Physical role: Low. Immediate identification of 28 distinct groups: 28 of 28 with dates averaged (same-group r 0.83--0.99); on individual dates, 26--28 when reference and evaluation used the same weight change (scaling, or the same random change) and 3--9 when scaling was compared with random changes. Year-long identification (all 35 groups, one initial condition): 31 and 30 of 35 at day 1 (18 and 19 uncentred), falling to 2--3 at month 1 and year 1. The median scaling-to-control ratio was 0.92 for singular-value-preserving random changes (34 matrix groups). Against the largest response found by the local search, the median ratio was 0.02 (all 35 groups). Median nonlinearity for a 10~\% weight reduction was 0.08. & Controlled follow-up only; neither a climate role nor an edit is established. \\
ACE2-ERA5 spectral filters\cite{r43} by degree band (0--10, 11--40, 41--179; 24 groups) & Within-band $\times$0.9 scaling compared with same-norm random controls. & 1 January and 1 July 1979\cite{r7,r30}, three draws; 4 of 10 predictions hold. & Band locality: Medium (same-band share largest in 24 of 24; within-band ratios 1.3--3.7). Role: Low. & Attributes spatial scale of an immediate change only. \\
ACE2-ERA5, direct input link & Year-long test against two controls. & Four initial conditions (1 year); four 2001 dates\cite{r1,r30}. & Signature: High. Tuning: Low. Own r is 0.81 / 0.65 / 0.49 at day 1, month 1 and year 1. & Diagnostic probe only; $\pm$10~\% shifts climate. \\
ACE2-ERA5, block-4 filter and block-2 inner skip & Year-long test against two controls. & Four initial conditions (1 year). & Signature: Medium. Role: Low. Own r is 0.71 / 0.35 / 0.32 (filter) and 0.70 / 0.15 / 0.28 (skip). For the block-4 filter, a random change matched the sign of the day-1 change in 9 of 10 variable--region comparisons. & Probe alongside a same-size random control. \\
ACE2-ERA5, whole block 8, its filter and position embedding & Day-1 match; 6-hour recognition lost by month 1. & Four initial conditions (1 year). & Signature: Medium. Role: Low. Own r is 0.68 / 0.15 / 0.10 (block 8), 0.59 / 0.01 / $-$0.01 (filter) and 0.73 / 0.01 / 0.06 (embedding). & Probe alongside random controls. \\
Direct input link and block 8 across six ACE atmospheres\cite{r1,r19,r32,r33,r46,r47} & Descriptive transfer of immediate ranking. & One initial condition per checkpoint. & Ranking: Medium. At $\times$0.9 and $\times$0.5 the direct input link ranked first overall and block 8 ranked first among processing blocks in all six atmospheres. At $\times$0 the link ranked first in three of six. & Measure each checkpoint directly. \\
NeuralGCM 0.7$^\circ$ deterministic, 1.4$^\circ$ stochastic and 2.8$^\circ$ deterministic\cite{r3,r4} (38, 27, 37 groups; JAX) & Immediate $\times$0.9 scaling versus three random directions. & January and July 1995 ERA5\cite{r7} initial conditions, 65 shared output fields. & Ratio: Medium. Role: Low. Ratios exceed 1 in 30 of 38, 22--23 of 27 and 27 of 37 groups (medians 1.58--1.63, 2.03--2.05, 1.15--1.19; 2.8$^\circ$ surface embedding 3.3$\times$). & Group-level screen; recounted twice. \\
GraphCast-small\cite{r13,r22} (24 weight groups; JAX) & Immediate $\times$0.9 scaling versus three random directions. & January and July 1995 ERA5\cite{r7} initial conditions, 65 shared output fields. & Ratio: Medium. Role: Low. Ratios exceed 1 in 5--6 of 24 (median 0.68--0.69): grid-to-mesh and edge embedders 5--22$\times$, decoder 2.2$\times$, 16 mesh layers 0.45--0.90, grid embedder 0.37, process layers $\sim$0.95. & Group-level screen; recounted twice. \\
Output stages of the ocean components of SamudrACE\cite{r32,r33} & Descriptive single-step sweeps. & Two oceans, one initial condition each. & Role: Low. The $\times$0.5 output-to-input ratio is 1.26--1.28. & Compares stages within each ocean component. Only two output fields had matching names across the models. \\
ACE2-ERA5 forced El Ni\~no paths (direct link, main blocks, block 8) & Replacing intermediate layer outputs along one pathway with those from the warm run (activation patching\cite{r10,r31}). & 120-day runs from four January 1979 ERA5\cite{r7} initial conditions. & Pathway share: Medium (Methods, `Activation patching'). Separate weight edits are reported in Supplementary Table 3. & Locates a distributed response only. \\
DLESyM v1.0\cite{r5,r34,r42} first ocean encoder, last decoder, output layer & Descriptive year-long coupled sweep. & One 1995 ERA5\cite{r7} initial condition, no random directions. & Low. Polar SST shifts 2--3 K, and 24 of 63 coupled tensors fail at $\pm$10~\%. & Selects candidate layers only. \\
DLESyM v1.0 atmosphere, 12 stages & Immediate $\times$0.9 scaling versus three random types and worst case. & 1995 and 1998 ERA5\cite{r7} initial conditions, three draws. & Role: Low. Scaling gives 2.5--2.6$\times$ the response of random controls (worst-case and linearity values in Supplementary Methods); group rankings agree with those for the linear approximation to the forecast response. & Immediate screen; non-linear at $\pm$10~\%. \\
DLESyM v1.0 ocean edit (12 tensors recalculated by least squares, plus the SST offset at factor 0.65) & Five-run evaluation, three full-offset random controls from the 1998 initial condition and comparison of edited and unedited runs from the same eight initial conditions. & Five evaluation runs and eight 48-month comparisons, all within the training years. & Global 2-m RMSE cut of 35--38~\%: Medium. Regional safety: Low (9--26 of 80 checks fail; of all combinations of variable, region and climate statistic, 37~\% differ, 21~\% are equivalent and 42~\% are inconclusive; European SST, Sahel water vapour and the North American coastal SST cycle worsen). & Calibrates one coupled v1.0 configuration (atmosphere network 0 with ocean network 0\cite{r34,r42,r53}) with regional costs; no general optimum. \\
NeuralGCM 1.4$^\circ$ six upper-level wind columns, $\times$0.5 & Controls run from the same saved state, plus ensembles with pressure correction\cite{r4}. & Saved 1995 state, 1995 and 2015 runs, 20-year six-seed 1995 ensemble. & Monthly global means within limits, with pressure correction: Medium (6 of 6 edited runs within the limits for 20 years, against 0 of 6 unedited runs past year 4). Uniform climate: Low (one seed cools to near 200 K at 50 hPa). & Research runs with pressure correction and 50-hPa checks. \\
NeuralGCM 1.4$^\circ$ wind columns of levels 4--6, 11--13, and $\times$0.75 on levels 1--3 & Controlled localisation, control on deeper levels and weaker multiplier. & State saved on 2 March 1995 and simulation initialised on 1 January 1995. & Levels 4--6: Medium (survives 1 year). Levels 11--13: Low (onset on day 39). $\times$0.75 on levels 1--3: Low (fails in month 20). & Vertical and multiplier bounds around $\times$0.5. \\
DLESyM v0\cite{r2,r51} ocean memory; DLWP\cite{r29} edit; 24-hour Pangu-Weather\cite{r27} static offset; ACE2-ERA5 precipitation and Arctic summer tests & Tests in which the edit failed its criterion, a control matched the edit or the baseline bias was too small to evaluate an improvement (Supplementary Table 3; Methods, `Statistics'). & Limited initial conditions. & Low. Uncoupled DLESyM v0 ocean memory is 51 versus 161 days coupled (the calculated mean SST shift for 25~\% longer memory was 1.6--3.6 K RMS over 60$^\circ$ S--60$^\circ$ N). In DLWP, one random control cut the year-2 bias from $-$8.11 to $-$1.29 K, matching the edit ($-$1.28 K); a second reached $-$11.39 K. Pangu static offset left trajectories unchanged; other values in Methods (`Statistics'). & Negative or unresolved cases. \\
FourCastNet v3\cite{r28} output-stage collapse direction & Breakdown-mode projection, not an accepted edit. & Screened at 0.25--0.75 (four initial conditions); validated at 0.03 (six seeds, two new initial conditions). & Lifetime: Medium (about 2.49$\times$ across both initial conditions, more than random controls of the same operator norm). Regional climate: Low (for four additional initial conditions, the edit increased the first-October warm bias north of 60$^\circ$ N by 4.9--6.8 K relative to the unedited runs; relaxation damping adds a 108~\% sea-level-pressure penalty; Supplementary Table 3). & Breakdown-mode study with regional costs. \\
Pangu-Weather surface output layer, 16 2-m rows (shared 384-vector, 6,144 weights) & Edit estimated from forecasts with original and smoothed inputs, without using forecast errors. Controls were fixed warming and random vectors matched for the size of the first-day temperature response. & 24 training-period (1979--1987) and 24 out-of-sample initial conditions (12 each in 2020--2023, 1957--1960; 216 runs). & Against random weight changes in training years: Medium (Methods, `Pangu-Weather surface-head edit'). Cause of the cooling: Low (fixed warming matches or exceeds the 10-day drift reduction; Supplementary Fig. 11). & Temperature calibration applied at every step of chained 24-h forecasts. \\
\bottomrule\end{longtable}}
\subsection*{Supplementary Table 3. Weight edits, controls and diagnostic comparisons}

The table summarises the tests cited in the text. For the 10-day weather benchmark, a negative score is a reduction in forecast error. Its field and lead-time limits and its ensemble criteria are defined in Methods, `Forecast edits in weather models'.

{\footnotesize\setlength\tabcolsep{3pt}\begin{longtable}{>{\raggedright\arraybackslash\hspace{0pt}}p{0.142\linewidth}>{\raggedright\arraybackslash\hspace{0pt}}p{0.202\linewidth}>{\raggedright\arraybackslash\hspace{0pt}}p{0.202\linewidth}>{\raggedright\arraybackslash\hspace{0pt}}p{0.372\linewidth}}\toprule
\textbf{Model} & \textbf{What we tested} & \textbf{Criterion or comparison} & \textbf{Result} \\ \midrule\endhead
ACE2-ERA5\cite{r1} & Edits to the final-block skip, position embedding and final-block spectral weights & Compare climatological RMSE with the unedited model & Climatological RMSE rose to 1.9--10 times its unedited value. \\
ACE2-ERA5\cite{r1} & Block 1 $\times$0.75 for tropical precipitation & Reduce error more than the half-dose and all random controls, within regional limits & Error fell by 0.094 mm d$^{-1}$, against 0.177 and 0.172 for two random controls. Antarctic winter error rose by 0.95 K. \\
ACE2-ERA5\cite{r1} & Arctic summer 850-hPa temperature & Unedited bias above the 0.451 K detection threshold & The 0.299 K baseline bias was too small to judge an improvement. \\
ACE2-SHiELD\cite{r1} & Hold CO$_2$ fixed, 1979--2014 & Compare 1995--2014 trends with time-varying CO$_2$ & The 500-hPa geopotential trend fell from 32.51 to 4.48 m$^{2}$ s$^{-2}$ per decade. \\
DLESyM v0\cite{r2,r51} & Ocean memory and linear ocean input edits & Increase eastern-Pacific memory without shifting mean SST & Memory was 51 days uncoupled and 161 days coupled. For linear edits sized to increase memory by 25~\%, the calculated shift in mean SST was 1.6--3.6 K RMS over 60$^\circ$ S--60$^\circ$ N. \\
DLESyM v1.0\cite{r5,r34,r42} & Rank-one ocean output edits and SST lower bounds & Stay within physical limits and meet climate checks & The three rank-one edits ran away. SST lower bounds failed 27, 13 and 22 of 80 checks. \\
DLESyM v1.0\cite{r42} & Recalculated ocean 0 with atmosphere 1 or 0 & Compare climatological errors and 80 climate checks & The planned pairing with atmosphere 1 failed. With atmosphere 0, chosen afterwards, all five errors fell in 1986--1989 and 1993--1996, with 16 and 23 checks failed. \\
DLESyM v1.0 and v1.1\cite{r42} & Vary the SST offset, then transfer the 0.65 factor & Look for a monotone error reduction and test Southern Ocean SST & The v1.0 tests did not identify an optimal factor. In v1.1 absolute annual Southern Ocean SST bias rose by 0.57--1.43 K. \\
NeuralGCM 1.4$^\circ$ stochastic\cite{r3,r4} & Change upper-level wind outputs & Suppress divergence while keeping realistic stratospheric winds & Zeroing all 18 upper-level columns caused runaway. Zeroing the six wind columns raised wind to 153 m s$^{-1}$ (limit 150). Halving those six met both one-year limits. \\
NeuralGCM entry in AIMIP\cite{r8} & Compare two releases of the AIMIP\cite{r8} output & Check whether excess precipitation variability depends on the release & The ratio to reference variability was 3.69 and 4.12 in the later release, versus 1.21 and 1.22 in the earlier release, for 1995--2014 and 2015--2024. The two data releases contain different simulations, leaving the cause of their difference unresolved. \\
Pangu-Weather, 24 h\cite{r27} & Static temperature offset and alternative weight edits & Change forecast evolution and reproduce drift reduction on separate dates & A static input--output offset left forecast trajectories unchanged. A perceptron edit gave no measurable reduction, and an open-ocean fit did not transfer between development subsets. \\
Pangu-Weather, 24 h\cite{r27} & Edit the surface temperature output weights & Compare with random edits in training years and with a constant correction outside them & The 60-day test passed within the training years. The edit reduced drift in both periods outside them, but the constant correction matched or exceeded it at 2 m and 850 hPa (Supplementary Fig. 11c; Supplementary Methods, `Pangu-Weather surface-head edit: details'). \\
FourCastNet v3\cite{r28} & Relax output scales & Extend lifetime without increasing climate errors & The run reached 500 days against 380 unedited. Climatological RMSE increased by 108~\% for sea-level pressure and 26~\% for 500-hPa geopotential. \\
FourCastNet v3\cite{r28} & Remove part of the output direction associated with breakdown & Extend lifetime beyond size-matched controls while preserving regional climate & At a fraction of 0.03 lifetime reached about 2.49 times the unedited value at two initial conditions. The first-October warm bias north of 60$^\circ$ N grew by 4.9--6.8 K on four other dates. \\
DLWP\cite{r29} & Advance solar time with forecast time & Separate a forcing-time error from a weight error & Holding solar time at the initial value gave a +1.2 K global 2-m bias in years 2--10. Advancing it gave $-$7.99 K. \\
DLWP\cite{r29} & Rank-one edit for cold bias & Outperform size-matched random edits & Year-2 bias changed from $-$8.11 to $-$1.28 K. One control reached $-$1.29 K and the other $-$11.39 K. \\
GraphCast-small\cite{r13,r22} & Test linear growth directions & Compare growing modes with the sea-level-pressure drift & The modes concentrated in upper-air temperature and geopotential. Their cosine similarity with the pressure drift was at most 0.13. \\
Aurora 1.5\cite{r17,r35} & Scale attention and offset pressure drift & Meet the weather benchmark and compare with forecast relaxation & Error fell 0.79~\% across 42 evaluation dates. Relaxation alone cut error 4.15~\% on 12 held-out dates. On development dates error fell 0.85~\% (1~\% required) and two of the 28 numbers exceeded the 0.5~\% limit. \\
GraphCast operational\cite{r13,r22} & Scale 2-m temperature output weights & Same benchmark and relaxation comparison & Error fell 0.44~\% on 36 dates. On 12 held-out dates relaxation alone cut error 2.70~\%, with a further 0.24~\% from the edit. \\
AIFS Single 2.0\cite{r36,r37} & Scale the output increment and check missing wave inputs & Same benchmark under consistent inputs & The increment edit cut error 0.66~\% on held-out dates. After relaxation its development change was $-$0.27~\% (interval $-$0.57 to +0.02~\%). Two different edits that appeared to pass with substitute inputs raised error 1.5--3.6~\% using operational analyses\cite{r66}. \\
Pangu-Weather, 6 h\cite{r27} & Adjust the output biases using the day-10 development bias & Meet the overall and field-specific limits & Error fell 1.31~\% on held-out dates, but day-10 meridional wind error rose 1.04~\%, above the 0.5~\% limit. \\
Aurora 1.5 Ensemble\cite{r17,r35}, WeatherNext 2\cite{r41} and AIFS ENS v2\cite{r38,r39,r40} & Change noise amplitude or noise-conditioning weights & Improve fair ensemble score\cite{r58} on held-out dates within the error limits & Both tested amplitudes worsened Aurora. WeatherNext 2 changed the held-out score by $-$0.26~\% (90~\% interval $-$1.19 to +0.64~\%). AIFS ENS gained on ERA5 initial conditions but lost 1.17~\% on six operational ensemble initial conditions at 24--72 h. \\
\bottomrule\end{longtable}}
\subsection*{Supplementary Table 4. What the cross-audits caught}
Each row is a failed check or an error found by the review steps of Methods (`Workflow, preregistration and verification').

{\footnotesize\setlength\tabcolsep{3pt}\begin{longtable}{>{\raggedright\arraybackslash\hspace{0pt}}p{0.316\linewidth}>{\raggedright\arraybackslash\hspace{0pt}}p{0.299\linewidth}>{\raggedright\arraybackslash\hspace{0pt}}p{0.316\linewidth}}\toprule
\textbf{Issue caught} & \textbf{How it was caught} & \textbf{Action taken} \\ \midrule\endhead
29 of 41 year-scale predictions failed, and only 4 of 12 day-1 fingerprint checks with fixed correlation thresholds passed. & Preregistered scoring script applied the fixed rules. & All failures are kept in Supplementary Figs. 6b and 7a, b; closest-match identification without a correlation threshold is reported separately. \\
5 of 10 offset-factor curve predictions failed for the 1998 initial condition, and the unedited model's error in the 2008 run was below the minimum required for the test. & Scoring script evaluated the pre-set conditions in order. & No optimal factor or factor band is claimed; Supplementary Fig. 10b shows all tested factors. \\
The initial ACE2-ERA5\cite{r1,r43} benchmark failed the native spectral-layout check (expected maximum order dimension 180, measured 181). & Run-time check and reviewer audit. & We updated the layout check and reran the benchmark. \\
Two NeuralGCM\cite{r3,r4} runs sharing seed 1 diverged within the first month. & Independent recount of raw output files by two reviewers. & Methods and Results treat them as separate stochastic trajectories across GPU jobs. \\
Two issues were not weight-edit targets: a solar-time routine and a difference between two releases of one model's output. & Code and input inspection during the campaign. & Both are logged in Supplementary Table 3 outside the weight-edit claims. \\
\bottomrule\end{longtable}}

We ran no agent-versus-human comparison. This table records what the review rules caught, not agent capability.

\subsection*{Supplementary Table 5. Model checkpoints and settings}

The SHA-256 checksum identifies the weight file used. Rows 1--30 cover the atlas, including separate release files for individual DLESyM networks. Rows 31--35 cover reference models and edit tests. Rows 36--42 identify the seven models in the 10-day weather benchmark.

{\footnotesize\setlength\tabcolsep{3pt}\begin{longtable}{>{\raggedright\arraybackslash\hspace{0pt}}p{0.041\linewidth}>{\raggedright\arraybackslash\hspace{0pt}}p{0.275\linewidth}>{\raggedright\arraybackslash\hspace{0pt}}p{0.301\linewidth}>{\raggedright\arraybackslash\hspace{0pt}}p{0.301\linewidth}}\toprule
\textbf{No.} & \textbf{Checkpoint} & \textbf{Grid, step or forecast setting} & \textbf{SHA-256} \\ \midrule\endhead
1 & ACE2-ERA5\cite{r1,r30} & 180 $\times$ 360 Gaussian ($\sim$1$^\circ$), 6 h & \texttt{\seqsplit{0e2522df5a8db087f18bcebea6e88277cea7c3b71892b868c427aca27673edc2}} \\
2 & ACE2-SHiELD\cite{r1} & 180 $\times$ 360 Gaussian ($\sim$1$^\circ$), 6 h & \texttt{\seqsplit{2debfee21a4cee81f34631c81539a7f312b9ea67a73cebd0f9b00cbe3cc23482}} \\
3 & ACE2-SOM\cite{r45} & 180 $\times$ 360 Gaussian ($\sim$1$^\circ$) & \texttt{\seqsplit{b191174e45ac16d86f3a5ccfa99b8952866ec6482b19e08edbd0f05114f5f9ab}} \\
4 & ACE2-EAMv3\cite{r46} & 180 $\times$ 360 Gaussian ($\sim$1$^\circ$), 6 h & \texttt{\seqsplit{99c0bc44966394aab2a7bc91241893a115c163b57965af1717c5ac0e9bbf0e06}} \\
5 & ACE2S-SHiELD+\cite{r19} & 180 $\times$ 360 Gaussian ($\sim$1$^\circ$), 6 h & \texttt{\seqsplit{4d11bf3ceb714d2c0ca971a5a7c1304efeab01f70edfe8a318e7816af0fbab25}} \\
6 & HiRO-ACE v1 (ACE2S coarse emulator)\cite{r47} & 180 $\times$ 360 Gaussian ($\sim$1$^\circ$), 6 h & \texttt{\seqsplit{0b39d182074c02c92d6700088dfe5f6f1ae9488d50eb4a260564121ec2dcbf0e}} \\
7 & HiRO-ACE v2 (downscaler checkpoint)\cite{r48} & 5760 $\times$ 11520 static input grid & \texttt{\seqsplit{1b99f84b31f43151b306d32dc7d0dca68cc6bb131050938ec70172b6cfa140b2}} \\
8 & SamudrACE-CM4 (coupled atmosphere and ocean)\cite{r32} & Atmosphere 180 $\times$ 360 ($\sim$1$^\circ$), 6 h, ocean 180 $\times$ 360 (1$^\circ$), 120 h (5 d) & \texttt{\seqsplit{cc20dcc8ec4254729f42acad7c2d55efc1b61a118f4370f730e537cdae8c2b79}} \\
9 & SamudrACE-E3SMv3 atmosphere\cite{r33} & 180 $\times$ 360 Gaussian ($\sim$1$^\circ$), 6 h & \texttt{\seqsplit{166c7dee86bf03801445ae3cb5ed8e875aa5605ff8d80596f5154c1d8bdf2f59}} \\
10 & SamudrACE-E3SMv3 ocean\cite{r33} & 180 $\times$ 360 (1$^\circ$), 120 h (5 d) & \texttt{\seqsplit{4a28ffa0f16e08632a642121c1c647fef2df1839a2704dc174af14cd72350e02}} \\
11 & DLESyM v0 atmosphere (ISCCP-ERA5)\cite{r2,r51} & HEALPix NSIDE = 64 (12 $\times$ 64 $\times$ 64), 6 h substep (96 h coupled call) & \texttt{\seqsplit{a930bbfbcbe6a1e0c7d76b493bdff33b564d734f40a1f735b1709c0603ad0f5f}} \\
12 & DLESyM v0 ocean (ISCCP-ERA5)\cite{r2,r51} & HEALPix NSIDE = 64 (12 $\times$ 64 $\times$ 64), 48 h output slot (96 h call) & \texttt{\seqsplit{57340a519ecf760cac70c32be42c6fb44b01b3f82f3139d1e6b2f97b36f84363}} \\
13 & DLESyM v1.0 atmosphere, member 0 (default)\cite{r5,r34,r42} & HEALPix NSIDE = 64 (12 $\times$ 64 $\times$ 64), 6 h substep (96 h coupled call) & \texttt{\seqsplit{e2b4064f202997eafcf751754466e19475a3900f03fb3647d3163721f64d6cf0}} \\
14 & DLESyM v1.0 atmosphere, member 1\cite{r42} & HEALPix NSIDE = 64 (12 $\times$ 64 $\times$ 64), 6 h substep (96 h coupled call) & \texttt{\seqsplit{c85ae4367c0bbdd869ede7724291e718d30edb4b7c0ab82870f82d08af710c1c}} \\
15 & DLESyM v1.0 atmosphere, member 2\cite{r42} & HEALPix NSIDE = 64 (12 $\times$ 64 $\times$ 64), 6 h substep (96 h coupled call) & \texttt{\seqsplit{e700a370d2626cee5f5b3f8fdaa3d8f74d173b968e8894fa8eaca94913fb4c15}} \\
16 & DLESyM v1.0 atmosphere, member 3\cite{r42} & HEALPix NSIDE = 64 (12 $\times$ 64 $\times$ 64), 6 h substep (96 h coupled call) & \texttt{\seqsplit{10386d365fd79ed3a84a09afe9bf983486763ae19886afee5f62d3be9bac2915}} \\
17 & DLESyM v1.0 ocean, member 0 (default, edit site)\cite{r42} & HEALPix NSIDE = 64 (12 $\times$ 64 $\times$ 64), 48 h (96 h coupled call) & \texttt{\seqsplit{42ef6740811ef75246c712dc2532caade55b865aace0f5139fd4f508e55b4e0b}} \\
18 & DLESyM v1.0 ocean, member 1\cite{r42} & HEALPix NSIDE = 64 (12 $\times$ 64 $\times$ 64), 48 h (96 h coupled call) & \texttt{\seqsplit{7db0ec70a70f58c1d3a39b71df6613d1db8081958ddb4837864ceafd5dc3b807}} \\
19 & DLESyM v1.0 ocean, member 2\cite{r42} & HEALPix NSIDE = 64 (12 $\times$ 64 $\times$ 64), 48 h (96 h coupled call) & \texttt{\seqsplit{472f42408f5f0c6f37b6472a27c15778b8e0ae010a0ee7436b2200d321b31cbc}} \\
20 & DLESyM v1.0 ocean, member 3\cite{r42} & HEALPix NSIDE = 64 (12 $\times$ 64 $\times$ 64), 48 h (96 h coupled call) & \texttt{\seqsplit{3b605790051e87caf1a9c6a9a71bea54ff368a9c8ac741c96bbd4dbf5fde3582}} \\
21 & DLESyM v1.1 atmosphere\cite{r42} & HEALPix NSIDE = 64 (12 $\times$ 64 $\times$ 64), 6 h substep (96 h coupled call) & \texttt{\seqsplit{d3752ccd163c0157263f1dcdb722824cb4a2cca7368b07050172a8ba632a0950}} \\
22 & DLESyM v1.1 ocean\cite{r42} & HEALPix NSIDE = 64 (12 $\times$ 64 $\times$ 64), 48 h (96 h coupled call) & \texttt{\seqsplit{0f88299adef56d7fbfba500c2656ebba3bc13da70b150a6dbb00fed82f8715e4}} \\
23 & CAMulator\cite{r21} & 192 $\times$ 288 ($\sim$1$^\circ$, 32 levels), 6 h & \texttt{\seqsplit{3c0b5f89784c1e0456cfd930f416e49b96adb0eb55d9063c11c2a3e6eb30a547}} \\
24 & Samudra2 (1$^\circ$)\cite{r23} & 1$^\circ$ global ocean (5-day OM4 data) & \texttt{\seqsplit{ff6160cb7e2a2b5b8ee2c5731c16bd114a1bb5a267245cc5ed766fb00c7f797a}} \\
25 & NeuralGCM 0.7$^\circ$ deterministic\cite{r3,r4} & 0.7$^\circ$ Gaussian, 1 h & \texttt{\seqsplit{443999a3baee835518a49cf955c17ad89d2bdb18c54ede1f1fd2301df79588dd}} \\
26 & NeuralGCM 1.4$^\circ$ stochastic (edit site)\cite{r3,r4} & 1.4$^\circ$ Gaussian, 1 h & \texttt{\seqsplit{e766c193b12bdf2544681eac51e7ffb6aef560856cb6c683f779fb0eae202ca8}} \\
27 & NeuralGCM 2.8$^\circ$ deterministic\cite{r3,r4} & 2.8$^\circ$ Gaussian, 1 h & \texttt{\seqsplit{bdec1b4612c7385fc492aa031db252c66fb74b788a7efbb118b8b60b06644d3e}} \\
28 & GraphCast-small\cite{r13,r22} & 1.0$^\circ$ (181 $\times$ 360, 13 pressure levels), 6 h & \texttt{\seqsplit{e9438d8ad31ca6e1d3397a33b2508f4bbb6ec16aed84629f9a64712bf756bc29}} \\
29 & FloeNet-OM4\cite{r24} & --- & \texttt{\seqsplit{4b1b3e82218374f84c4311febb616246551a8d68b8025cdebb85e338f39e955e}} \\
30 & Monthly diffusion (MD-1.5 v0.9, legacy checkpoint)\cite{r25,r26} & 1.5$^\circ$ (121 $\times$ 240), 1 month & \texttt{\seqsplit{7a217274bcb83d2d262412ccf21358d45dac4d4bac6fdb37c42da7f7339e2da2}} \\
31 & NeuralGCM 2.8$^\circ$ stochastic-precipitation\cite{r49,r50} & 2.8$^\circ$ Gaussian & \texttt{\seqsplit{59f65398d4481af0a9fd3d94e07e6ee5c818c614da07940a162d97070af17153}} \\
32 & NeuralGCM 1.4$^\circ$ deterministic\cite{r3,r4} & 1.4$^\circ$ Gaussian & \texttt{\seqsplit{a4f298af6e712fcb0da3c31df2b04a66e748eadb19f00e9e5fbeab90fc9edacd}} \\
33 & Pangu-Weather\cite{r27} (24 h) & 0.25$^\circ$ (721 $\times$ 1440, 13 pressure levels + 4 surface fields), 24 h & \texttt{\seqsplit{613a5c140a1399abcaffb4dbce32af373a1f5f56c515704f5be61925bb9fdcfd}} \\
34 & FourCastNet v3\cite{r28} & 0.25$^\circ$ (721 $\times$ 1440), 6 h & \texttt{\seqsplit{5e0d7cc69f00e04dfc3cf0a97736abef578b7b2437702a5888edac869f724b2e}} \\
35 & DLWP\cite{r29} & 6 $\times$ 64 $\times$ 64 cubed sphere (0.25$^\circ$ 721 $\times$ 1440 interface), 6 h & \texttt{\seqsplit{d5fca730f37fd2b5d439a5cc0a17d728d351b785c6fc865305fb8f945ed8b670}} \\
36 & Aurora 1.5\cite{r17,r35} & 0.25$^\circ$, deterministic & \texttt{\seqsplit{25aaf2bccc04e2df30006ebc539ebb2f9aee5ac64cbfbe814f6c0a441850d285}} \\
37 & Pangu-Weather\cite{r27} (6 h) & 0.25$^\circ$, 40 steps & \texttt{\seqsplit{8a4c022a7dd65fdc235757ba907574bc9a1fc0223a2ea4a4800961a52f475e97}} \\
38 & AIFS Single 2.0\cite{r36,r37} & Deterministic & \texttt{\seqsplit{4e23ef40157b9e99b5228e5dcffb0d0d5a789f019ffcc89033ee9caae5d7ff6b}} \\
39 & GraphCast operational\cite{r13,r22} & 0.25$^\circ$, 13 pressure levels & \texttt{\seqsplit{8f485e84343072785522f6d0696c198c7712e9f0f14070a88c26a9ff35db75a9}} \\
40 & Aurora 1.5 Ensemble\cite{r17,r35} & Four forecast members & \texttt{\seqsplit{d99b9192a39a32d79ee5ffa2bbebefce8abaff1c395399cc07cc72132bef3b71}} \\
41 & AIFS ENS v2\cite{r38,r39,r40} & Eight forecast members & \texttt{\seqsplit{e383027ee9a20469b920ed11f9af2ecc9268ebcf826447638c5c9fa296fbbbdb}} \\
42 & WeatherNext 2\cite{r41} & Eight forecast members & \texttt{\seqsplit{0cb95d92e45cccc698a10b5ca7cc339e3e15a450d689792351eb6044ced018bb}} \\
\bottomrule\end{longtable}}
\subsection*{Supplementary Table 6. The 41 predictions of the year-scale test and their outcomes}

Each row summarises one prediction about ACE2-ERA5\cite{r1}. A row passed only if every check passed. Field rankings use global changes on day 1 unless another window is given. Regional rankings also test which field--region changes stand out after removing the pattern shared across edits. Lower spherical-harmonic degrees mean larger spatial scales. A detectable change is at least two standard errors from zero across four initial conditions. Sign checks required at least 80~\% agreement on day 1 and 70~\% in month 1 for rows 4, 5, 11, 17, 23, 28 and 33. Response ratios compare $\times$1.1 with $\times$0.9, and paired time thresholds refer to day 1 and month 1, in that order.

{\footnotesize\setlength\tabcolsep{3pt}\begin{longtable}{>{\raggedright\arraybackslash\hspace{0pt}}p{0.049\linewidth}>{\raggedright\arraybackslash\hspace{0pt}}p{0.341\linewidth}>{\raggedright\arraybackslash\hspace{0pt}}p{0.343\linewidth}>{\raggedright\arraybackslash\hspace{0pt}}p{0.113\linewidth}>{\raggedright\arraybackslash\hspace{0pt}}p{0.059\linewidth}}\toprule
\textbf{No.} & \textbf{Weight group and multiplier} & \textbf{What was predicted} & \textbf{Checks passed / total} & \textbf{Outcome} \\ \midrule\endhead
1 & Direct input link, $\times$0.9 & Five moisture fields supply day 1's top three and $\geq$2 of month 1's top five. & 1/2 & failed \\
2 & Direct input link, $\times$0.9 & Antarctic response $\geq$1.3 times southern mid-latitudes, with surface pressure and temperature standing out. & 3/4 & failed \\
3 & Direct input link, $\times$0.9 & Day 1 peaks at degrees 1--3. Degrees 0--3 carry $\geq$60~\% of month-1 power. & 3/6 & failed \\
4 & Direct input link, $\times$0.9 & Tropics cool and dry, poles warm and moisten. Moisture responds most near the surface. & 2/3 & failed \\
5 & Direct input link, $\times$0.9 & Top-level winds shift eastward in tropics, westward in southern mid-latitudes. Precipitation increases mainly in mid-latitudes. & 2/4 & failed \\
6 & Direct input link, $\times$0.9 and $\times$1.1 & Month-1 tropical 850-hPa air $\geq$1 K cooler at $\times$0.9 and $\geq$1 K warmer at $\times$1.1. & 5/5 & passed \\
7 & Direct input link, $\times$0.9 and $\times$1.1 & Response ratios of 0.93--1.07 then 0.85--1.15, with day-1 fingerprint correlation $\geq$0.8. & 1/3 & failed \\
8 & Whole block 8, $\times$0.9 & Day-1 top three include $\geq$2 among upper-level water (two levels) and surface latent heat flux. & 1/1 & passed \\
9 & Whole block 8, $\times$0.9 & Moisture transport ranks high, selected radiation fields low. Nonpolar response $\geq$1.4 times polar response. & 2/3 & failed \\
10 & Whole block 8, $\times$0.9 & Day-1 power peaks at degrees 4--15, which carry 31--51~\% of the total. & 4/6 & failed \\
11\textsuperscript{\dag} & Whole block 8, $\times$0.9 & Lower air cools, humidity and precipitation rise, and moisture transport adds water across regions. & 2/12 & failed \\
12 & Whole block 8, $\times$0.9 and $\times$1.1 & Month-1 global 850-hPa air $\geq$0.3 K cooler at $\times$0.9 and $\geq$0.3 K warmer at $\times$1.1. & 7/7 & passed \\
13 & Whole block 8, $\times$0.9 and $\times$1.1 & Comparable response sizes. Fingerprint correlations $\geq$0.8 on day 1 and $\geq$0.5 in month 1. & 3/5 & failed \\
14 & Block-8 spectral filter, $\times$0.9 & Day-1 top three include $\geq$2 among upper-level water (two levels) and surface latent heat flux. & 1/1 & passed \\
15 & Block-8 spectral filter, $\times$0.9 & Moisture or Antarctic height and pressure rank high, winds low. Tropical response $\geq$1.2 times Arctic. & 3/3 & passed \\
16 & Block-8 spectral filter, $\times$0.9 & Day-1 power peaks at degrees 4--15 (42--62~\%). The global mean contributes $\leq$5~\%. & 4/6 & failed \\
17\textsuperscript{\dag} & Block-8 spectral filter, $\times$0.9 & Lower air cools outside polar regions. Moisture transport adds water. Upper-level moisture responds most. & 3/13 & failed \\
18 & Block-8 spectral filter, $\times$0.9 & Month-1 tropical 850-hPa cooling $\geq$0.1 K, with $\geq$30 detectable field--region changes. & 1/2 & failed \\
19 & Block-8 spectral filter, $\times$0.9 and $\times$1.1 & Response ratios below the direct input link's both times, fingerprint correlations $\geq$0.8 then $\geq$0.5. & 4/6 & failed \\
20 & Block-4 spectral filter, $\times$0.9 & Day-1 top three include $\geq$2 among upper-level water, surface sensible heat and latent heat fluxes. & 0/1 & failed \\
21 & Block-4 spectral filter, $\times$0.9 & Northern reflected sunlight and sensible heat flux stand out. Northern mid-latitude response $\geq$1.3 times Antarctic. & 2/3 & failed \\
22 & Block-4 spectral filter, $\times$0.9 & Degrees 4--63 carry $\geq$65~\% of day-1 power. Degrees 0--3 carry $\leq$15~\%. & 5/5 & passed \\
23 & Block-4 spectral filter, $\times$0.9 & 850-hPa air cools across regions. Tropical latent heat flux rises and near-surface moisture falls. & 2/2 & passed \\
24 & Block-4 spectral filter, $\times$0.9 and $\times$1.1 & $\leq$30 detectable month-1 changes, day-1 response ratio 0.78--0.95, fingerprint correlations $\geq$0.7 then $\geq$0.3. & 3/4 & failed \\
25 & Block-2 inner skip, $\times$0.9 & At least two upper-level moisture fields (levels 1--3) among the three largest day-1 changes. & 1/1 & passed \\
26 & Block-2 inner skip, $\times$0.9 & Largest regional response $\leq$1.25 times smallest. Antarctic wind, moisture and polar radiation stand out. & 1/3 & failed \\
27 & Block-2 inner skip, $\times$0.9 & Day-1 power peaks at degrees 16--63 (31--51~\%), with $\leq$15~\% at degrees 0--3. & 4/5 & failed \\
28 & Block-2 inner skip, $\times$0.9 & Surface and 850-hPa air warm, mid-tropospheric air cools and global surface pressure falls. & 2/2 & passed \\
29 & Block-2 inner skip, $\times$0.9 and $\times$1.1 & $\leq$20 detectable month-1 changes, day-1 response ratio 0.90--1.00, fingerprint correlations $\geq$0.7 then $\geq$0.3. & 2/4 & failed \\
30 & Position embedding, $\times$0.9 & Day-1 top three include $\geq$2 among upper-level water (two levels) and surface latent heat flux. & 1/1 & passed \\
31 & Position embedding, $\times$0.9 & Mid-latitude response $\geq$1.15 times the tropical--Arctic average. The 500-hPa height response stands out. & 2/3 & failed \\
32 & Position embedding, $\times$0.9 & Day-1 power shares meet three band ranges, with $\leq$15~\% at degrees 0--3. & 1/4 & failed \\
33 & Position embedding, $\times$0.9 & Latent heat flux, precipitation and upper-tropospheric temperature rise. Polar 2-m air cools. & 2/2 & passed \\
34 & Position embedding, $\times$0.9 and $\times$1.1 & $\geq$30 detectable month-1 changes, response ratios 0.82--0.98 then $\leq$1, fingerprint correlations $\geq$0.8 then $\geq$0.5. & 2/5 & failed \\
35 & All six groups, $\times$0.9 & Day-1 fingerprint correlations reach 0.7 or 0.8, depending on group, and 0.9 without centring. & 4/12 & failed \\
36 & All six groups, $\times$0.9 & Month-1 fingerprints meet their correlation thresholds (0.3 or 0.5) and identify all six groups. & 4/12 & failed \\
37 & All six groups, $\times$0.9 & Year-1 fingerprints identify $\geq$4 groups, two retain correlations $\geq$0.3, two show few detectable changes. & 1/5 & failed \\
38 & Direct input link, whole block 8, block-8 spectral filter and position embedding, $\times$0.9 & $\geq$70~\% of detectable signed month-1 changes retain sign and 0.5--3$\times$ magnitude at year 1. & 4/4 & passed \\
39 & All six groups, $\times$0.9 and $\times$1.1 & Opposite signs and opposing month-1 850-hPa temperature patterns, with smaller $\times$1.1 responses in four groups. & 13/22 & failed \\
40 & All six groups vs comparison controls, $\times$0.9 & Edits outperform comparison groups by $\geq$0.30 in fingerprint correlation. Control signs agree 30--70~\%. & 19/24 & failed \\
41 & Five comparison groups with single-step fingerprints, $\times$0.9 & Each comparison group matches its single-step fingerprint with correlations $\geq$0.7 then $\geq$0.5. & 3/10 & failed \\
\multicolumn{2}{l}{\textbf{Total}} & \textbf{12 of 41 predictions passed} & \textbf{130/226} &  \\
\bottomrule\end{longtable}}

\textit{Five control predictions passed 27 of 36 checks. One prediction passed, three failed and one was descriptive. \textsuperscript{\dag} Rows 11 and 17 drew on probes without the model's moisture-budget correction\cite{r1,r44}. Their moisture-transport checks count as failures. Row 39's denominator includes one check with too few eligible changes to score.}

\end{document}